\documentclass{aa}
\usepackage{graphicx,txfonts,natbib}

\newcommand{\chisq}{\ensuremath{\mathrm{\chi^2}}}
\newcommand{\chisqred}{\ensuremath{\mathrm{\chi^2_\mathrm{red}}}}
\newcommand{\Av}{\ensuremath{A_\mathrm{V}}}
\newcommand{\BH}{\ensuremath{\mathrm{BH}}}
\newcommand{\MBH}{\ensuremath{M_\mathrm{BH}}}
\newcommand{\rBH}{\ensuremath{r_\mathrm{BH}}}
\newcommand{\rhoBH}{\ensuremath{\rho_\mathrm{BH}}}
\newcommand{\Msun}{\ensuremath{\mathrm{M}_\odot}}
\newcommand{\vsys}{\ensuremath{v_\mathrm{sys}}}
\newcommand{\ML}{\ensuremath{M/L}}
\newcommand{\vel}{\ensuremath{\langle v\rangle}}
\newcommand{\velsq}{\ensuremath{\langle v^2\rangle}}
\newcommand{\Rcore}{\ensuremath{R_\mathrm{core}}}
\renewcommand{\deg}{\ensuremath{\,\mathrm{deg}}}
\newcommand{\ERG}{\ensuremath{\,\mathrm{erg}}}
\newcommand{\CM}{\ensuremath{\,\mathrm{cm}}}
\newcommand{\KM}{\ensuremath{\,\mathrm{km}}}
\newcommand{\MIC}{\ensuremath{\,\mu\mathrm{m}}}
\newcommand{\SEC}{\ensuremath{\,\mathrm{s}}}
\newcommand{\DEG}{\ensuremath{\,\mathrm{deg}}}
\newcommand{\1}{\ensuremath{^{-1}}}
\newcommand{\2}{\ensuremath{^{-2}}}
\newcommand{\3}{\ensuremath{^{-3}}}
\newcommand{\KMS}{\KM\SEC\1}
\newcommand{\ten}[1]{\ensuremath{10^{#1}}}
\newcommand{\xten}[1]{\times\ensuremath{10^{#1}}}
\newcommand{\MPC}{\,\ensuremath{\mathrm{Mpc}}}
\newcommand{\PC}{\,\ensuremath{\mathrm{pc}}}
\newcommand{\YR}{\,\ensuremath{\mathrm{yr}}}
\newcommand{\HA}{\ensuremath{\mathrm{H}\alpha}}
\newcommand{\PaA}{\ensuremath{\mathrm{Pa}\alpha}}
\newcommand{\PaB}{\ensuremath{\mathrm{Pa}\beta}}
\newcommand{\NII}{\ensuremath{\mathrm{[N\,II]}}}
\newcommand{\CI}{\ensuremath{\mathrm{[C\,I]}}}
\newcommand{\OII}{\ensuremath{\mathrm{[O\,II]}}}
\newcommand{\SII}{\ensuremath{\mathrm{[S\,II]}}}
\newcommand{\FeII}{\ensuremath{\mathrm{[Fe\,II]}}}
\newcommand{\SIII}{\ensuremath{\mathrm{[S\,III]}}}
\newcommand{\OIII}{\ensuremath{\mathrm{[O\,III]}}}
\newcommand{\SIX}{\ensuremath{\mathrm{[S\,IX]}}}
\newcommand{\parfrac}[2]{\ensuremath{\left(\frac{#1}{#2}\right)}}

\begin{document}
\title{The supermassive black hole in Centaurus A:\\
a benchmark for gas kinematical measurements
\thanks{Based on observations obtained at the  Space  Telescope Science
Institute,  which  is  operated by the Association of  Universities for
Research  in Astronomy, Incorporated, under NASA contract NAS 5-26555
(HST Program ID 8119). Also
based on observations collected at the European Southern Observatory, Paranal,
Chile (ESO Program ID 63.P-0271A).} }
\titlerunning{The Black Hole in Centaurus A}

\author{A. Marconi \inst{1}\and
G. Pastorini\inst{2}\and
F. Pacini\inst{2}\and
D.~J. Axon\inst{3,4}\and
A. Capetti\inst{5}\and
D. Macchetto\inst{6,7}\and
A.~M. Koekemoer\inst{6}\and
E.~J. Schreier\inst{8}
}

\offprints{A. Marconi}

\institute{
INAF - Osservatorio Astrofisico di Arcetri
Largo E. Fermi 5, I-50125 Firenze, Italy\\
\email{marconi@arcetri.astro.it}
\and
Dipartimento di Astronomia e Scienza dello Spazio,
Universit\`a degli Studi di Firenze
Largo E. Fermi 2, I-50125 Firenze, Italy\\
\email{guia@arcetri.astro.it,pacini@arcetri.astro.it}
\and
Department of Physics,
Rochester Institute of Technology, 85 Lomb Memorial Drive,
Rochester, NY 14623, USA\\
\email{djasps@rit.edu}
\and
Department of Physical Sciences, University of Hertfordshire, Hatfield
AL10 9AB, UK
\and
INAF - Osservatorio Astronomico di Torino, Strada
  Osservatorio 20, I-10025 Pino Torinese, Italy\\
\email{capetti@to.astro.it}
\and
Space Telescope Science Institute
3700 San Martin Drive, Baltimore, MD 21218, USA\\
\email{macchetto@stsci.edu,koekemoe@stsci.edu}
\and
Affiliated with  ESA's Space Telescope Division
\and
Associated Universities, Inc.
Suite 730 1400 16th Street, NW Washington, DC 20036\\
\email{ejs@aui.edu}
}

\date{Received; accepted}

\abstract{
We present new HST Space Telescope Imaging Spectrograph observations of the
nearby radio galaxy NGC 5128 (Centaurus A). The bright emission line with
longest wavelength accessible from HST, $\SIII\lambda 9533\AA$, was used to
study the kinematics of the ionized gas in the nuclear region with a 0\farcs1
spatial resolution.
The STIS data were analized in conjunction with the ground-based near-infrared
Very Large Telescope ISAAC spectra used by \cite{marconi:cenabh} to infer the
presence of a supermassive black hole and measure its mass.  The two sets of data have spatial resolutions differing by almost a factor of five but provide
independent and consistent measures of the BH mass, which are in agreement with
our previous estimate based on the ISAAC data alone.  The gas kinematical
analysis provides a mass of $\MBH=(1.1\pm0.1)\xten{8}\Msun$ for an assumed disk
inclination of $i=25$\deg\ or $\MBH= (6.5\pm0.7)\xten{7}\Msun$ for
$i=35$\deg, the largest $i$ value allowed by the data.
We performed a detailed analysis of the effects on \MBH\ of the intrinsic
surface brightness distribution of the emission line, a crucial ingredient in
the gas kinematical analysis. We estimate that the associated systematic errors
are no larger than $0.08$ in $\log\MBH$, comparable with statistical errors and
indicating that the method is robust. However, the intrinsic surface brightness
distribution has a large impact on the value of the gas velocity dispersion. A
mismatch between the observed and model velocity dispersion is not
necessarily an indication of non-circular motions or kinematically hot gas,
but is as  easily due to an inaccurate computation arising from too course a
model grid, or the adoption of an intrinsic brightness distribution which is
too smooth.
The observed velocity dispersion in our spectra can be
matched with a circularly
rotating disk and also the observed line profiles and the higher order moments
in the Hermite expansion of the line profiles, $h_3$ and $h_4$, are consistent
with emission from such a disk.  
To our knowledge, Centaurus A is the first external galaxy for which
reliable BH mass measurements from gas and stellar dynamics are available and,
as in the case of the Galactic Center, the \MBH\ gas kinematical estimate is in
good agreement with that from stellar dynamics.
The BH mass in Centaurus A is in excellent agreement with the
correlation with infrared luminosity and mass of the host spheroid but is a
factor $\sim 2-4$ above the one with the stellar velocity dispersion. But this
disagreement is not large if one takes into account the intrinsic scatter of
the $\MBH-\sigma_\mathrm{e}$ correlation.
Finally, the high HST spatial resolution allows us to constrain the size of any
cluster of dark objects alternative to a BH to $r_\bullet<0\farcs035$ ($\simeq
0.6\PC$).  Thus Centaurus A ranks among the best cases for supermassive Black Holes in
galactic nuclei.
}

\maketitle

\section{Introduction}

One of the drivers of current astrophysical research is understanding the birth
and evolution of galaxies but it is becoming increasingly clear that the central
supermassive black hole plays a major role in the evolution of the host galaxy. 

During the last few years it has been realized that most, if not all, nearby
luminous galaxies host a supermassive  black hole (BH) in  their nuclei with
masses in the $\ten{6}-\ten{10}\Msun$ range (e.g.~\citealt{ferrarese:bhreview}
and references therein). The BH mass (\MBH) is closely related with mass or
luminosity of the host spheroid ($L_{sph}$;
e.g.~\citealt{kormendy:bhreview,marconi:mbhk,haering:mbhm}) and with the
stellar velocity dispersion ($\sigma$;
\citealt{ferrarese:mbhsigma,gebhardt:mbhsigma}).  These facts indicate that the
formation of a massive BH is an essential ingredient in the process of galaxy
formation and that there is a tight relation between host galaxy and the
central BH. Like for the X-ray background, BHs are relics of past AGN activity
\citep{yu:bhmodel,marconi:bhmodel}, and it is being realized that an accreting
supermassive BH has a major impact on the star formation rate in the host
galaxy: the feedback from the accreting BH (i.e.~from the AGN) is thus
responsible for setting the close relations between BH and host galaxy
properties (e.g.~\citealt{menci:galmodel,granato:galmodel,dimatteo:galsimul}).

BHs are detected and their mass directly measured in nearby galaxies using gas
or stars as tracers of the kinematics in the nuclear region.  The gas
kinematical method has the advantage of being relatively simple, of requiring
relatively short observation times and being applicable to AGNs.  On the
contrary, non gravitational or non circular motions might bias or completely
invalidate the method. Moreover, detectable emission lines are not always
present in galactic nuclei.  The stellar dynamical method has the advantage
that star motions are always gravitational and that stars are present in all
galactic nuclei.  However, long
observation times are required to obtain high quality observations and stellar dynamical models are very complex leading to potential indeterminacy \citep{valluri:bhstarkin,cretton:bhstarkin}.

The weakest points of both methods - assumption of circular gravitational
motions for gas kinematics, complexity of modeling for stellar dynamics - have
shed some doubts on the reliability of BH mass estimates
(e.g.~\citealt{tremaine:mbhsigma,cappellari:ic1459bh,
verdoes:n4335bh,valluri:bhstarkin,cretton:bhstarkin}).  This is a very
important issue because the correlations between BH mass and host galaxies
properties, which are based mostly on \MBH\ from gas kinematic and stellar
dynamics, are pivotal in understanding the coevolution of galaxies and BHs over
cosmic time.  Securing the correlations means establishing without any doubt
the reliability of gas kinematics and stellar dynamics \MBH\
measurements. 

Another important open issue is that the so-called supermassive BHs are in
reality massive dark objects, i.e.\ clumps of dark mass unresolved at the
spatial resolution of current observations.  Only in the case of our own
galaxy, we can safely state that the massive dark object is a supermassive
black hole. In all other cases the massive dark object could very well be a
cluster of dark stellar remnants (stellar mass black holes, neutron stars, hot
jupiters etc.). Such clusters, due  to friction and collisions, are doomed to
collapse to a supermassive BH but timescales can be longer than the age of the
universe \citep{maoz:MDO}. In order to exclude that a massive dark object is a
cluster of dark objects one needs to constrain its size such that its life
time is much shorter than the age of the universe, making a BH a very
likely alternative (e.g.~\citealt{vandermarel:m32bh,
macchetto:m87bh,schodel:galcenbh2}).

Centaurus A (NGC5128) is a giant elliptical galaxy and hosts a powerful radio source and an AGN revealed by the presence of a powerful radio and X-ray jet (see \citealt{israel:cenareview} for a review; see also \citealt{tingay:cenajetVLBA,grandi:cenasax,hardcastle:cenaradiox,evans:cenax} and references therein).
The AGN in Centaurus A is low luminosity and is consistent with a misaligned BL-Lac
\citep{capetti:cenapol,chiaberge:cenabllac}. 
The most striking feature of Centaurus A is the dust lane, rich in molecular and ionized gas \citep{quillen:cenadisknir,quillen:cenadiskmol,leeuw:cenascuba,
karovska:cenamidir,mirabel:cenaiso}, which crosses the whole galaxy hiding the nuclear region under at least 7 magnitudes of extinction in $V$ \citep{schreier:cenahst,marconi:cenahst}.
A recent estimate of the distance of Centaurus A places the galaxy at $\sim
3.8$ Mpc \citep{rejkuba:cenaD} but in this paper we adopt $D=3.5\MPC$
($1\arcsec\simeq 17\PC$, \citealt{hui:cenaD,soria:cenaD}) to be consistent with our
previous papers. 

\cite{schreier:cenadisk} found evidence for a disk-like feature in 
ionized gas disk from \PaA\
HST/NICMOS images. \cite{marconi:cenabh} used near infrared spectroscopy to
perform a gas kinematical analysis of such feature finding evidence for a BH with mass
$\MBH=2^{+3.0}_{-1.4}\xten{8}\Msun$. The large error associated with \MBH\
takes into account the degeneracy due to the unknown inclination of the gas
disk for which \cite{marconi:cenabh} could only assume $i>15\deg$.  Observations were
performed with a spatial resolution of about 0\farcs5 ($\simeq 8.5\PC$). Even
at that moderate spatial resolution the BH sphere of influence is well
resolved: with $\sigma = 138\KMS$ \citep{silge:cenabh} the radius of the BH
sphere of influence is $\rBH= G\MBH/\sigma^2 = 1.3\arcsec$
.  Centaurus A is thus the perfect benchmark
to assess the validity of the gas kinematical method because the results will
be little affected by spatial resolution or by uncertainties on the stellar
contribution to the gravitational potential.  The HST/STIS observations
presented here have a spatial resolution which is a factor of 5 better than our
previous ground-based observations allowing a unique test on the reliability of
the gas kinematical method. A comparison between rotation curves taken at
ground based and HST resolution was performed in Cygnus A and showed that after
allowing for the different instrumental setups and spatial resolution, the best
fit ground based model could nicely reproduce the observed STIS rotation curves
\citep{tadhunter:cygnusabh}.

Recently \cite{silge:cenabh} have presented a stellar dynamical measurement of
the BH mass in Centaurus A. To our knowledge this is the first time that a BH
mass can be reliably measured in an external galaxy both with gas kinematics and stellar dynamics and
the agreement of the two measurements can greatly strengthen the reliability of
both methods.  \cite{cappellari:ic1459bh} presented gas and stellar
dynamical measurements of the BH mass in IC1459 finding a disagreement of
almost a factor 10 between the two. However in IC1459 the rotation curve of the
gas is not consistent with that of a rotating disk and the \MBH\ measurement
cannot be considered reliable.

Last, but not least, the proximity of Centaurus A, combined with HST resolution
will allow us to put a tight constraint on the size of any nuclear dark cluster
with the possibility of establishing whether the putative BH is really a BH and
not a cluster of dark stellar remnants.

In summary,
Centaurus A is an ideal benchmark for direct BH mass measurements due to (i)
its proximity, (ii) the well-resolved BH sphere of influence, (iii) the
availability of data at different spatial resolution both of which resolve the
BH sphere of influence and (iii) the availability of both gas and stellar
dynamical \MBH\ measurement.
\begin{figure*}[!]
\centering
\includegraphics[angle=90,width=0.99\textwidth]{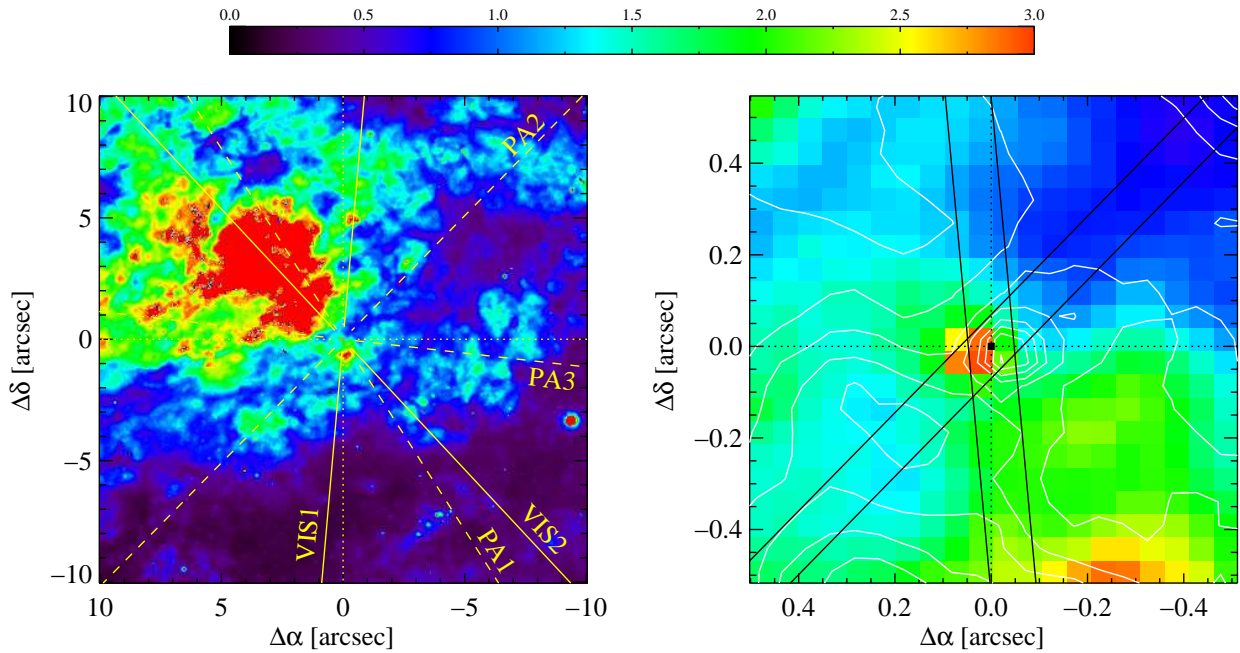}
\caption{\label{fig:slitpos} Left: slit positions overlaid on the
Centaurus A STIS image. VIS1 and VIS2 identify the STIS slit positions and PA1, PA2 and PA3 the ISAAC ones. The 0,0 position is the target position 
assumed by STIS and derived by the location of the bright star on the right
(located at $\simeq -9\arcsec, -3\arcsec$). Right: Zoom on the central $1\arcsec\times1\arcsec$.}
\end{figure*}

\begin{figure*}[!]
\centering
\includegraphics[angle=90,width=0.9\textwidth]{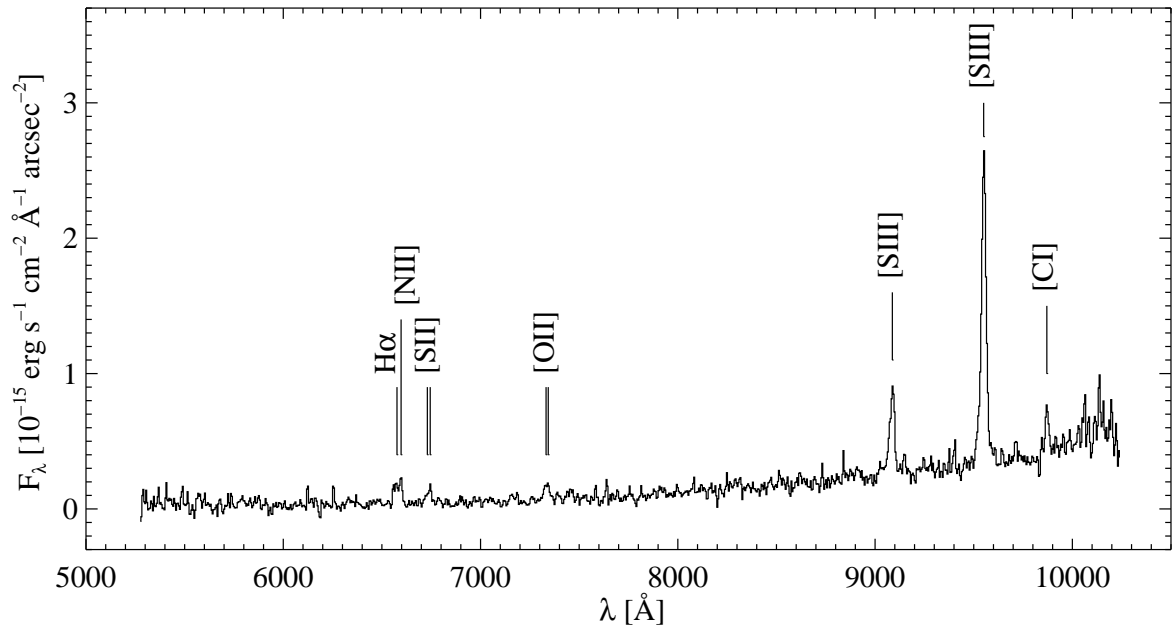}
\caption{\label{fig:nucspectra} Average STIS spectrum from VIS1 and VIS2
extracted from a $0\farcs1\times 0\farcs6$ aperture centered on the assumed
nucleus position.}
\end{figure*}

In Sec.~\ref{sec:obs} we present the HST/STIS observations and data reduction,
and the re-analysis of our old VLT/ISAAC data.  In Sec.~\ref{sec:results} we
describe the kinematical analysis, including the use of the Hermite expansion
of the line profiles, and we present the kinematical parameters measured from
all data.  In Sec.~\ref{sec:models} we describe the modeling of the kinematical
data and how it is possible to obtain \MBH\ measurements from STIS and ISAAC
data independently.  In Sec.~\ref{sec:discuss} we discuss the results from the
modeling of STIS and ISAAC data, we evaluate the influence of the intrinsic
surface brightness distribution of the disk on the derived \MBH\ value, and we
present the problems which might be encountered when comparing observed and
model velocity dispersions of the gas, the commonly adopted test of the
reliability of the gas kinematical analysis.  We compare our \MBH\ value with
the stellar dynamical measurement, we discuss \MBH\ within the context of the
correlation with host galaxy parameters and we then estimate an upper limit on
the size of any dark cluster alternative to a \BH.  Finally, in
Sec.~\ref{sec:conclusions} we summarize our work and present our conclusions.

\begin{figure*}[!]
\centering
\includegraphics[width=0.48\textwidth]{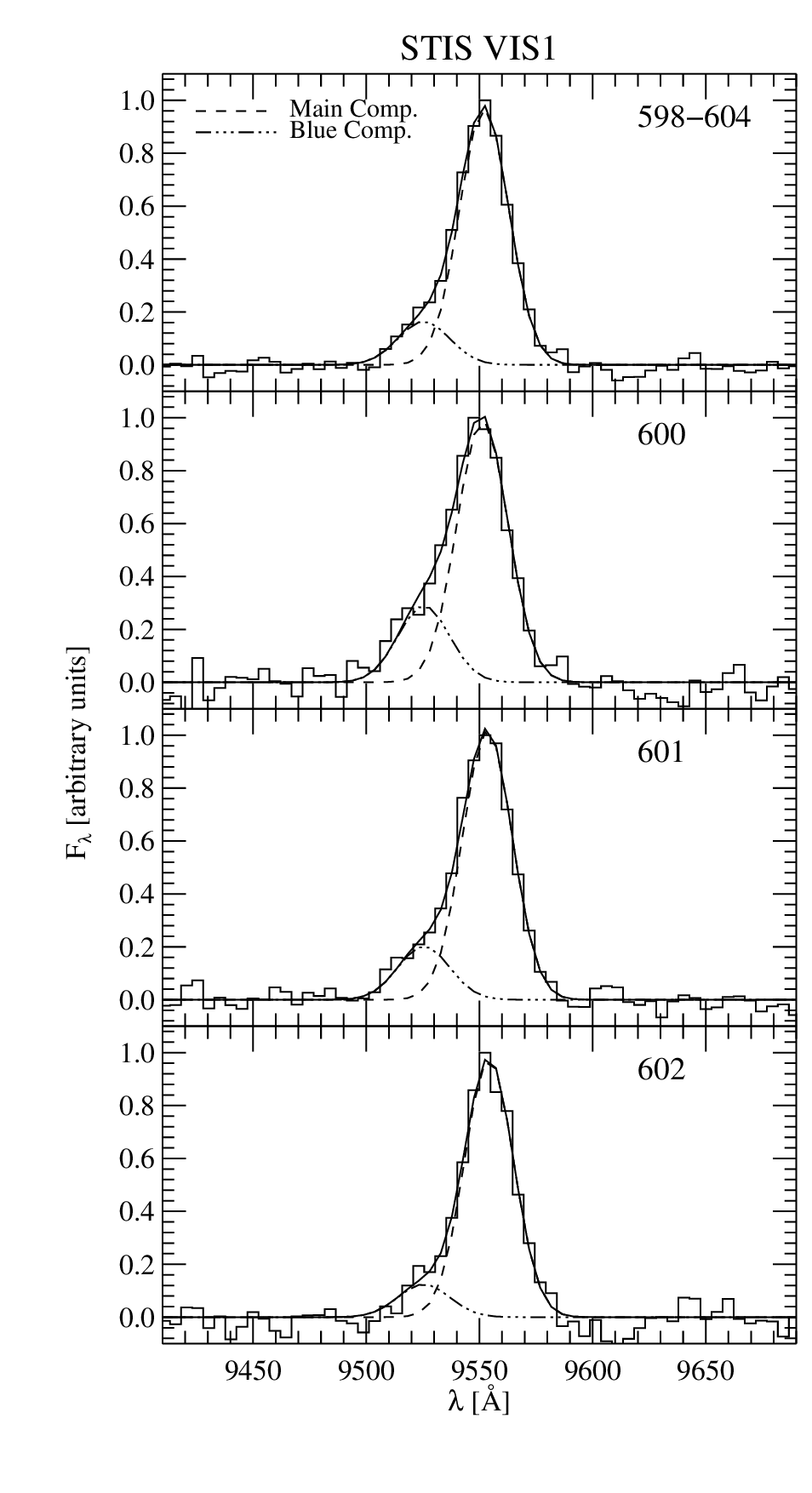}
\includegraphics[width=0.48\textwidth]{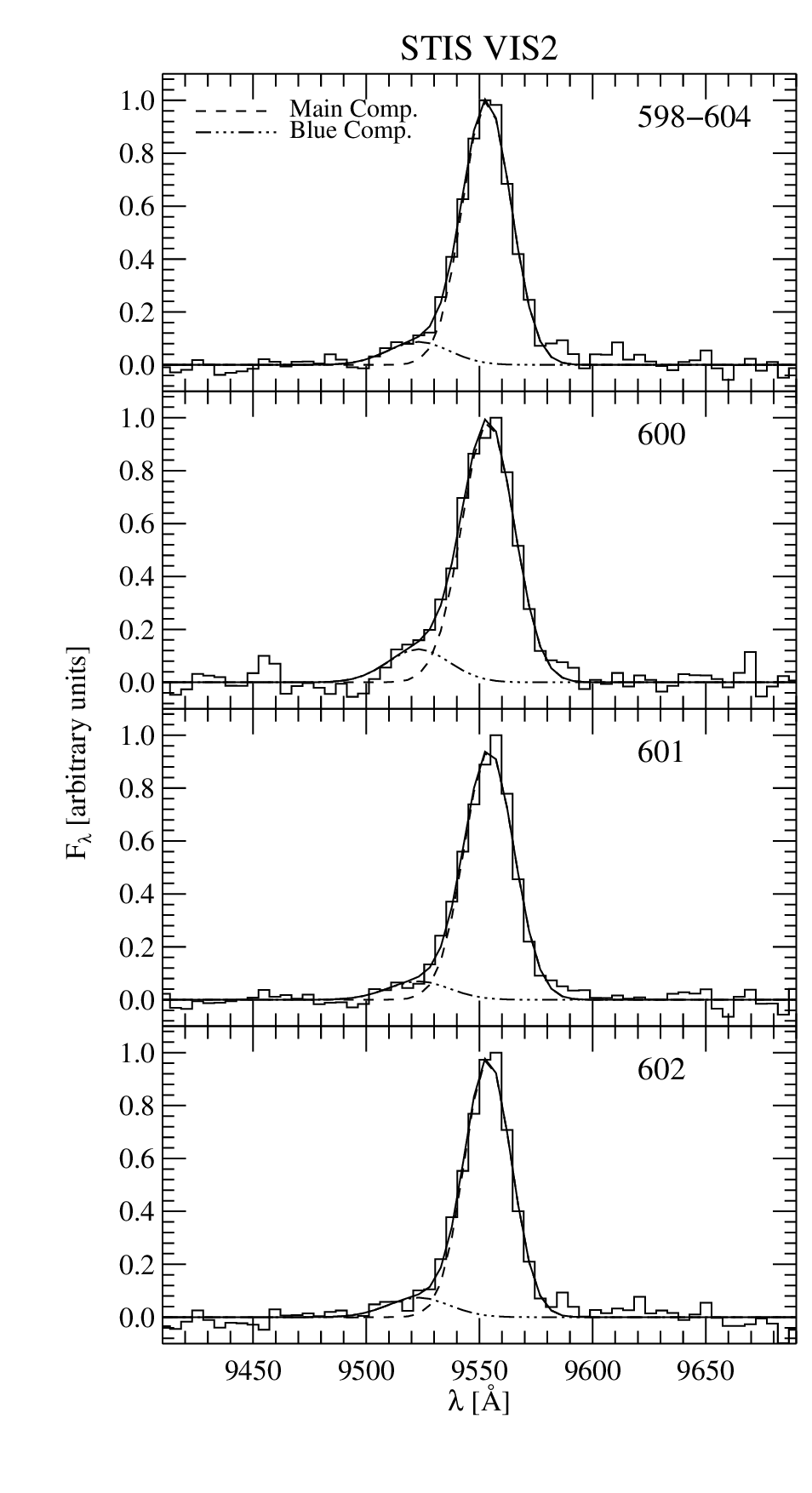}
\caption{\label{fig:stis_fits} Examples of the fits of the line profiles for the STIS observations of VIS1 and VIS2. The dashed and dot-dashed lines identify the two components of the fit, called 'main' and 'blue'. For every slit position the blue component has the same velocity and width at each row and their values are determined in the fit of the overall nuclear spectrum shown in top panels. The numbers in the upper right corners of each panel are the row and range of rows where the fit was performed.}
\end{figure*}

\section{\label{sec:obs}Observations and Data Reduction}

\subsection{STIS observations\label{sec:obs_stis}}
Centaurus A was observed with STIS on the HST in 2000 May 9 and July 1.  The
nucleus location in Centaurus A is not well defined in optical ($<1\MIC$)
images thus telescope acquisition was performed on a foreground star $\sim 10\arcsec$
away from the galaxy nucleus in order to register the coordinate
system of the telescope 
($\Delta\alpha,\Delta\delta= -9.3\arcsec,-3.3\arcsec$
in Fig.~\ref{fig:slitpos}a).  During each
visit, a $\sim 5\arcsec\times5\arcsec$ acquisition image was obtained with the
F28X50LP filter on the foreground star which  was subsequently centered and
re-imaged following the ACQ and ACQ/PEAK procedures.  The exposure times of the
acquisition images were 3 and 10\SEC\ respectively.
After registering the coordinates system on the star, the telescope was
subsequently pointed on the expected position of the Centaurus A nucleus
derived from the analysis of WFPC2 images performed by \cite{marconi:cenahst}.  The
observational strategy then consisted in obtaining an image in the F28X50LP
filter to check the telescope pointing and slit locations with respect to the
nucleus.  Spectra were then obtained with the 52X0.1 aperture (0\farcs1 slit
width) centered on the expected nucleus position. Given the different roll
angle of the telescope during the two visits, the slits were located at PA
-5\DEG\ (hereafter VIS1) and 43\DEG\ (hereafter VIS2).  The slit positions
are overlaid on the STIS image in Fig.~\ref{fig:slitpos}.  At each slit
position we obtained three spectra with the G750L grating centered at 7751\AA,
in order to observe the \SIII$\lambda\lambda$ 9071,9533\AA\ doublet.  The
second and third were
spectra shifted along the slit by 0\farcs5 and 1\arcsec\ in
order to remove cosmic-ray hits and hot pixels.  The spectra were obtained with
the 0\farcs1 slit and no binning of the detector pixels, yielding a spatial
scale of 0\farcs0507/pix along the slit, a dispersion per pixel of
$\Delta\lambda = 4.882$~\AA\ and a spectral resolution of ${\cal R} =
\lambda/(2\Delta\lambda) \simeq 1000$ at 9500\AA.  The log of the observations
is shown in Table~\ref{tab:logobs}.
\begin{table}
\caption[]{\label{tab:logobs}Log of observations.}
\begin{tabular}{lcccccr}
\hline\hline
\\
ID & Date  & Instr. & PA    & ${\cal R}^\mathrm{ a}$ & Res$^\mathrm{ b}$ \\
   &       &        & [deg] &             & [\arcsec]          &         \\
\hline
\\
VIS1	    & May 9, 2000	& STIS	 & -5.0      & 1000       & 0.1	\\
VIS2	    & July 1, 2000	& STIS	 & ~43.0      & 1000       & 0.1	\\
PA1	    & July 21, 1999	& ISAAC	 & ~32.5    & 10000      & 0.5	\\
PA2	    & July 22, 1999	& ISAAC	 & -44.5   & 10000      & 0.5	\\
PA3	    & July 22, 1999	& ISAAC	 & ~83.5    & 10000      & 0.5	\\
\\
\hline
\end{tabular}\\
$^\mathrm{a}$ Spectral resolution.\\
$^\mathrm{b}$ Spatial resolution: FWHM of Point Spread Function (PSF) or seeing.
\end{table}
\begin{figure*}[!]
\centering
\includegraphics[width=0.45\textwidth]{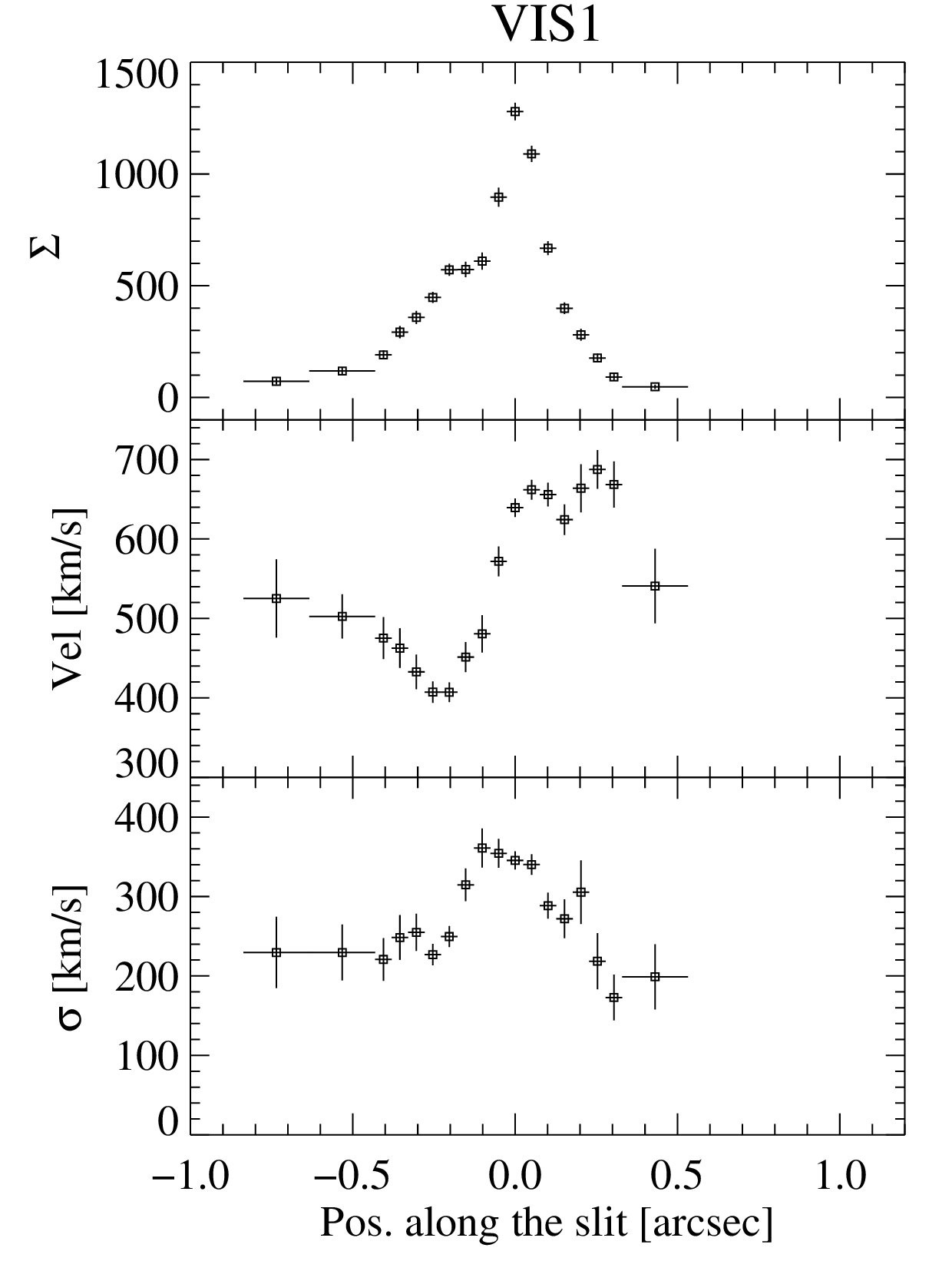}
\includegraphics[width=0.45\textwidth]{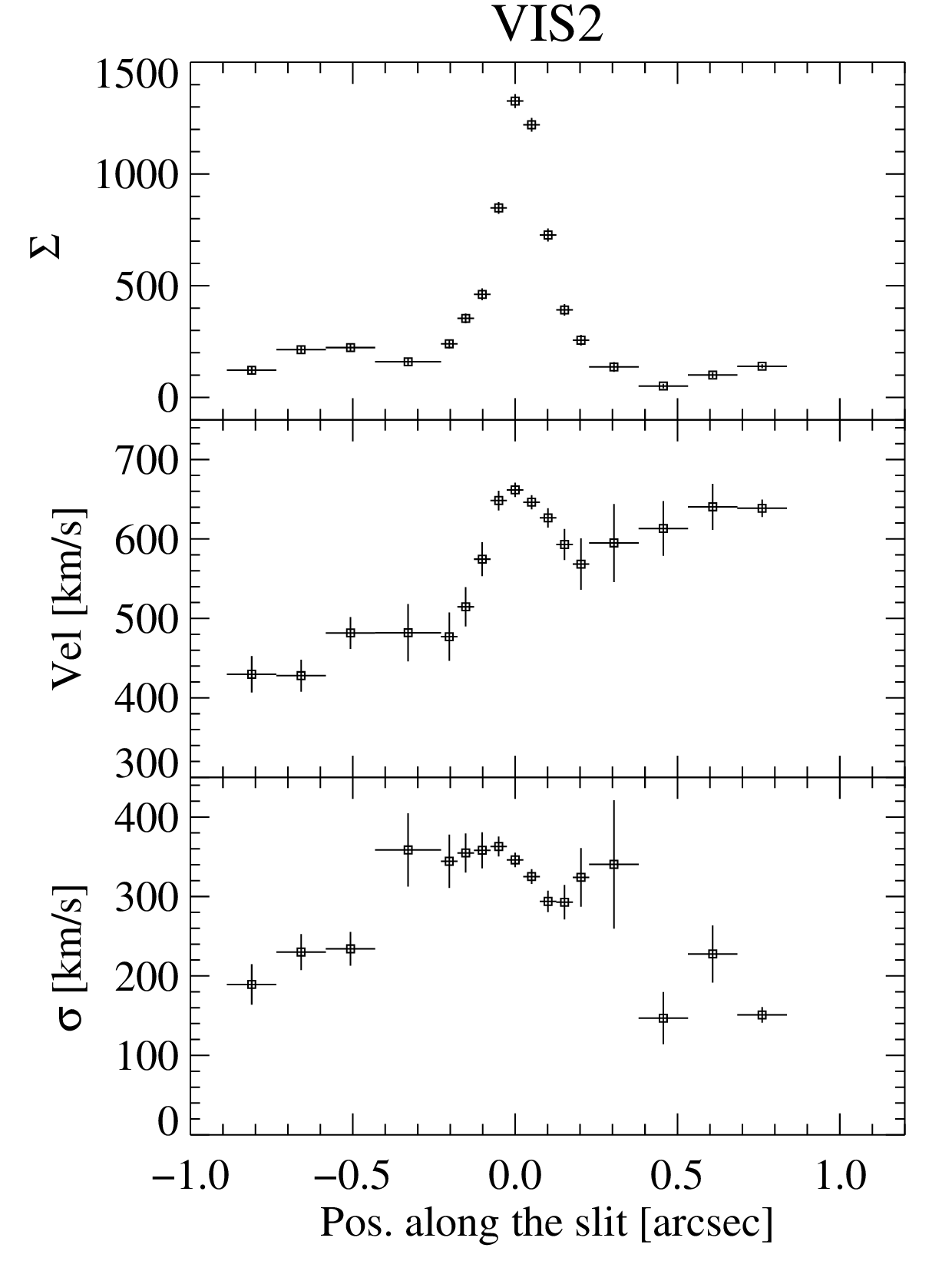}
\caption{\label{fig:stiscurves} From top to bottom: Average emission line
surface brightnesses (units of \ten{-16}\ERG\SEC\1\CM\2\ arcsec\2), velocities, and velocity dispersion observed along the
slit in STIS observations. Columns 1 to and 2  refers to VIS1 and VIS2,
respectively.}
\end{figure*}
The raw STIS images were processed with the {\it calstis} pipeline
\citep{stis_handbook}
using the best reference files available as of December 2003.
The raw STIS spectra were processed with standard pipeline tasks but
following a different procedure to correct the fringing exhibited by
the STIS CCD in the far
red ($>7500\AA$). In order to correct for fringing, we obtained contemporaneous
flat fields with the same setup as the scientific observations.  We then
followed the prescriptions of \cite{stis_fringe}.
The on-orbit flats were normalized using the task
\texttt{normspflat} in the \texttt{stsdas.hst\_calib.stis} package of
IRAF\footnote{IRAF is distributed by the National Optical Astronomy
Observatories, which are operated by the Association of Universities for
Research in Astronomy, Inc., under cooperative agreement with the National
Science Foundation.}. Spectra were first prepared with \texttt{prepspec} and
defringed with \texttt{defringe}. We did not use the task \texttt{mkfringeflat}
to rescale and shift the flat fringes to those observed in the object because
the signal-to-noise ratio on the continuum was very poor.
However, using two flats obtained in different spacecraft orbits during each visit, we verified that variations of the fringe positions, if any, do not affect the final rotation curves.
The defringed spectra were then wavelength calibrated (\texttt{wavecal}) and
corrected for 2D distorsion (\texttt{x2d}).  The spectra taken at different
slit positions were then raligned using as reference the \SIII\ line and
combined rejecting deviant pixels.  The expected accuracy of the wavelength
calibration is $0.1 - 0.3$ pix within a single exposure and $0.2 - 0.5$
pix among different exposures \citep{stis_handbook}
which converts into $\sim 15 -
45\KMS$ (relative) and $\sim 30 - 70\KMS$ (absolute).  The relative error on
the wavelength calibration is negligible for the data presented here because
our analysis is restricted to the small detector region including \SIII\
($\Delta\lambda<300$\AA).
From an analysis of the arc exposures it is found that the lamp lines at $\sim 9000-9500$\AA\ have a FWHM of $\sim 2.2$ pix consistently with the expected spectral resolution \citep{stis_handbook}. Therefore,
 the degradation of the spectral resolution at $\sim 9000-9500$\AA\ produced by the grating corresponds to a velocity dispersion of $\sigma=140\KMS$. When also the degradation introduced by the slit width is considered, the final instrumental resolution corresponds to $\sigma=160\KMS$, i.e.\ the width of the arc lines at $\sim 9000-9500$\AA.

The target acquisition strategy followed in these observations ensures that the
STIS slits are centered on the same point, regardless of its distance from the
Centaurus A nucleus. This can be verified from the images obtained just after the telescope offset from the acquisition star and before the
spectra.  The aperture center
coordinates on the images are given by the \texttt{HOSIAX} and \texttt{HOSIAY}
keywords. We used these nominal aperture centers as pivots for rotation and
realignment of the images taken in the 2 different visits. By using 9 point
sources common to both images we then verified that the images are aligned to
better than 0.1 and 0.25 pixels rms in RA and DEC, i.e.~0\farcs006 and 0\farcs014 respectively.  This is then
the accuracy with which the STIS slits are centered on the same point which is
shown as a filled square in Fig.~\ref{fig:slitpos}b.
\begin{figure*}[!]
\centering
\includegraphics[width=0.33\textwidth]{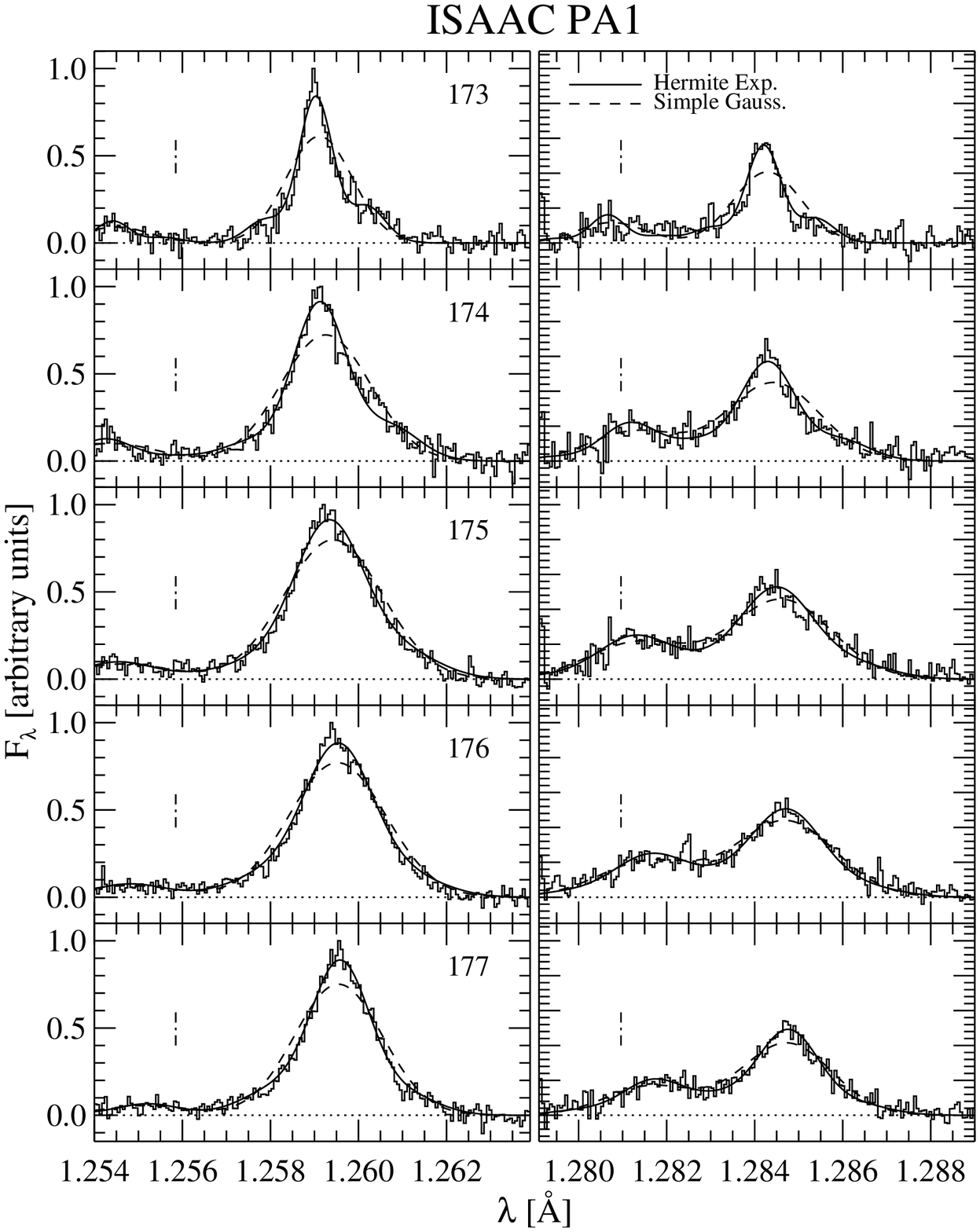}
\includegraphics[width=0.33\textwidth]{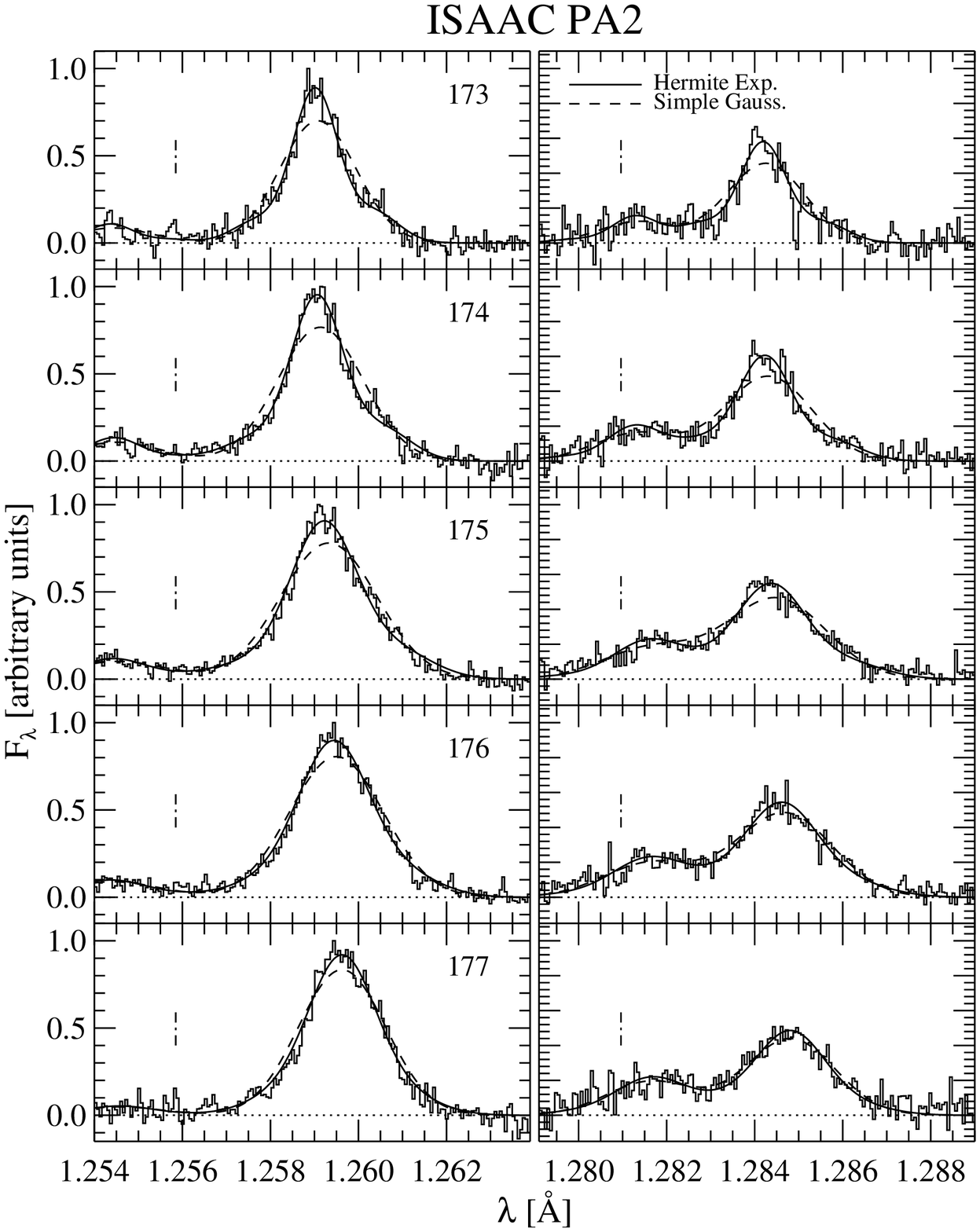}
\includegraphics[width=0.33\textwidth]{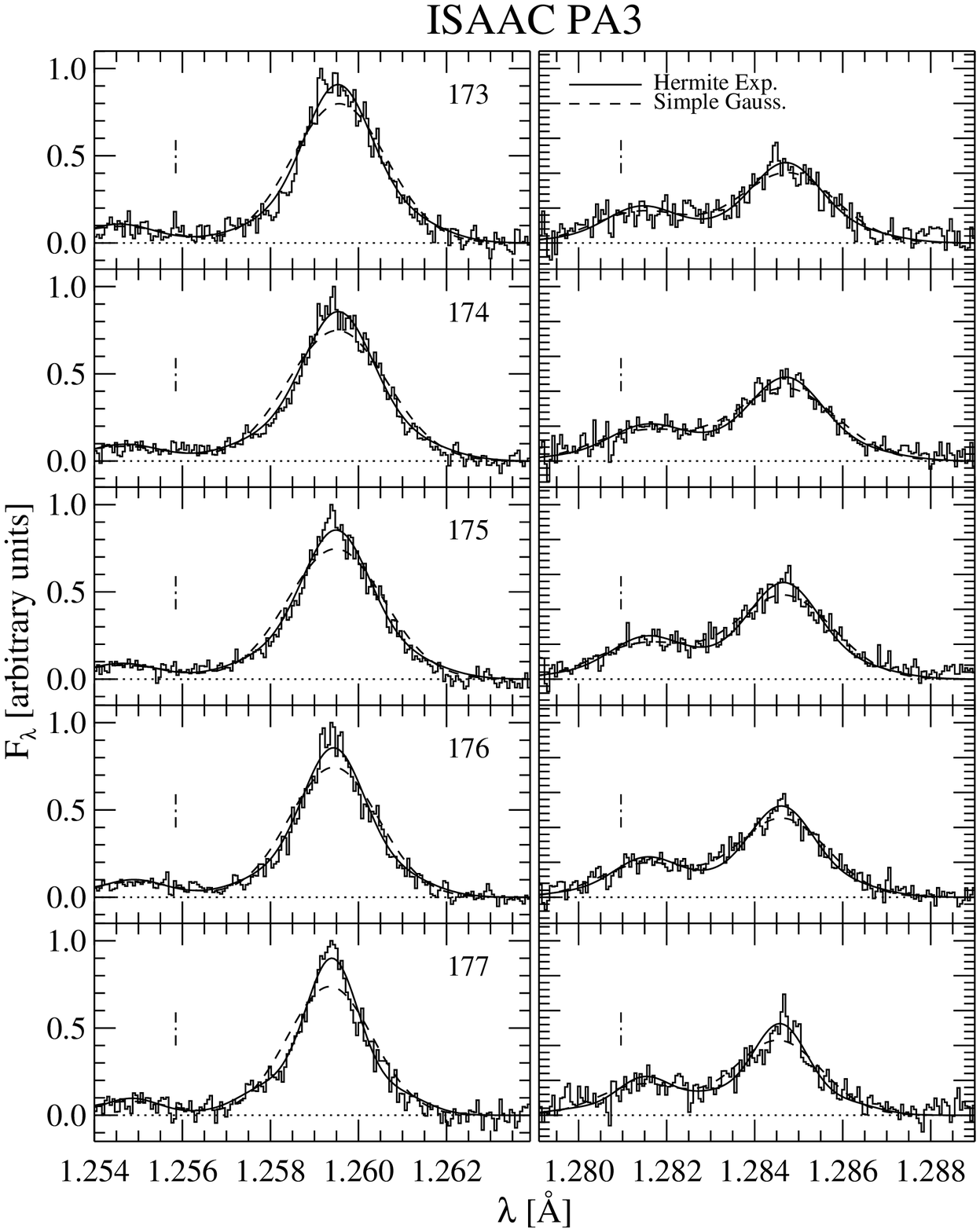}
\caption{\label{fig:isaac_fits} Examples of the fits of the line profiles for
the ISAAC observations of PA1, PA2 and PA3. The solid lines represents the case
of the Hermite expansion while the dashed line is the simple gaussian fit. The line at shorter wavelengths is \FeII\ while at longer wavelengths the strongest line is \PaB. All emission lines have the same average velocity, velocity dispersion and Hermite coefficients}. At the blue side of \FeII\ a faint line is visible, this is \SIX. The line at the blue side of \PaB\ is another \FeII. The
dot dash short line mark the expected position of the blue component found in
the \SIII\ line.
\end{figure*}

\subsection{ISAAC Observations}
\begin{figure*}[!]
\centering
\includegraphics[width=0.33\textwidth]{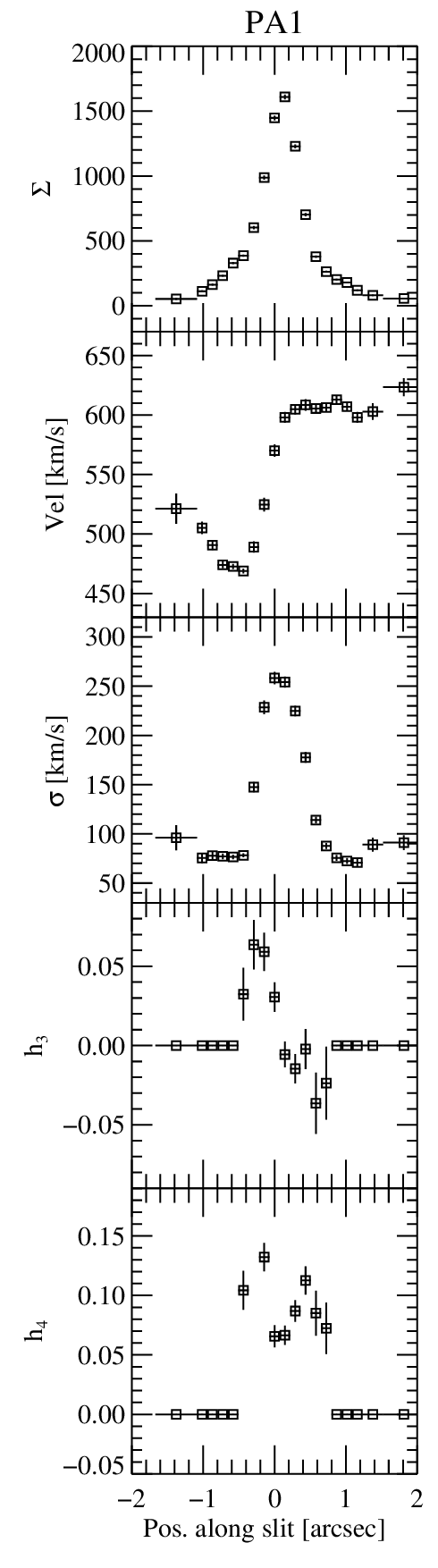}
\includegraphics[width=0.33\textwidth]{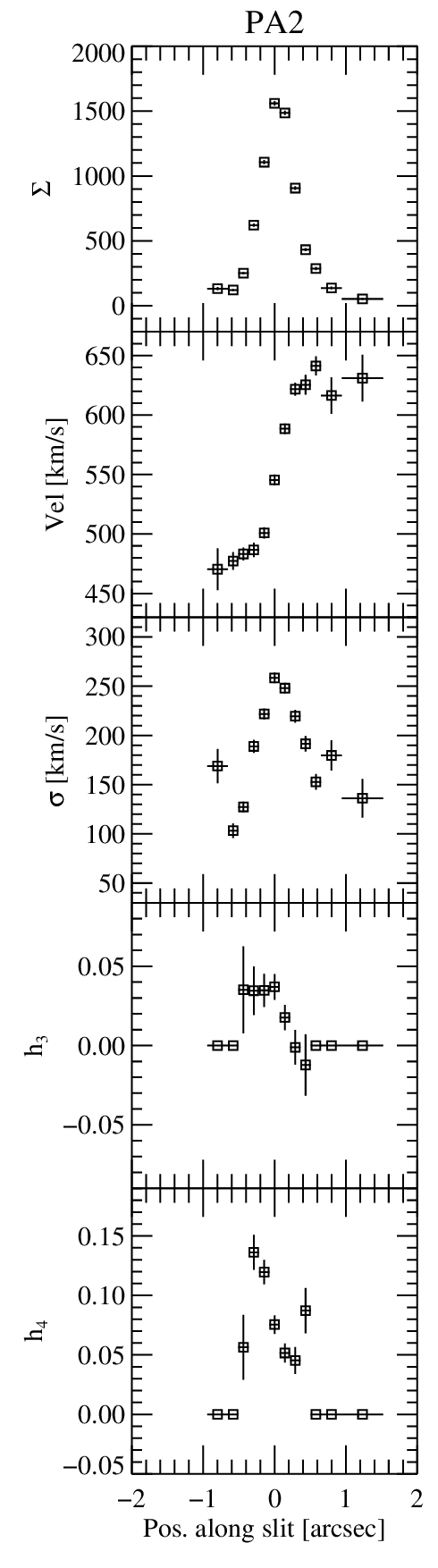}
\includegraphics[width=0.33\textwidth]{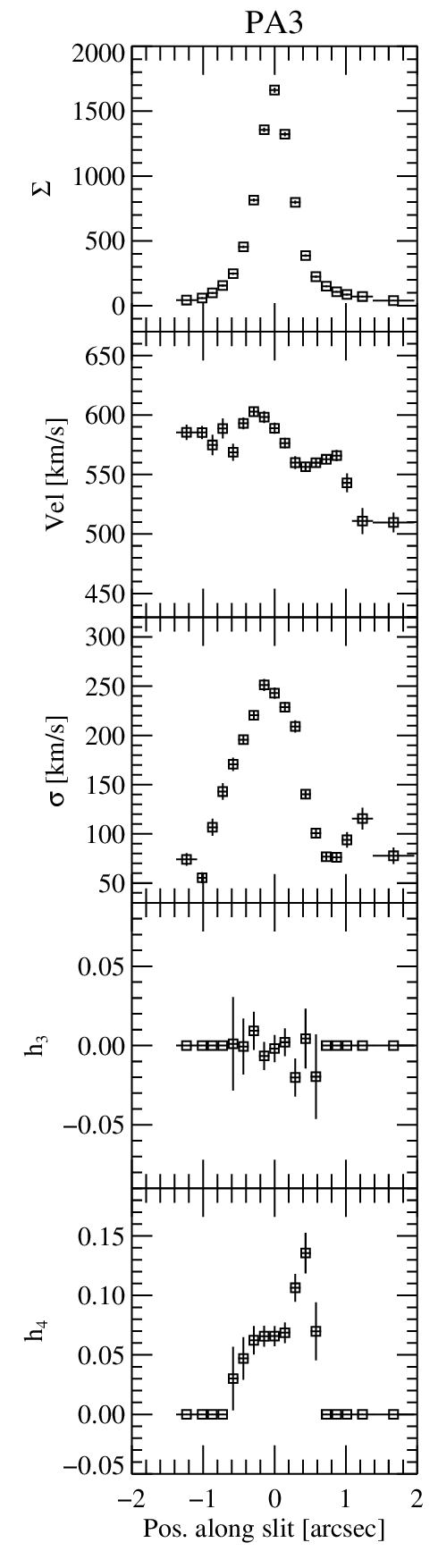}
\caption{\label{fig:isaaccurves} From top to bottom: Average emission line
fluxes, velocities, velocity dispersion, $h_3$ and $h_4$ observed along the
slit in ISAAC observations. Columns 1 to 3 refers to PA1, PA2 and PA3,
respectively. The kinematical quantities were derived from a simultaneous fit of \PaB\ and \FeII.} 
\vspace{3cm}
\end{figure*}

We complement the new HST STIS Observations with the J-band, ground-based
spectroscopic data already presented and analyzed by \cite{marconi:cenabh}.  Briefly, the data were
obtained in 1999 using ISAAC at the ESO VLT and consist of medium resolution
spectra obtained with the 0\farcs3 slit and the grating centered at
$\lambda=1.27\MIC$. The dispersion is 0.6 \AA/pix, yelding a resolving power of
10000. Seeing during the observations was in the range 0\farcs4-0\farcs6, with
photometric conditions. The observation procedure was to obtain an acquisition
image in the K band and then center the slit on the prominent K band peak in the nuclear region of Centaurus A. Spectra
were then obtained with the slits placed at three different position angles, summarized
in Table \ref{tab:logobs}. More details on observations and data reduction can
be found in \cite{marconi:cenabh}.  We have re-reduced the ISAAC spectra
following the same strategy as \cite{marconi:cenabh} but with a difference:
before coadding all the wavelength calibrated frames, we have traced the
position of the nuclear continuum peak along the dispersion direction
and used this trace to rectify the dispersion direction.
This ensures that there is no variation of the location of the continuum
peak along the slit in the wavelength range covered by the observations.

\section{Results\label{sec:results}}
\subsection{STIS Line Kinematics}
Fig.~\ref{fig:nucspectra} shows the average STIS spectrum from VIS1 and VIS2
extracted from a $0\farcs1\times 0\farcs6$ aperture centered on the assumed
nucleus position.  The lines \HA, \NII, \SII, \OII, \SIII\ and \CI\ are clearly
detected and their relative strenght are an indication of the strong reddening
which affects the nuclear region of Centaurus A ($\Av>7$ mag e.g.,
\citealt{schreier:cenahst,marconi:cenahst}): typically, \HA+\NII\ is stronger
than \SIII\ while here $\SIII/\HA+\NII> 10$. Therefore, only $\SIII\lambda
9500$\AA\ has enough signal-to-noise ratio (hereafter SNR) for kinematical
analysis.  We thus focus only on the 9300-9800 \AA\ spectral region.

After selecting the spectral region, we subtracted the continuum by fitting a
linear polynomial row by row along the dispersion direction.  The continuum
subtracted lines were then fitted row by row with gaussian functions using the
task {\it specfit} in the IRAF {\it stsdas.contrib} package. When the
signal-to-noise ratio (SNR) was insufficient (faint line) the fitting was
improved by co-adding two or more pixels along the slit direction.  

Fitting single gaussians is acceptable for most of the positions along the slit
but it does not produce good results for the points in the nuclear region where
a careful analysis of the line profiles shows that they are persistently
asymmetric with the presence of a blue wing (Fig.~\ref{fig:stis_fits}).  A fit
row by row with two gaussian components, the main component and the blue one,
shows that, within the large uncertainties, the ``blue wing'' has always the
same velocity and width.  We have thus deblended the ``blue'' component in the
spectrum obtained by co-adding the central 7 rows.  The velocity and width of
the blue component ($v=-240\KMS$ and $\sigma=360\KMS$ for VIS1, $v=-320\KMS$
and $\sigma=430\KMS$ for VIS2) were then held fixed in the row by row fit
significantly improving the fitting of the line profiles
(Fig.~\ref{fig:stis_fits}).  A similar blue wing was detected and deblended in
\HA+\NII\ in NGC 4041 \citep{marconi:n4041bh} and, as in that
paper, the present data does not allow to identify the nature of the blue
wing. We remark that the final rotation curve does not depend on the exact values of $v$ and $\sigma$ used to parameterize the blue wing
since that component is faint compared to the main one. However, the rotation curves obtained without taking into account the blue wing are not consistent with a circularly rotating disk, like in the case of NGC4041 \citep{marconi:n4041bh}.
 
The fitting procedure provides line-of-sight velocities, velocity dispersions and surface brightnesses along each slit for the $\SIII\lambda 9500$\AA\ emission
line which are shown in Fig.~\ref{fig:stiscurves}.

The observed line surface brightness in VIS1 and VIS2 are strongly peaked at the
expected nucleus positions, similarly to what was observed in Pa$\alpha$ 
\citep{schreier:cenadisk} and \FeII\ \citep{marconi:cenahst,marconi:cenabh}.
The SNR of the data does not allow the detection of extended emission beyond 1\arcsec\ from the nucleus. The rotation curve in VIS1 is "S-shaped" as expected in the case
of a circularly rotating disk observed with a slit close to its major axis. In VIS2 the S-shaped pattern seems superimposed on a constant velocity gradient. The amplitude of the rotation curves is $\sim 300\KMS$ (VIS1) and $\sim 200\KMS$
(VIS2) suggesting that the VIS1 slit is closer to the orientation of the disk line of nodes. The behaviour of the velocity dispersion is similar to what is qualitatively expected from a rotating disk, i.e.\ a constant $\sigma$ which increases toward the nucleus location reaching a peak value of $\sim 350\KMS$.

\subsection{ISAAC Line Kinematics}
\begin{figure}[!]
\centering
\includegraphics[width=0.95\linewidth]{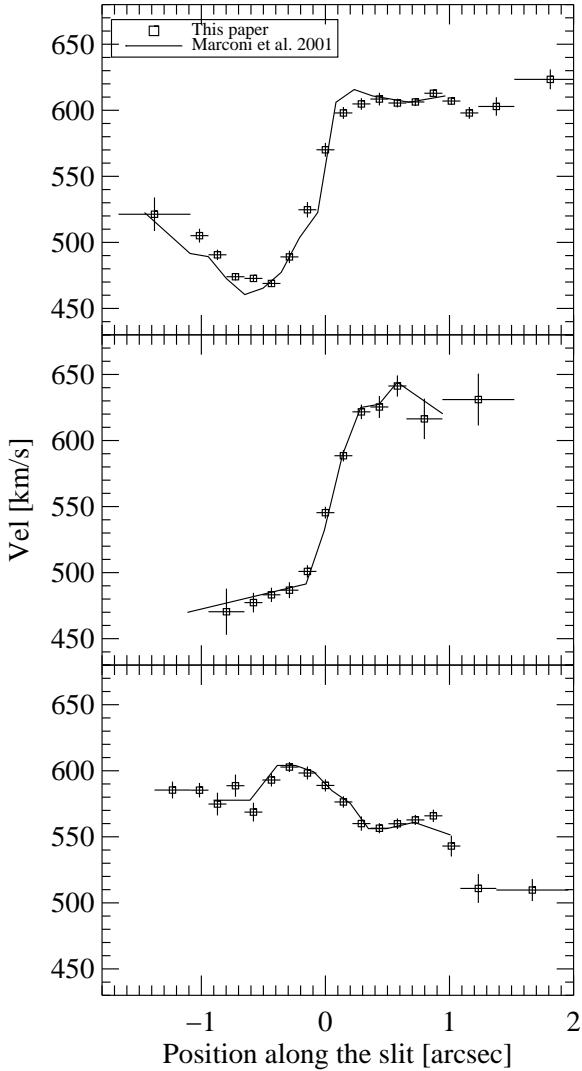}
\caption{\label{fig:isaaccurves_paper1} Comparison of average velocities from \FeII\ and \PaB\ emission lines with those from \protect\cite{marconi:cenabh}. }
\end{figure}
\begin{figure}[!]
\centering
\includegraphics[width=0.95\linewidth]{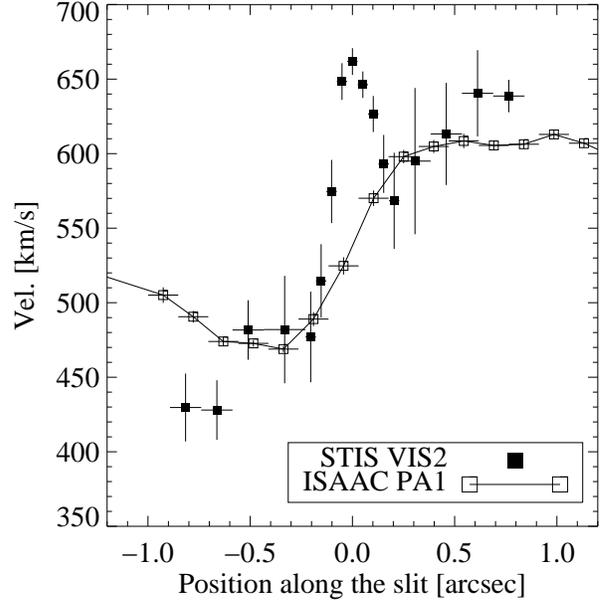}
\caption{\label{fig:compare} Comparison of average velocities from STIS VIS2 and ISAAC PA1.}
\end{figure}

The fit with one or two gaussians - in case of a significant blue wing -
produces good results for the STIS data because of the moderately low SNR and
spectral resolution ($\sim 1000$).  However, the ISAAC data have a much higher
SNR and
spectral resolution and the line profiles present a significant structure in
the nuclear region similar to the expected one in case of unresolved rotation.
\cite{marconi:cenabh} estimated kinematical parameters by fitting single
gaussians to \FeII\ and \PaB\ lines. While the fitting was acceptable for the
purposes of that paper, it had significant residuals.  In order to estimate
accurately the average velocity and velocity dispersion, i.e.~the same
parameters
computed by our model (see below), the residuals should not have any structure.
When a simple gaussian is a too simple model for the observed line profile, a
possibility is to fit 2 gaussians, but the fitting is unstable and gaussian
parameters are strongly correlated.  Therefore we decided to model the line
profile with the expansion in Hermite polynomial introduced by
\cite{marel:hermexp} coupled with the penalized pixel fitting technique
\citep{merritt:penlike,cappellari:ppf}.  Such parameterization of the line
profile coupled with the penalized fitting technique has been used to describe
the Line of Sight Velocity distribution of stellar kinematical data but it is
easily applicable to the analysis of gas kinematical data. Of course, this
refinement is necessary only because of the signal to noise and spectral
resolution of the data is moderately high, otherwise a single gaussian would
be more than adequate to describe the data.
Briefly, following \cite{marel:hermexp} we parameterize the line profile
$L(v)$ as:
\begin{equation} 
L(v) = \gamma\,\alpha(w)\,\left[1+h_3 H_3(w)+h_4 H_4(w)\right]
\end{equation}
where 
\begin{equation}
w = \frac{v-v_0}{\sigma_0}
\end{equation}
\begin{equation}
\alpha(w) = \frac{1}{2\pi}\,\mathrm{e}^{-\frac{1}{2}w^2}
\end{equation}
and $H_3(w)$, $H_4(w)$ are the third and fourth order Hermite polynomials with
the normalization adopted by \cite{marel:hermexp}.  $v_0$ and $\sigma_0$ are
the
central velocity and velocity dispersion of the gaussian component and
correspond to the average velocity and velocity dispersion of the line profile
when it is a perfect gaussian. $h_3$ and $h_4$ quantify the deviations from the
gaussian line profile and $h_3=h_4=0$ corresponds to the perfect gaussian case.
Though the Hermite expansion can be extended to orders higher than
4$^\mathrm{th}$,
we did
not find significant improvements in the profile fitting, given the SNR of the
data.
The above line profile was fitted by minimizing the penalized $\chi^2$ where
\begin{equation}
\chi_\mathrm{p}^2=\chi^2[1+\beta^2 (h^2_3+h^2_4)]
\end{equation}
see \cite{cappellari:ppf} for more details. $\chi^2$ is the classical Chi-squared and $\beta$ is a "bias" 
factor which favours best fits with low $h_3$ and $h_4$ values.
We found that $\beta=1$ is adequate to the spectral resolution
and SNR of our data.
We investigated comprehensively the possibility that lines did not share common kinematics but we did not find evidence for this. On this basis we proceded the subsequent analysis assuming 
that all lines in the spectrum have the same average velocity, velocity dispersion,
$h_3$ and $h_4$.
\begin{figure*}[!]
\centering
\includegraphics[angle=90,width=0.95\linewidth]{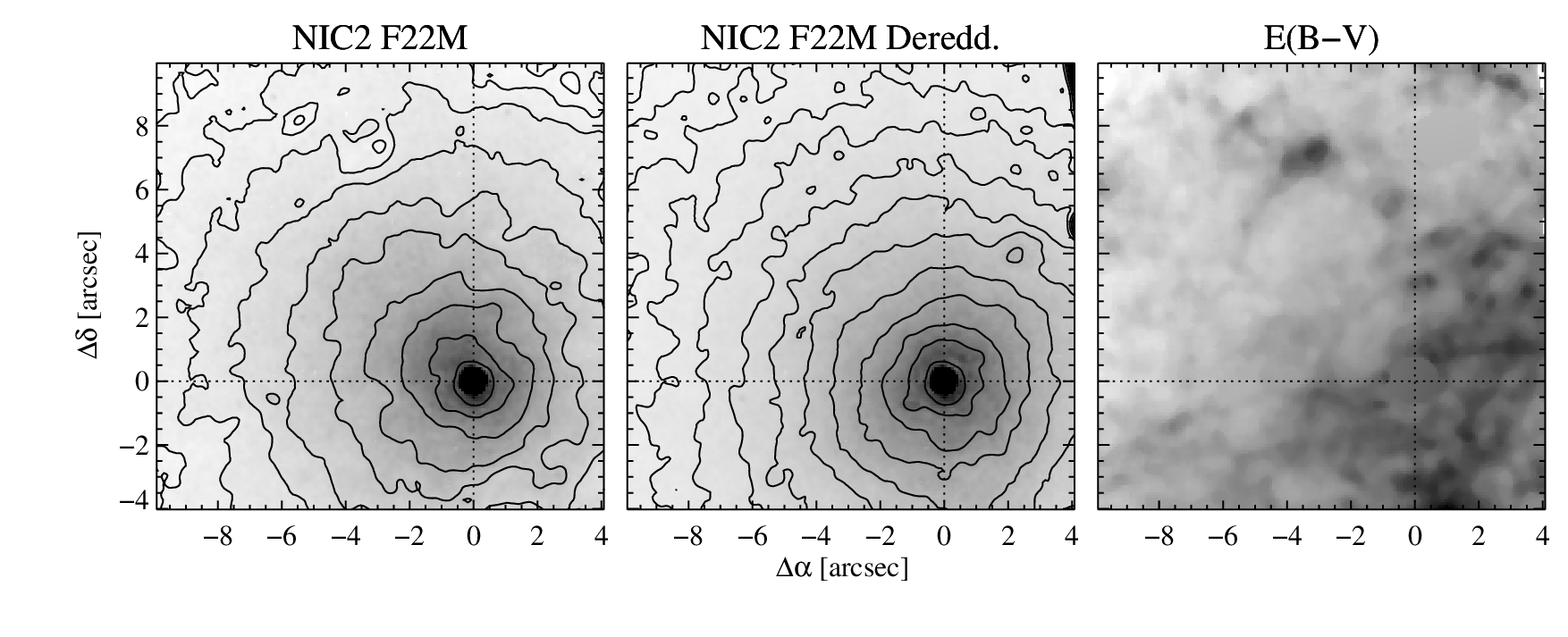}
\includegraphics[angle=90,width=0.95\linewidth]{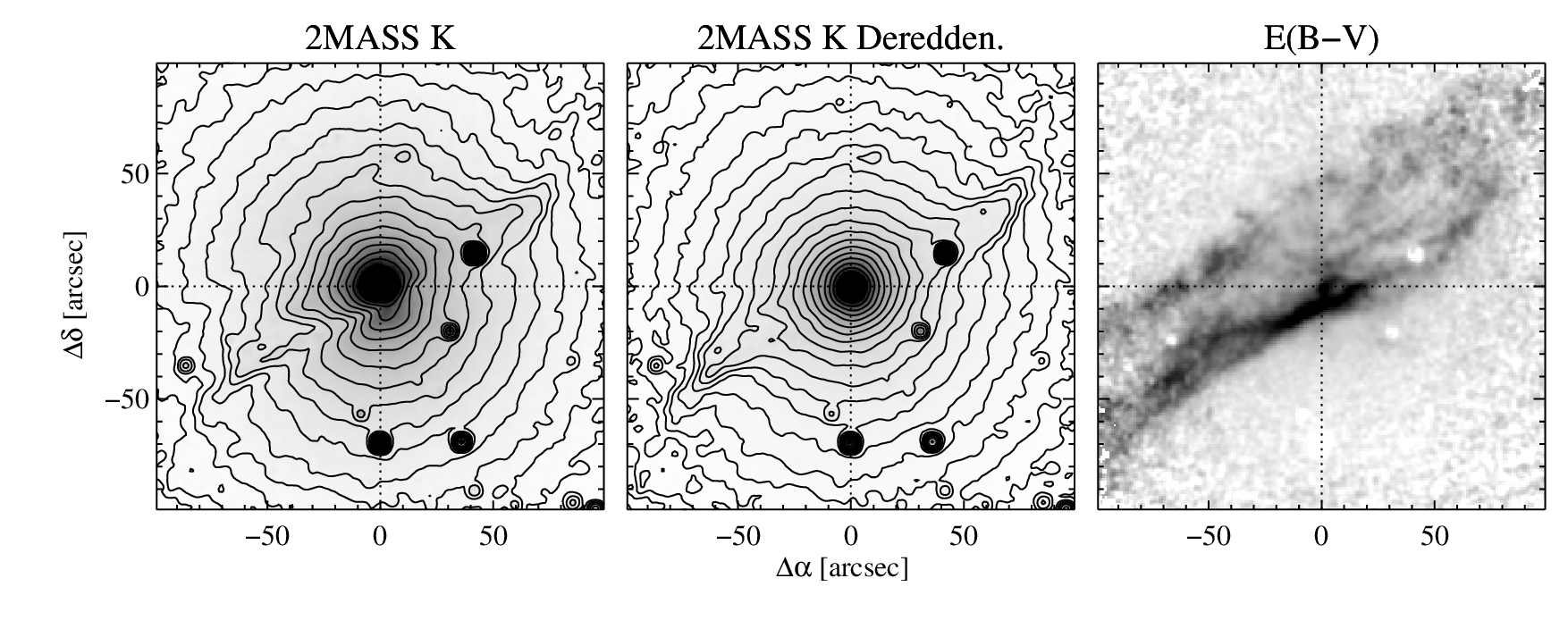}
\caption{\label{fig:dered} Top, from left to right: observed NICMOS F222M
image, dereddened F222M image, derived E(B-V) map.  Bottom, from left to right:
observed 2MASS K band image, dereddened K image, derived E(B-V) map.  }
\end{figure*}

Examples of the fits of the line profiles for the ISAAC observations
of PA1, PA2 and PA3 are shown in Fig.\ref{fig:isaac_fits} where
we also compare the Hermite expansion with the case of the pure gaussian
fit (dashed line).
\begin{figure*}
\centering
\includegraphics[width=0.45\linewidth]{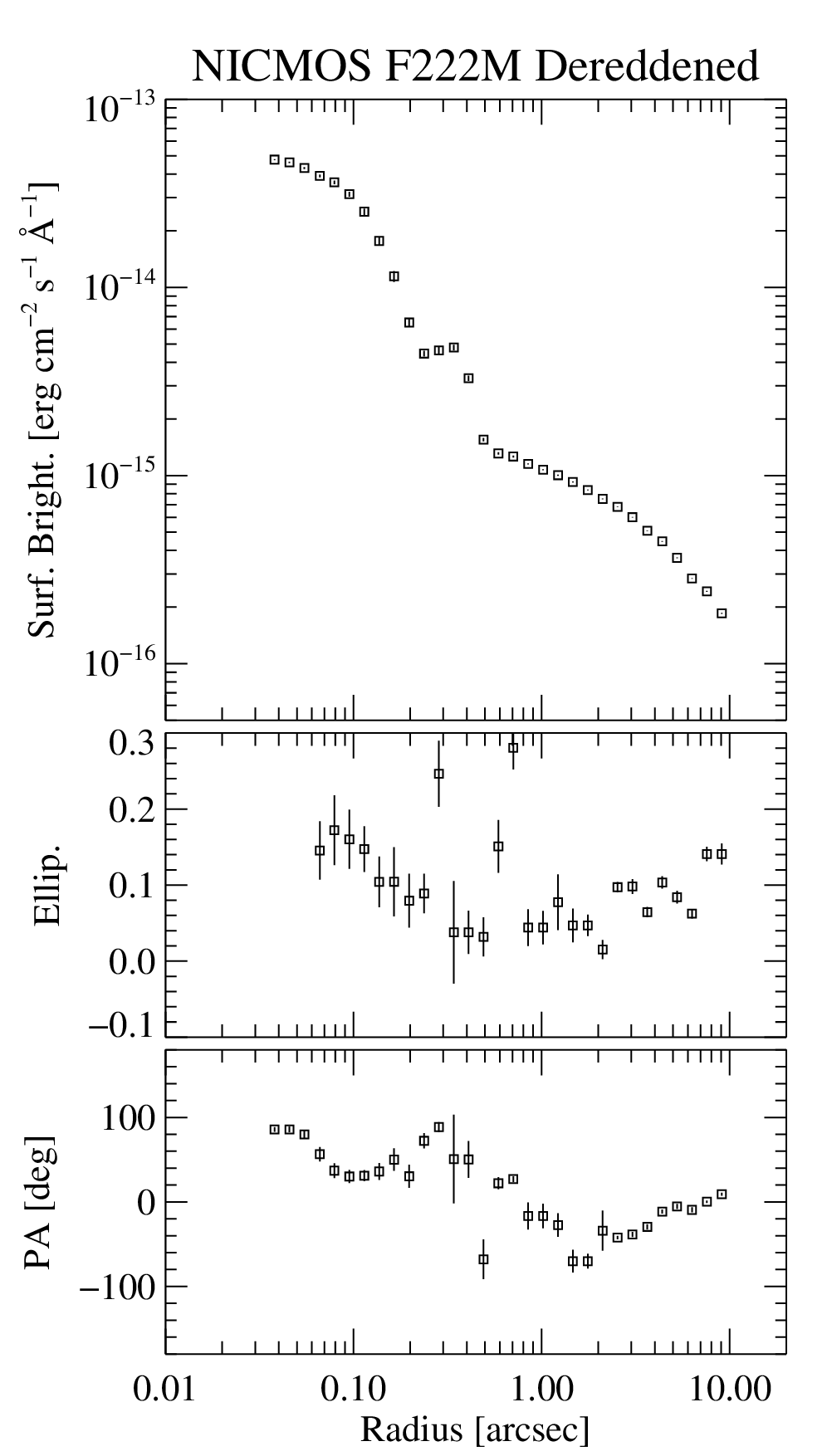}
\includegraphics[width=0.45\linewidth]{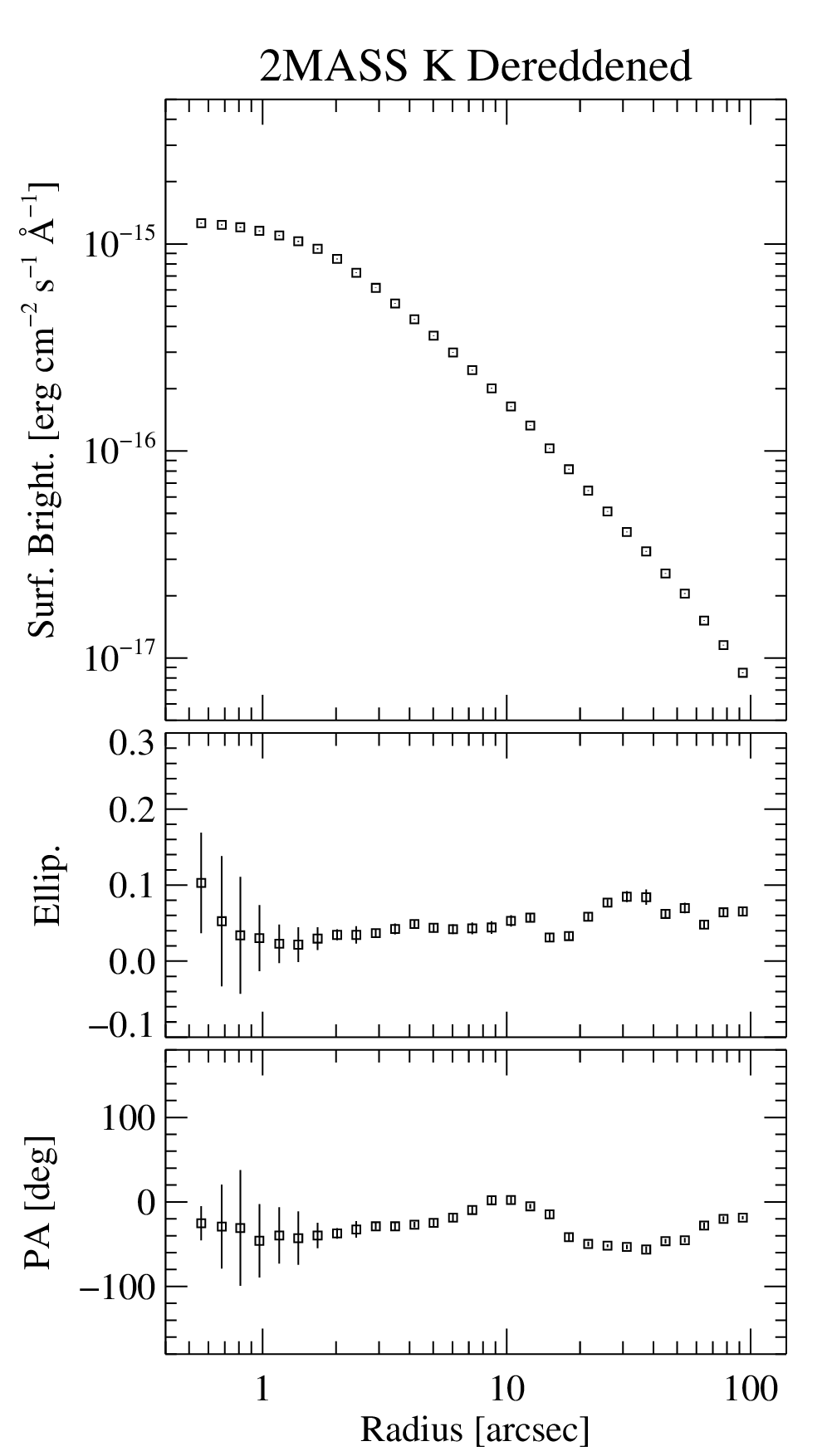}
\caption{\label{fig:ellipsefit}Results of ellipse fitting on dereddened NICMOS F222M (left) and 2MASS K images (right). The top panels show the observed surface brightness as a function of radius. Middle panels and lower panels show ellipticity and position angle of the ellipse major axis respectively. }
\end{figure*}
 
The fitting procedure provides line-of-sight velocities, velocity dispersions and surface brightnesses along each slit for the \FeII\ and \PaB\ emission
lines which are shown in Fig.~\ref{fig:isaaccurves}.
In Fig.~\ref{fig:isaaccurves_paper1} we compare our new determination of the
average velocity with that by \cite{marconi:cenabh}. The new rotation curves
does not differ significantly from the old ones apart from being less noisy.
ISAAC kinematical data are described and discussed at length in
\cite{marconi:cenabh} and we refer to that paper for more details.  Here we
concentrate exclusively in the nuclear region, i.e.\ $\le 2\arcsec$ from the
position of the K band peak. In the nuclear region the ISAAC data exhibit the
same behaviour as the STIS ones, i.e.\ strongly peaked surface brightness
profiles, S-shaped rotation curves with amplitudes that vary with slit PA
($\sim 150\KMS$ at PA1, $\sim 200\KMS$ at PA2, and $\sim 50\KMS$ at PA3).  The
observed velocity dispersion is characterized by a well defined constant
plateau, expecially in PA1, and a smoth increase toward the nucleus location
where it reaches an amplitude of $\sim 250\KMS$ at all slit orientations.
$h_3$ and $h_4$ are significantly different from 0 only in within 1\arcsec\ of
the nucleus where the SNR of the data is higher. In all other locations they
have been fixed  to 0.

In Fig.~\ref{fig:compare} we compare the rotation curve from STIS at VIS2 and
the rotation curves from ISAAC at PA1.  This comparison is not affected by
geometrical effects because the two slits have a similar orientation, differing
only by 5\deg.  The ISAAC data are characterized by a much higher SNR and a by
a smoother rotation curve. The effect of the different spatial resolution and
slit width is clearly that of smoothing away the high velocity increase
observed in the STIS data. This effect is expected and was discussed at length
by \cite{macchetto:m87bh}. The STIS rotation curve is thus in good agreement
with the ISAAC data apart for the two couples of points 
$\pm 0\farcs7$ away from the center. It is difficult to establish
whether the discrepancy is due to low SNR of the data and/or to the presence of
non circular motions in the \SIII\ emitting medium. Since we have verified that the inclusion of those points does not change the final results apart for incresing \chisq, we have decided not to consider them
in the following analysis.

\section{\label{sec:models}Modeling the rotation curves}

The aim of our new HST/STIS observations is to confirm at higher angular
resolution the detection of the supermassive \BH\ in Centaurus A by
\cite{marconi:cenabh}.  Both STIS and ISAAC data are analized to verify if
they provide consistent results.

To model the kinematical data we follow the procedure first described by
\cite{macchetto:m87bh} and subsequently refined by several authors
\citep{vandermarel:n7052bh,marconi:cenabh,barth:n3245bh,marconi:n4041bh}.  Our
modeling code, described in detail by \cite{marconi:n4041bh},
was used to fit the observed rotation curves. Very briefly the code computes
the rotation curves of the gas assuming that the gas is rotating in circular
orbits within a thin disk in  principal plane of the galaxy potential. We neglect any hydrodynamical effect like gas pressure. The
gravitational potential is made of two components: the stellar potential,
characterized by its mass-to-light ratio \ML, and a dark mass concentration
(the putative black hole), spatially unresolved at HST+STIS resolution and
characterized by its total mass \MBH.  In computing the rotation curves we take
into account the finite spatial resolution of the observations, the intrinsic 
surface brightness distribution of the emission lines (hereafter ISBD)
and we integrate over the slit and pixel area.
The  free parameters characterizing the best fitting
model are found by standard $\chi^2$ minimization.

In Sec.~\ref{sec:stellamass} we determine the radial light profile of the
stellar component which, multiplied, by \ML\ directly provide the enclosed mass
at a given distance from the nucleus.  In Sec.~\ref{sec:lineflux} we determine
the ISBD which is the
weight for averaging the kinematical quantities.  Finally, in
Sec.~\ref{sec:kinfitting} we present the results of the kinematical fitting
obtained using the stellar mass distribution and intrinsic surface brightness
previously determined. In the same section we also present a new way to compute
the kinematical model quantities in order to avoid subsampling problems.

\subsection{\label{sec:stellamass}The stellar mass distribution}
\begin{figure}[!]
\centering
\includegraphics[width=0.95\linewidth]{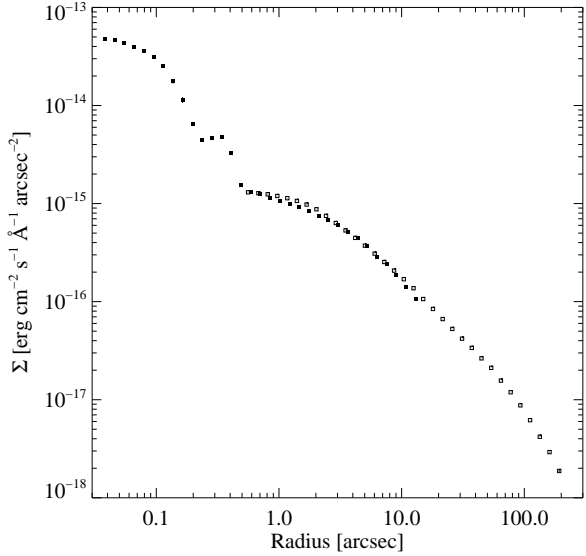}
\caption{\label{fig:profcompare}
Comparison of the surface brightness profiles obtained
from reddening corrected NICMOS F222M (filled
squares) and 2MASS K images (empty squares).  }
\end{figure}
\begin{figure}[!]
\centering
\includegraphics[width=0.95\linewidth]{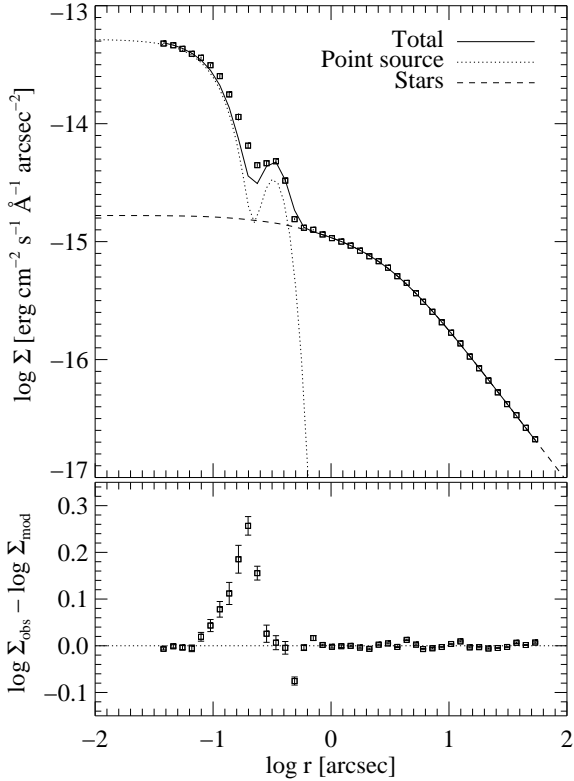}
\caption{\label{fig:stellarfit} Upper panel: brightness profile of Centaurus A
in the K band resulting from merging the NICMOS F222M and 2MASS K profiles.
Surface brightness is in units of \ERG\SEC\1\CM\2\AA$^{-1}$\,arcsec$^{-2}$. The
solid line is the fit to the profile obtained from a spherical stellar density
distribution with an added nuclear point source. The dotted and dashed lines show the relative contributions of point source and stellar density, respectively. }
\end{figure}
aspect

The inversion procedure to derive the stars distribution from the surface
brightness is not unique if the gravitational potential does not have a
spherical symmetry, revealed by circular isophotes.  Assuming that the
gravitational potential is an oblate spheroid, the inversion depends on the
knowledge of the potential axial ratio $q$, and the inclination of its
principal plane with respect to the line of sight, $i$.
As these two quantities are
related by the observed isophote ellipticity, we are left with the freedom of
assuming different galaxy inclinations to the line of sight.  
Following \cite{vandermarel:n7052bh}, we 
assumed an oblate spheroid density distribution parameterized as:
\begin{equation}
\rho(m) = \rho_0\left(\frac{m}{r_b}\right)^{-\alpha} \left[1+\left(\frac{m}{r_b}\right)^2\right]^{-\beta}
\end{equation}
$m$ is given by $m^2 = x^2+y^2+z^2/q^2$ where $xyz$ is a reference system with
the $xy$ plane corresponding to the principal plane of the potential and $q$ is
the intrinsic axial ratio.  This three-dimensional density model is converted
to an observed surface brightness distribution in the plane of the sky by
integrating along the line of sight, convolving with the Point Spread Function
(PSF) of the telescope+instrument system and averaging over the detector pixel
size. Then, the derived model light profile can be directly compared
with observed ones.  A detailed description of the
relevant formulas and of the inversion and fit procedure is presented in the
\cite{marconi:n4041bh}. 

We reconstructed the galaxy light profile combining a NICMOS F222M (K band)
image obtained with the NIC2 camera (pixel size 0\farcs075; see
\citealt{schreier:cenadisk} and \citealt{marconi:cenahst} for details)
with a 2MASS K band image from the Large
Galaxy Atlas (pizel size 1\arcsec; \citealt{jarrett:2massLGA}).
The first image,
with smaller pixel size and better spatial resolution was used for the central
regions (r $<$ 6\arcsec) while for the more extended emission we took advantage
of the larger field of view of the 2MASS image.

The nuclear region of Centaurus A is strongly reddened due to the well known
dust lane. This effect can be seen also in the K band and significantly affects
the profile determination. Therefore we applied a reddening correction to the
images, following \cite{marconi:cenahst} where the reader should refer to for more details. Briefly, using available NIC2 F160W and 2MASS J
band images of Centaurus A, we computed the $H-K$ and $J-K$ colors. Assuming as intrinsic
colors $H-K=0.2$ and $J-K=0.92$ we can then estimate reddening using the
extinction curve by \cite{cardelli:dust} with $A_V/E(B-V)=3.1$.
Fig.~\ref{fig:dered} shows the effects of this
reddening correction.
The isophotes from the NIC2 F222M image prior to correction are elongated
in a direction perpendicular to the dust lane, while after correction their
circular simmetry is recovered. Similarly, in the 2MASS K band image prior to
correction one can clearly see the distorsion caused by the lower edge of the dust lane running approximately at PA -45\deg. After correction the circular simmetry is well recovered in the central region. There is still a residual distortion due to an over correction in the region where the dust screen approximation fails. However this distortion can be easily taken into account in the ellipse fitting described below.
\begin{figure}[!]
\centering
\includegraphics[width=0.95\linewidth]{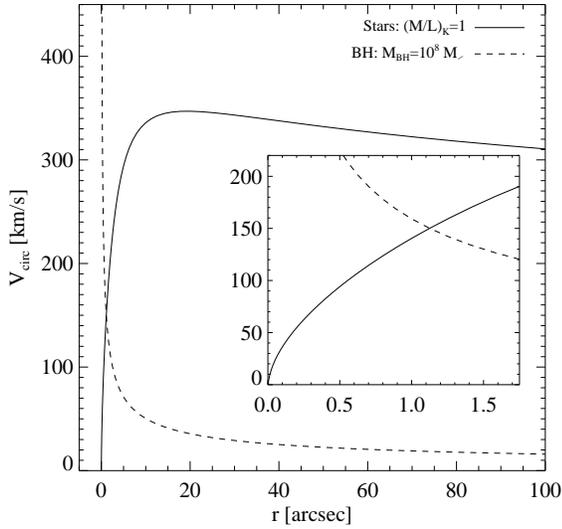}
\caption{\label{fig:vcirc} Circular velocity derived from the model fit in
Fig.~\ref{fig:stellarfit} and computed assuming $\ML=1$ (solid line). The dashed line represents the rotation curve expected in the case of a BH with mass \MBH=\ten{8}\Msun. 
The small inset shows the behaviour of the
circular velocity in the nuclear region. The radius of the BH sphere of influence, i.e.\ the radius at which the contribution of stars and BH to the circular velocity are equal is $\simeq 1\farcs1$. }
\end{figure}

We used the IRAF/STSDAS program \texttt{ellipse} to fit elliptical isophotes
to the galaxy after reddening correction and the results are shown in 
Fig.~\ref{fig:ellipsefit}.

The surface brightness profiles derived from NICMOS data show a central
unresolved source clearly identifiable by its first Airy ring at $\sim
0\farcs3$.  The ellipticity is small being $< 0.1$ if one excludes the inner
0\farcs4 where emission is dominated by the central unresolved source.  The
position angle shows significant variations but it is probably affected by
residual reddening. The surface brightness profile, ellipticity and PA show a
smoother behaviour in the 2MASS data. The ellipticity is still small and
increases to 0.1 only at scales larger than those probed by our spectra
($>20\arcsec$).

As shown in Fig.~\ref{fig:profcompare} the 2 dereddened profiles are in
excellent agreement in the overlap region taking into account that no rescaling
of the data was performed.

In order to determine the stellar contribution to the mass density we can then
safely assume a spherical simmetry since the isophotes, at least in the
nuclear region, are close to circular. Moreover, we can use
an observed light profile obtained by merging the NICMOS
($r\le 6\arcsec$) and the 2MASS ones ($r>6\arcsec$). Since the spatial resolution of the observations affects only the inner parts of the light profiles, 
the PSF to be used in the fitting is the one of NIC2 F222M,
which we take from TinyTim \citep{krist:tinytim}.
\begin{figure*}[!]
\centering
\includegraphics[angle=-90,width=0.85\linewidth]{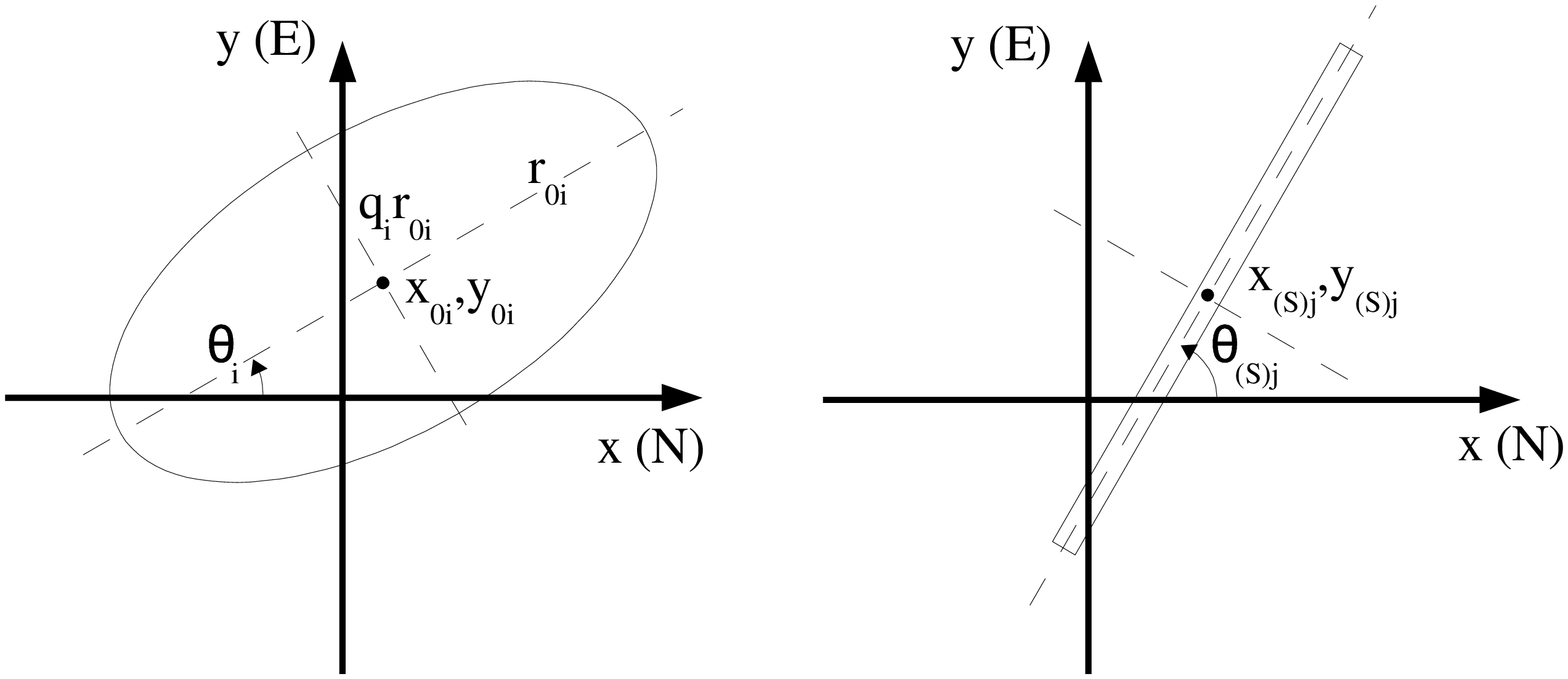}
\vskip 1cm
\caption{\label{fig:reference} Reference frame $xy$ in the plane of the sky.
$x$ and $y$ are aligned along the North and East directions, respectively.
Left: $xy$ characterization of the location of an elliptical isophote of the
$i$-th component of the ISBD. Right: $xy$
characterization of the location of the $j$-th slit.  }
\end{figure*}

The best fit of the observed stellar light profile is shown in
Fig.~\ref{fig:stellarfit} and is characterized by
$\rho_0=(1000\pm3)\,\Msun\PC\3$ (for $M/L=1$), $\alpha = 0.81\pm0.01$, $\beta=0.74\pm0.01$
and $r_b =3\farcs62\pm0\farcs01$ ($\sim 62\PC$ at the distance of Centaurus A).
The $1\sigma$ statistical errors were estimated with 100 montecarlo evaluations
of the fit.  The presence of an unresolved  nuclear point source,
and its associated Airy ring, are clearly visible in Fig.~\ref{fig:ellipsefit}.
Indeed a point source  with flux $(3.39 \pm 0.01)\times 10^{-15}
\ERG\SEC\1\CM\2\AA\1$ ($K=10.1$ mag) had to be added to the extended luminosity
distribution to  provide a  good fit  to the brightness profile. 
The fit is excellent everywhere except for large residuals in the region where
the central unresolved emission dominates. This is not a serious issue
since it is related to the detailed shape of the PSF, which is likely distorted
by the reddening correction procedure described in Sec.~\ref{sec:stellamass}, and does not
affect the stellar emission.
\begin{figure*}[!]
\centering
\includegraphics[angle=90,width=0.8\linewidth]{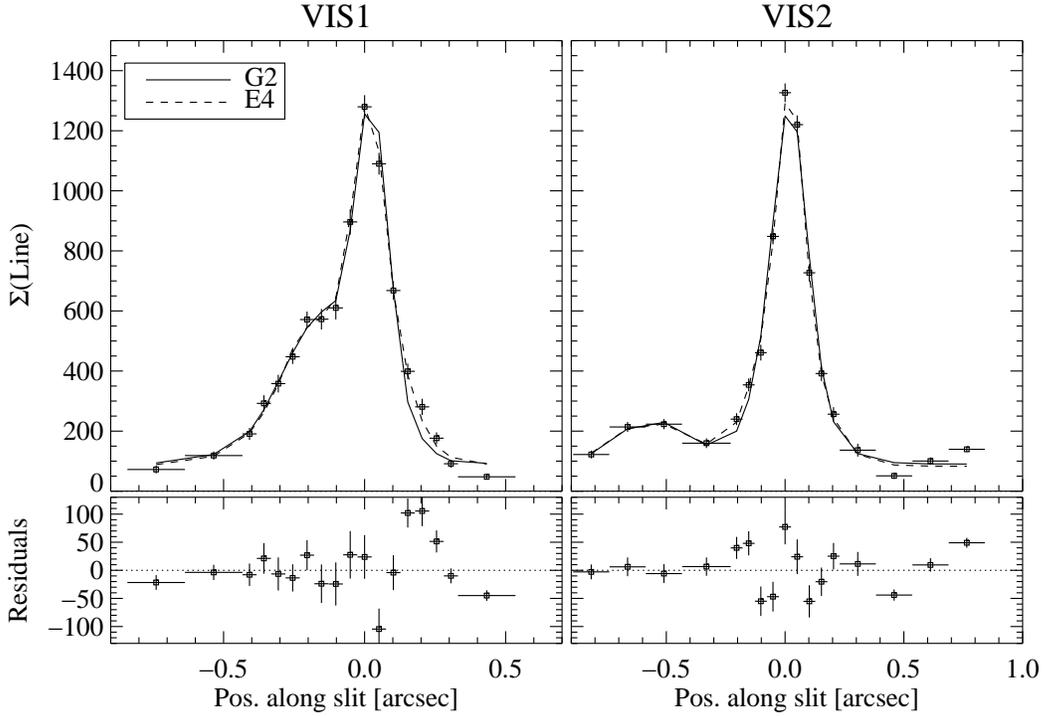}
\caption{\label{fig:stis_flux_fit} Fit of the STIS line surface brightness distribution along the slit for 2 sample cases.  }
\end{figure*}
\begin{figure*}[!]
\centering
\includegraphics[angle=90,width=0.99\linewidth]{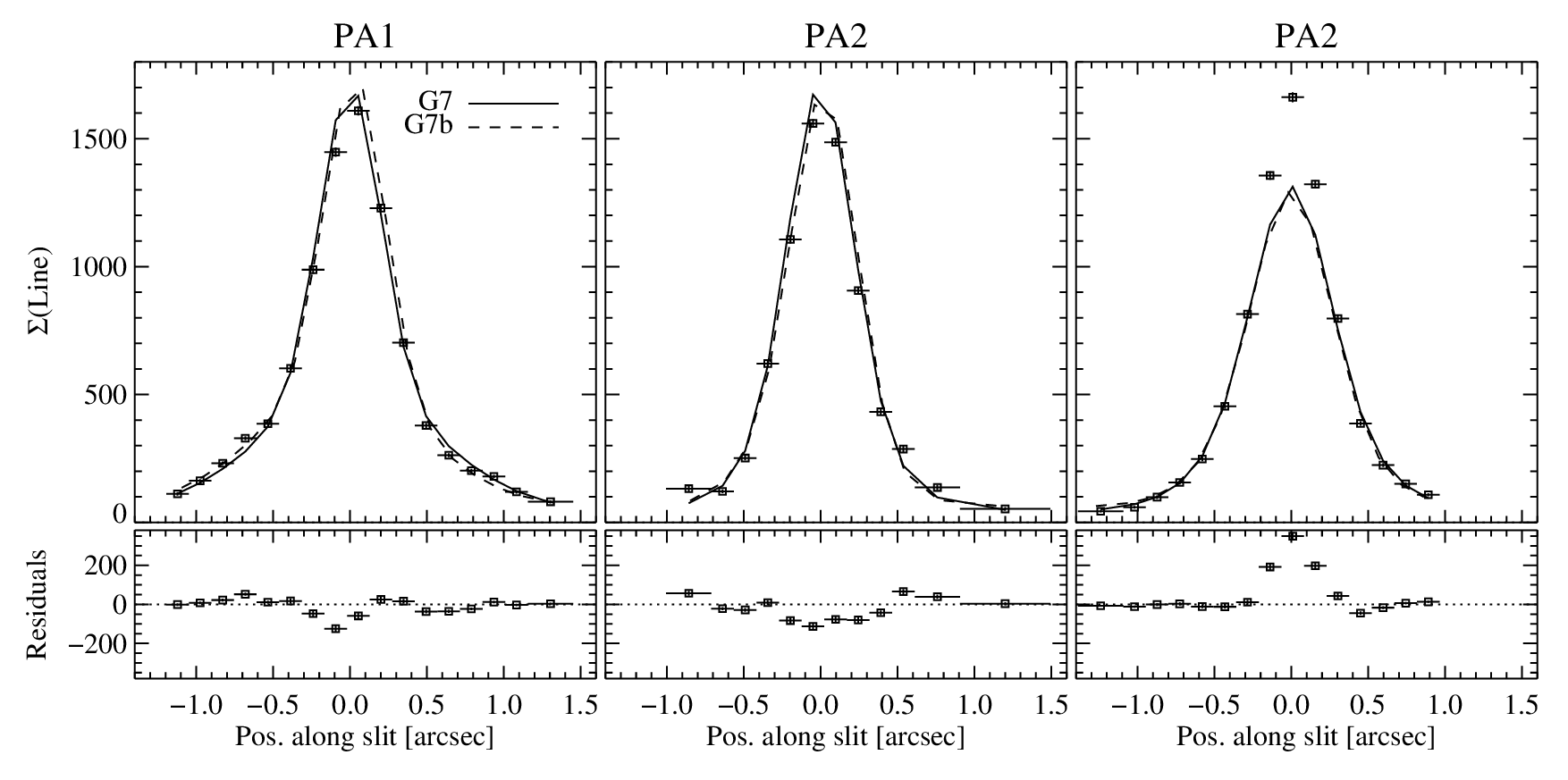}
\caption{\label{fig:isaac_flux_fit} Fit of the ISAAC line surface brightness distribution along the slit for 2 sample cases.  }
\end{figure*}
\begin{figure*}[!]
\centering
\includegraphics[angle=90,width=0.9\linewidth]{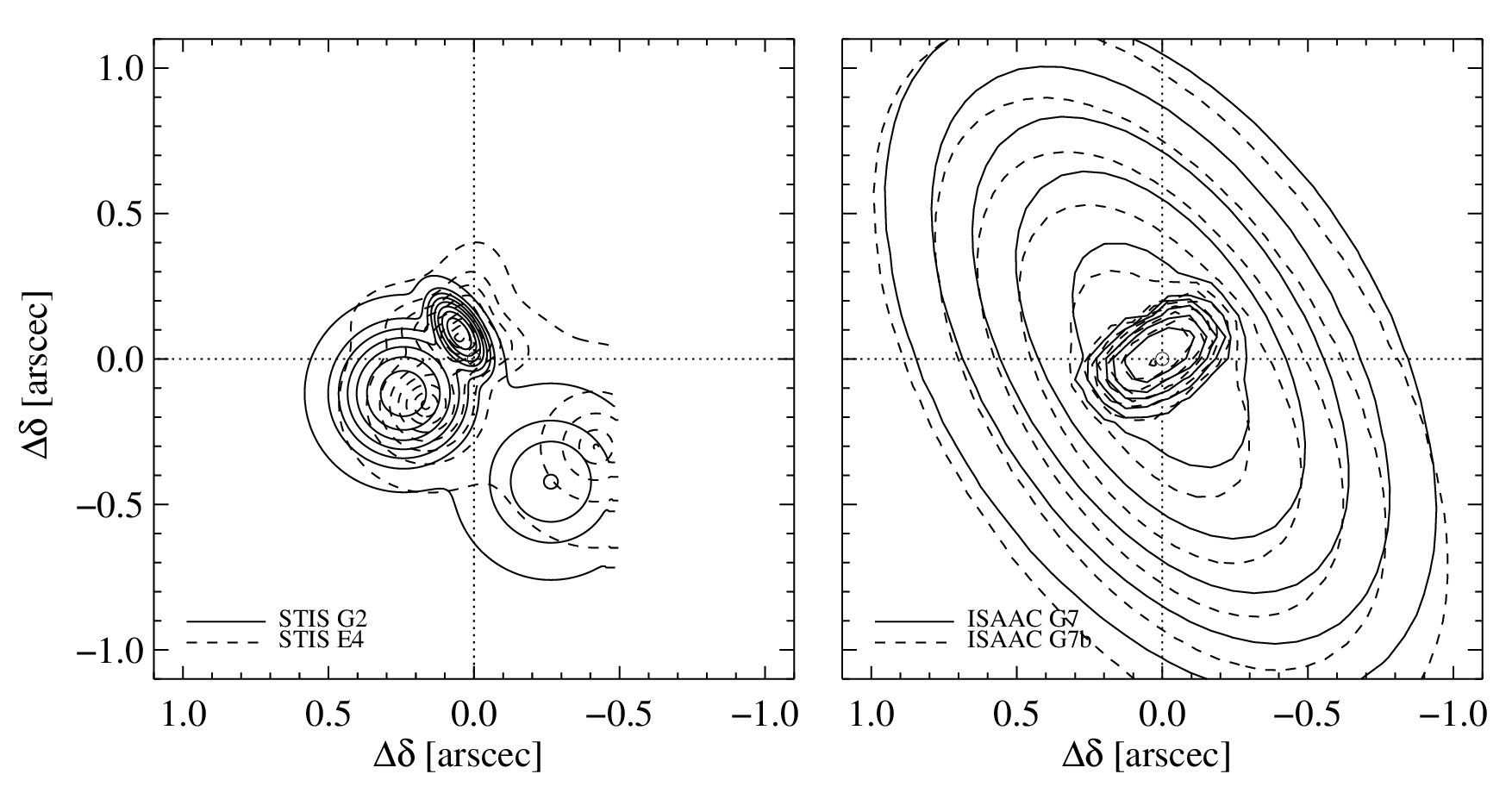}
\caption{\label{fig:fluxdist} Left: isophotal contour levels of the intrinsinc surface brightness distributions G2 and E4 used in Fig.~\protect{\ref{fig:stis_flux_fit}} to reproduce the STIS data. The ISBDs have not been convolved with the instrumental response. Contour levels starts from -2 and go to -0.2 in log of the peak of the ISBD. The filled circle indicate the location of the slit centers. Right: as in the left panel but for the ISBDs G7 and G7b used in Fig.~\protect{\ref{fig:isaac_flux_fit}} to reproduce the ISAAC data.  }
\end{figure*}
It  has been shown by \cite{quillen:hstpointsources} that unresolved
infrared sources are  found in the  great majority  of HST NICMOS images  of
Seyfert galaxies   and  that  their  luminosities  strongly correlate with
both the hard X-ray  and the \OIII\ line luminosity.  This result suggests  a
dominant AGN contribution to  the IR emission.  Moreover, we found that the
central unresolved source is significantly polarized $P=11.1\%$
\citep{capetti:cenapol} and this definitely excludes that it could be emission
from an unresolved star cluster. 

Finally, in Fig.~\ref{fig:vcirc} we plot the circular velocity due to the stars
as a function of the distance from the nucleus, computed assuming $\ML=1$
(solid line). The dashed line represents the circular velocity expected from a
BH with mass \MBH=\ten{8}\Msun. Below 1\arcsec, the BH completely dominates the
gravitational potential, a fact already noted by \cite{marconi:cenabh}. At
$r=1.1\arcsec$, the contribution of BH and stars to the gravitational potential
are equal. This size can be compared with the standard definition of the radius
of the BH sphere of influence, $\rBH=G\MBH/\sigma_e^2$, where $G$ is the
gravitational constant and $\sigma_e$ is the effective velocity dispersion of
the stars. In the case of Centaurus A, $\sigma_e = 138\pm10\KMS$
(\citealt{silge:cenabh}, see Sec.\ref{sec:correlations}) which combined with
\MBH=\ten{8}\Msun\ yelds $\rBH=1.3\pm0.3\arcsec$ in excellent agreement with
the estimate from our analysis of the stellar potential.

\subsection{\label{sec:lineflux}The intrinsic line surface brightness distribution}
\begin{table*}
\caption{Intrinsic line surface brightness distributions for STIS\label{tab:stis_flux}}    
\begin{tabular}{l l c r r r r r r }
\hline\hline       
id & function & $i$ & $I_{0i}$ & $r_{0i}$ & $x_{0i}$ & $y_{0i}$ & $\theta_i$ & $q_i$ \\
\hline                    
G2 & gauss & 1 & 3685 & 0.145 & 0.028 & 0.004 & 32.5$^\star$ & 0.5$^\star$ \\
   & gauss & 2 & 1950 & 0.231 & -0.178 & 0.196 & 0.00$^\star$ & 1.0$^\star$ \\
   & gauss & 3 & 295  & 0.278 & -0.481 & -0.311 & 0.00$^\star$ & 1.0$^\star$ \\
   & const & 4 & 90  &       &        &        &              &             \\
\hline
E4 & expo & 1 & 2438 & 0.066 & 0.016 & 0.036 & 77.5$^\star$ & 0.5$^\star$ \\
   & expo & 2 & 3333 & 0.056 & 0.023 & 0.006 & -12.5$^\star$ & 0.5$^\star$ \\
   & expo & 3 & 1948 & 0.058 & -0.217 & 0.103 & 0.00$^\star$ & 1.0$^\star$ \\
   & expo & 4 & 440  & 0.097 & -0.458 & -0.352 & 0.00$^\star$ & 1.0$^\star$ \\
   & const & 5 & 83  &       &        &        &              &             \\
\hline                  
\end{tabular}\\
$^\star$ fixed parameter.
\vspace{1cm}
\caption{Intrinsic line surface brightness distributions for ISAAC\label{tab:flux}}    
\begin{tabular}{l l c r r r r r r }
\hline\hline       
id & function & $i$ & $I_{0i}$ & $r_{0i}$ & $x_{0i}$ & $y_{0i}$ & $\theta_i$ & $q_i$ \\
\hline                    
G7  & gauss & 1 & 40500 & 0.119 & 0.00$^\star$ & 0.00$^\star$ &122.5$^\star$ & 0.5$^\star$ \\
    & gauss & 1 & 591   & 1.262 & 0.00$^\star$ & 0.00$^\star$ & 32.5$^\star$ & 0.5$^\star$ \\
    & gauss & 1 & 68    & 3.407 & 0.00$^\star$ & 0.00$^\star$ & 0.00$^\star$ & 1.0$^\star$ \\
\hline
G7b & gauss & 1 & 41184 & 0.121 & 0.106        & 0.009        & 12.5$^\star$ & 0.5$^\star$ \\
    & gauss & 1 & 627   & 1.190 & 0.00$^\star$ & 0.00$^\star$ & 32.5$^\star$ & 0.5$^\star$ \\
    & gauss & 1 & 64    & 9.478 & 0.00$^\star$ & 0.00$^\star$ & 0.00$^\star$ & 1.0$^\star$ \\
\hline                  
\end{tabular}\\
$^\star$ fixed parameter.
\end{table*}
The observed kinematical quantities are averages over apertures defined by the
slit width and the detector pixel size along the slit.  The ISBD is thus a fundamental ingredient
in the modeling of the kinematical quantities because it is the weight of the
averaging process (see, e.g., the discussion in \citealt{marconi:n4041bh}
and Appendix B).
\cite{barth:n3245bh} derived the ISBD by
deconvolving a continuum subtracted emission line image and were successful in
reproducing the microstructure of the rotation curves thus reducing the overall
\chisq\ of their best fitting model. However, apart for the microstructure of
the rotation curves which has little impact on the final \MBH\ estimate,
\cite{marconi:n4041bh} showed that it is crucial to adopt a good ISBD in the central region where the velocity gradients are the
largest. This has a strong impact on the quality of the fit but also
on the final \MBH\ estimate.  When the observed line surface brightness distribution along
the slit is strongly peaked in the nuclear region and has a profile which is
little different from that of an unresolved source, as in the present case, one
cannot simply use a deconvolved emission line image but should try to reproduce
the observed line surface brightness distribution with a parameterized intrinsic one and
taking into account the intrumental effects.
Therefore we decided to parameterize the ISBD with a combination of analytic functions.
We will show in the following analysis that the final result, i.e.\ the BH mass, is very little dependent on the assumed ISBD.

We adopt the reference system $xy$ in the plane of the sky where the $x$ and $y$ axes are
aligned with the North and East directions, respectively (see Fig.~\ref{fig:reference}).
The total emission line surface brightness at a point $x,y$ is represented in our model by
\begin{equation}\label{eq:surfbright}
I(x,y) = \sum_i\,I_{0i} f_i\parfrac{r_i}{r_{0i}}
\end{equation}
where $I_{0i}$ is the amplitude of the $i$-th component which is described by the
analytic function $f_i(r_i/r_{0i})$. $r_{0i}$ is a scale radius and
$r_i$ is the distance from a simmetry center defined by
\begin{eqnarray}
x_i^\prime & = & x - x_{0i}\nonumber \\
y_i^\prime & = & y - y_{0i}\nonumber \\
x_i^{\prime\prime} & = & x_i^\prime \cos\theta_i + y_i^\prime \sin\theta_i\nonumber \\
y_i^{\prime\prime} & = & (-x_i^\prime \cos\theta_i + y_i^\prime \sin\theta_i) / q_i\nonumber\\
r_i & = & \left[ (x_i^{\prime\prime})^2+(y_i^{\prime\prime})^2 \right]^{1/2}
\end{eqnarray}
Thus, the isophotes of the $i$-th surface brightness component are ellipses 
centered in $(x_{0i},y_{0i})$, with axial ratio $q_i$ (minor over major axis)
and major axis aligned along a direction with position angle $PA=\theta_i$
(Fig.~\ref{fig:reference}).
The adopted analytic functions are exponentials, gaussians and constants
defined as:
\begin{eqnarray}
f_i(r_i) & = & \exp\left(-r_i/{r_{0i}}\right)\nonumber\\
f_i(r_i) & = & \exp\left[-\frac{1}{2}(r_i/{r_{0i}})^2\right]\nonumber\\
f_i(r_i) & = & 1~~(r_i<=r_{01});~0~~(r_i>r_{01})
\end{eqnarray}

The ISBD of Eq.~\ref{eq:surfbright} is then
convolved with the PSF of the system and integrated over apertures defined
by the slit and the detector pixel size. The free parameters of the fit are determined
with a \chisq\ minimization by comparing observed and model data.  The position of the $j$-th in the
$xy$ reference system is characterized by $x_{(S)j},y_{(S)j}$, the slit center position
and $\theta_{(S)j}$ its position angle on the plane of the sky.
We now derive the ISBDs
for STIS and ISAAC data.

\begin{table*}[!]
\caption{\label{tab:stis_vel_fit} Fit results from the analysis of STIS data
with $i=25\DEG$.}
\begin{tabular}{lcccccccr}
\hline\hline
\\
Flux$\,^a$ & $x_0$ & $y_0$ & $\log\MBH$ & $\log\ML$ & $\theta$ & \vsys & \chisqred & $({\chisqred})_\mathrm{resc}\,^b$ \\
\\
\hline
\\
\multicolumn{9}{c}{Fit of velocity $(\Delta v_0=10.4\KMS)\,^c$} \\
G2 & -0.05 & -0.05 & 8.00 & 0.00$\,^\star$ & 161.0 & 554.9 & 1.27 & 1.00 \\ 
E4 & -0.06 & -0.04 & 7.89 & 0.00$\,^\star$ & 167.0 & 547.3 & 1.30 & 1.05 \\ 
G1 & -0.06 & -0.04 & 7.93 & 0.00$\,^\star$ & 162.6 & 548.7 & 1.32 & 1.05 \\ 
E6 & -0.06 & -0.03 & 7.89 & 0.00$\,^\star$ & 163.6 & 547.1 & 1.35 & 1.08 \\ 
G6 & -0.06 & -0.03 & 7.89 & 0.00$\,^\star$ & 164.3 & 545.5 & 1.37 & 1.09 \\ 
E7 & -0.06 & -0.03 & 7.88 & 0.00$\,^\star$ & 163.9 & 546.1 & 1.41 & 1.13 \\ 
E8 & -0.05 & -0.04 & 7.87 & 0.00$\,^\star$ & 168.7 & 543.4 & 1.41 & 1.14 \\ 
E9 & -0.05 & -0.03 & 7.86 & 0.00$\,^\star$ & 169.5 & 544.1 & 1.42 & 1.15 \\ 
G7 & -0.06 & -0.03 & 7.90 & 0.00$\,^\star$ & 162.9 & 545.1 & 1.48 & 1.17 \\ 
E2 & -0.06 & -0.04 & 7.95 & 0.00$\,^\star$ & 161.3 & 548.5 & 1.51 & 1.18 \\ 
C6 & -0.06 & -0.03 & 7.93 & 0.00$\,^\star$ & 161.1 & 546.2 & 1.51 & 1.20 \\ 
E1 & -0.06 & -0.04 & 7.91 & 0.00$\,^\star$ & 169.1 & 543.4 & 1.50 & 1.20 \\ 
E5 & -0.04 & -0.05 & 7.91 & 0.00$\,^\star$ & 178.9 & 541.2 & 1.49 & 1.21 \\ 
C1 & -0.05 & -0.02 & 7.80 & 0.00$\,^\star$ & 168.9 & 539.7 & 1.55 & 1.24 \\ 
E3 & -0.05 & -0.03 & 7.84 & 0.00$\,^\star$ & 173.8 & 538.7 & 1.60 & 1.29 \\ 
G4 & -0.04 & -0.04 & 7.89 & 0.00$\,^\star$ & 177.5 & 539.0 & 1.60 & 1.30 \\ 
G8 & -0.05 & -0.04 & 7.86 & 0.00$\,^\star$ & 172.9 & 539.2 & 1.65 & 1.33 \\ 
C8 & -0.05 & -0.03 & 7.81 & 0.00$\,^\star$ & 170.5 & 540.1 & 1.69 & 1.36 \\ 
C4 & -0.05 & -0.04 & 7.85 & 0.00$\,^\star$ & 174.8 & 544.4 & 1.72 & 1.40 \\ 
G3 & -0.05 & -0.02 & 7.82 & 0.00$\,^\star$ & 174.2 & 537.1 & 1.77 & 1.43 \\ 
G9 & -0.05 & -0.05 & 7.93 & 0.00$\,^\star$ & 174.3 & 537.1 & 1.80 & 1.44 \\ 
C7 & -0.05 & -0.01 & 7.79 & 0.00$\,^\star$ & 160.4 & 545.4 & 1.86 & 1.47 \\ 
\hline
\\
average$\,^d$  &  -0.05$\pm$0.01 & -0.03$\pm$0.01 & 7.88$\pm$0.05 &  & 168.4$\pm$5.7 & 543.7$\pm$4.3 \\
\\
\hline
\end{tabular}
\\
\\
$^a$ Adopted ISBD.\\
$^b$ Rescaled \chisq\ with errors computed as $\Delta v_i^\prime\,^2 =\Delta v_i\,^2 +\Delta v_0\,^2$.\\
$^c$ Systematic error adopted to renormalize \chisq.\\
$^d$ Average and rms of best fit parameter values.\\
$^\star$ Parameter was held fixed.\\
\end{table*}

\subsubsection{\label{sec:linefluxstis}STIS data}
The acquisition procedure followed during HST observations (i.e.
centering on a bright star and slew to the expected nucleus location)
ensures that 
the STIS slits are centered on the same point within an accuracy of a fraction
of a pixel.
Inspection
of Fig.~\ref{fig:stiscurves} then suggests that the ISBD
should be described at least by 4 components: one at the slit center (the
nuclear component), one offset by -0\farcs2 along the slit at VIS1, one at -0\farcs6 along the slit at
VIS2 and a constant base. The flattening and orientation of the offnuclear
components can be left as free parameters of the fit. However, 
in the case of a nuclear component with a size smaller than the spatial resolution it is safer to reduce the number of free parameters 
and therefore we have considered the following cases
for the nuclear component:
\begin{itemize}
\item [(1)] nuclear component circularly symmetric;
\item [(2)] nuclear component with $PA=32.5\DEG$ and $q=0.5$; 
\item [(3)] nuclear component with $PA=122.5\DEG$ and $q=0.5$; 
\item [(4)] 2 nuclear components with $PA=77.5,-12.5\DEG$ and $q=0.5$; 
\item [(5)] 2 nuclear components with $PA=77.5,-12.5\DEG$ and $q=0.17$; 
\item [(6)] 2 nuclear components circularly symmetric; 
\item [(7)] 2 nuclear components, one circularly symmetric and the other with
$PA=32.5\DEG$ and $q=0.5$; 
\item [(8)] 2 nuclear components, one circularly symmetric and the other with
$PA=122.5\DEG$ and $q=0.5$; 
\item [(9)] 2 nuclear components with $PA=32.5,122.5\DEG$ and $q=0.5$; 
\end{itemize}
$PA=32.5\DEG$ is the PA of the Centaurus A jet and is the direction along which
\PaA\ and \FeII\ emissions are elongated  and $q=0.5$ is their axial ratio
\citep{schreier:cenadisk,marconi:cenahst}.
$PA=122.5\DEG$ is the direction perpendicular to the previous one.
In the case of 2 components with $PA=77.5,-12.5\DEG$ we mimick an ionization
cone with axis at $PA=32.5\DEG$ and opening angle of 45\deg.

For each of the above cases we have used exponential (E), gaussian (G) and
constant (C) functions.  An ISBD is then labeled as, e.g, E5 which
means exponential functions in case 5 above.

Sample results of the fitting procedure are shown in
Fig.~\ref{fig:stis_flux_fit} for the ISBDs G2 and E4 and the best
fit parameters for these ISBDs are shown in
Tab.~\ref{tab:stis_flux}.  Inspection of Fig.~\ref{fig:stis_flux_fit} indicates
that ISBDs G2 and E4, though different,
result in very similar observed surface brightness
distributions.

\begin{figure*}[!]
\centering
\includegraphics[angle=90,width=0.85\linewidth]{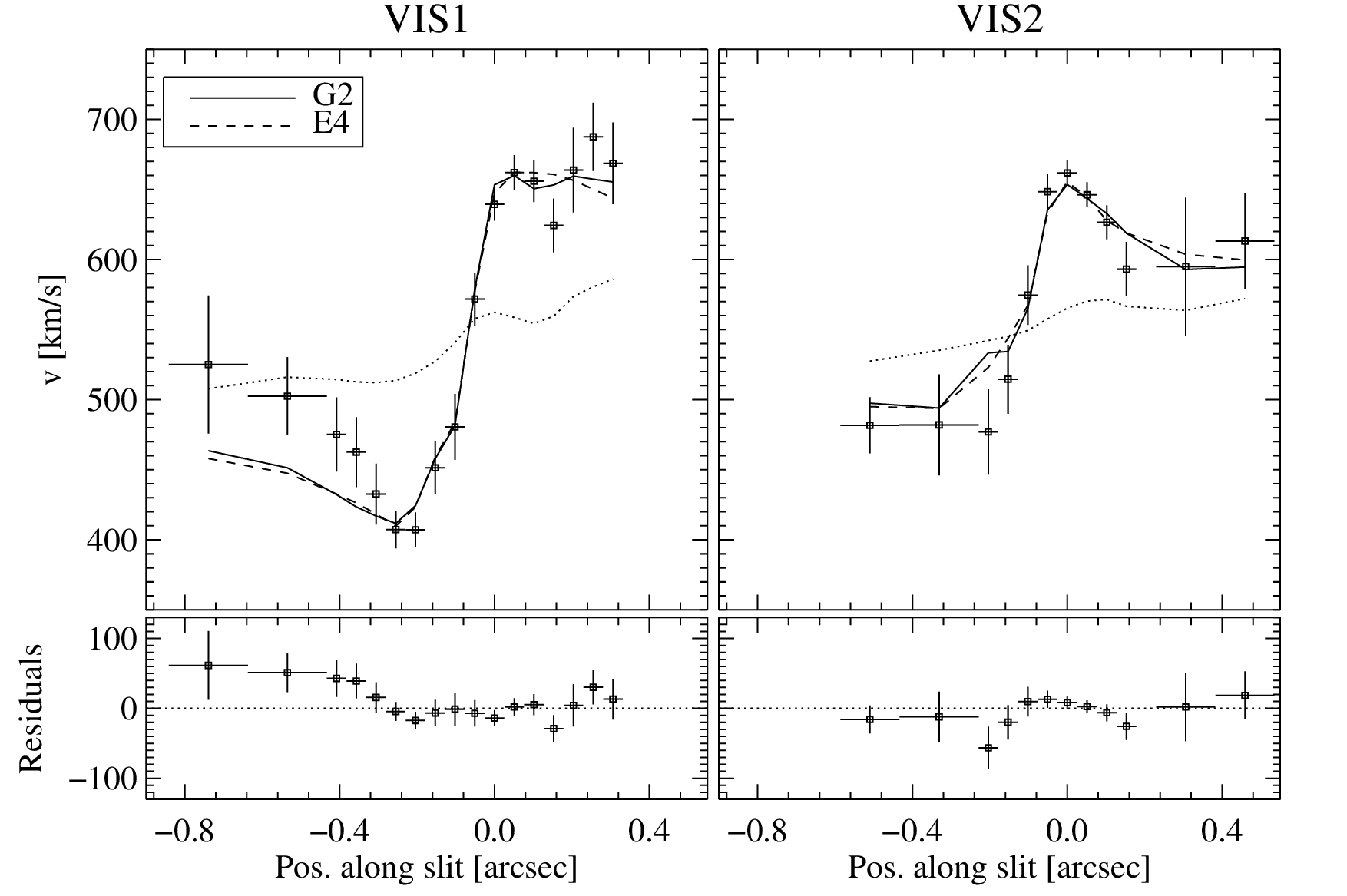}
\includegraphics[angle=90,width=0.85\linewidth]{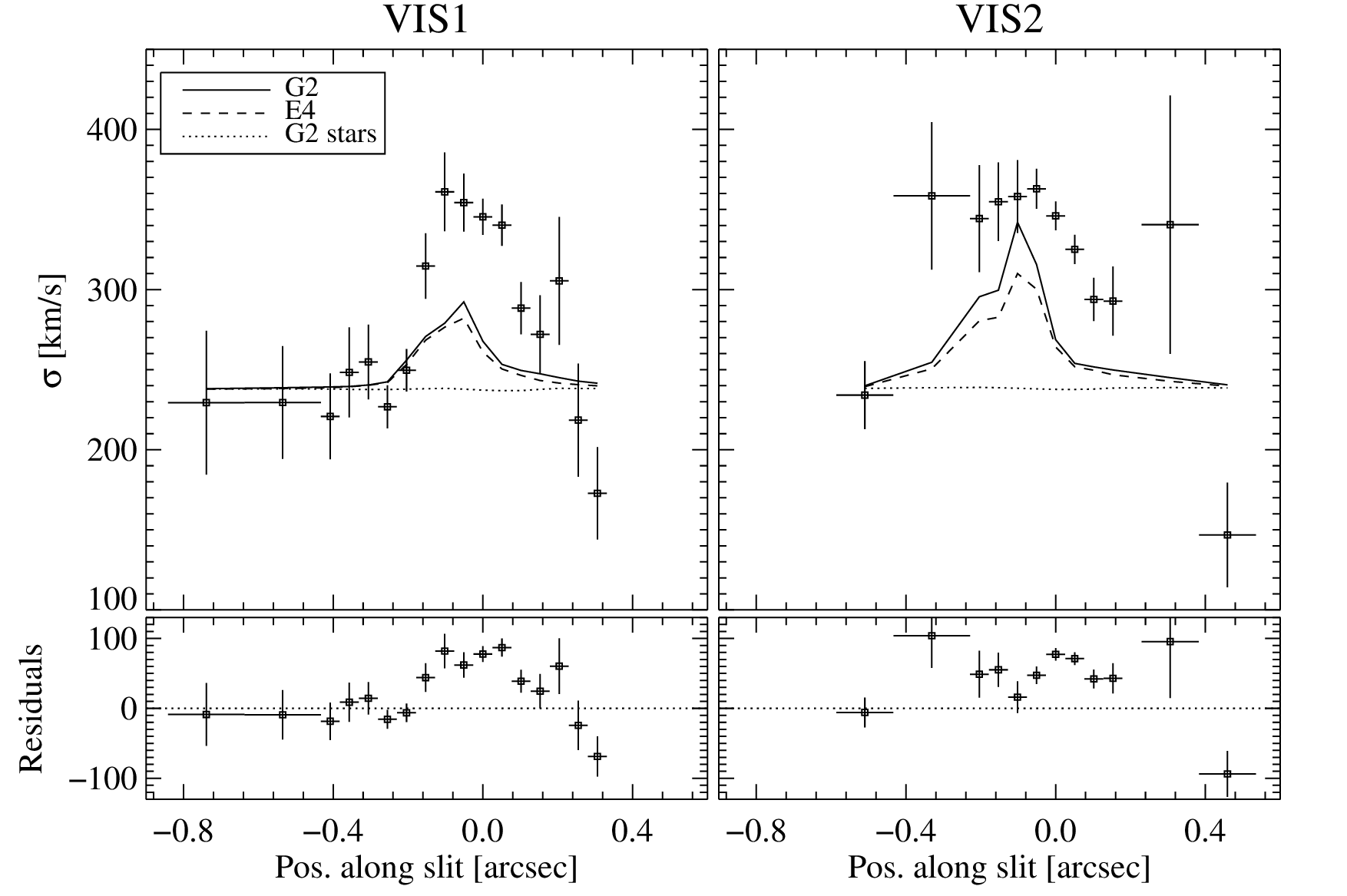}
\caption{\label{fig:stis_vel_fit}\label{fig:stis_sig} Top panels: Fit of the
STIS velocities along the slit. The empty squares with error bars represent the
observed values while the solid lines connect corresponding model values.
Dotted lines represent the contribution of the mass in stars to the rotation
curve, i.e.\ what would be observed without a BH.  'G2' and 'E4' are the two
ISBDs which provide the best fit models with the lowest \chisq.
Bottom panels: Velocity dispersions expected from the two best fit models
compared with the
observed ones. Notation as in the upper panels.}
\end{figure*}

\subsubsection{\label{sec:linefluxisaac}ISAAC data}

The observational procedure followed during ISAAC observations (i.e. centering
the slit on the K band peak, \citealt{marconi:cenabh}) ensures that the three
slits are centered on the same point within a fraction of the ISAAC pixel.
However it is not clear if the emission line peak observed in \PaB\ and \FeII\
is coincident with the continuum peak.  \cite{schreier:cenadisk} analyze a
\PaA\ image and find that the continuum subtracted line image has a point like
emission (at NICMOS resolution) which is concident with that of the continuum.
The same happens for \FeII\ at 1.6\MIC\ \citep{marconi:cenahst}.
Regardless of these indications in order to avoid biases in the final
results, we consider two cases, when the slits
are centered on the emission line peak (coincident with the continuum
peak) and when they are not.

The seeing of the observations can be estimated from the unresolved continuum
source observed along the slit in the ISAAC observations.  However, the
unresolved continuum source in the J band is not as strong as in the K band and
the estimate of the seeing can be uncertain by as much as 0\farcs1.
We parameterize the seeing as a gaussian with given FWHM (Full Width at Half Maximum) and
we consider 3 cases for the FWHM values:
\begin{itemize}
\item seeing 0\farcs4 for PA1, PA2 and  seeing 0\farcs5 for PA3;
\item seeing 0\farcs45 for PA1, PA2 and  seeing 0\farcs55 for PA3;
\item seeing 0\farcs5 for PA1, PA2 and PA3;
\end{itemize}

Inspection
of Fig.~\ref{fig:isaaccurves} suggests that the ISBDs
of \PaB\ and \FeII\
can be described with 4 components: a nuclear component, 
a disk component (the one associated with the extended nuclear feature observed in
\PaA\ by \citealt{schreier:cenadisk}), and an extended one.
The flattening and orientation of the disk components are fixed to $q=0.5$
and $\theta=32.5$, i.e. the values characterizing the Pa$\alpha$ feature,
and the extended component is assumed circularly symmetric.
We then parameterize the nuclear component according to the following
4 cases:
\begin{enumerate}
\item nuclear component circularly symmetric;
\item nuclear component with $PA=32.5\DEG$ and $q=0.5$; 
\item nuclear component with $PA=122.5\DEG$ and $q=0.5$; 
\item 2 nuclear components with $PA=77.5,-12.5\DEG$ and $q=0.5$.
\end{enumerate}
The adopted values of $PA$ and $q$ were chosen as described in the previous
section.
For each of the above cases we have used exponential (E), gaussian (G) and
constant (C) functions.  A surface brightness distribution is labeled by assigning a letter
E, G or C according to the functional form used, a number 1-4 (seeing case 1),
5-8 (seeing case 2) or 9-c (seeing case 3). The label is further characterized
by 'b' if we are in the case when the emission line peak is not coincident with
the slit centers.  Therefore E5 indicates, for instance, exponential functions
with surface brightness case 1 and seeing case 5 while G2b indicates gaussian functions in
surface brightness case 2 and seeing case 1.

\begin{figure*}[!]
\centering
\includegraphics[width=0.33\linewidth]{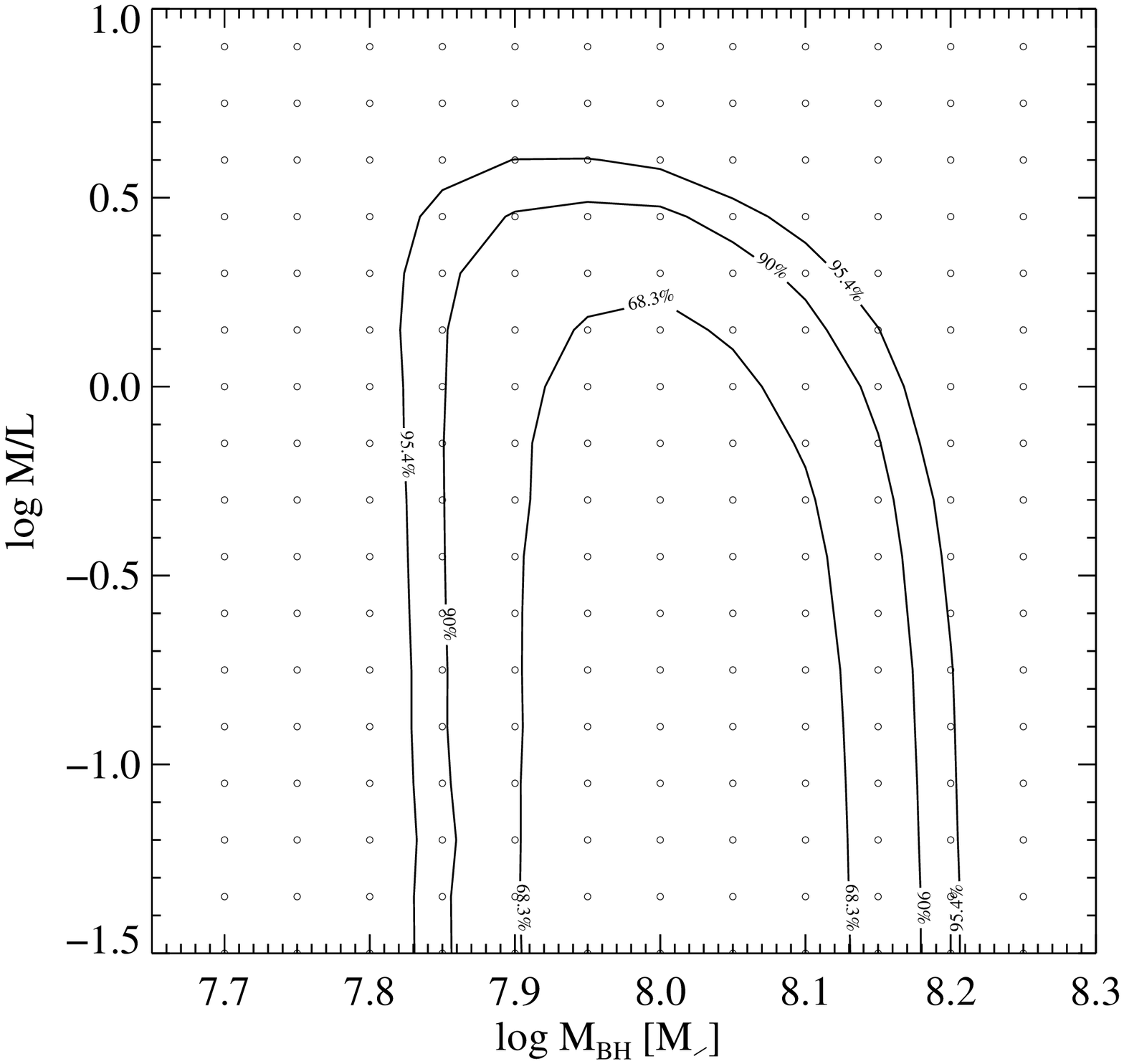}
\includegraphics[width=0.33\linewidth]{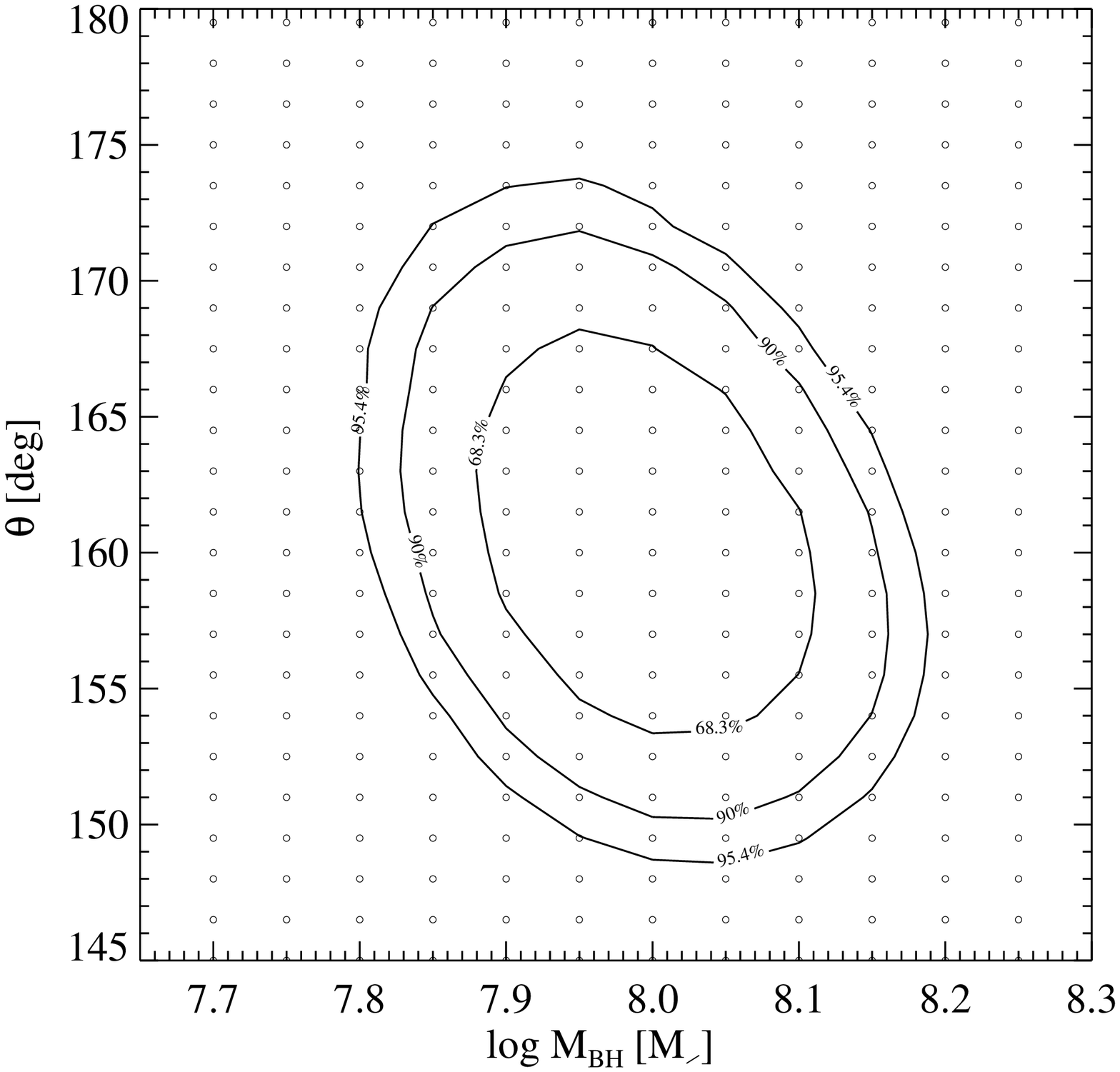}
\includegraphics[width=0.33\linewidth]{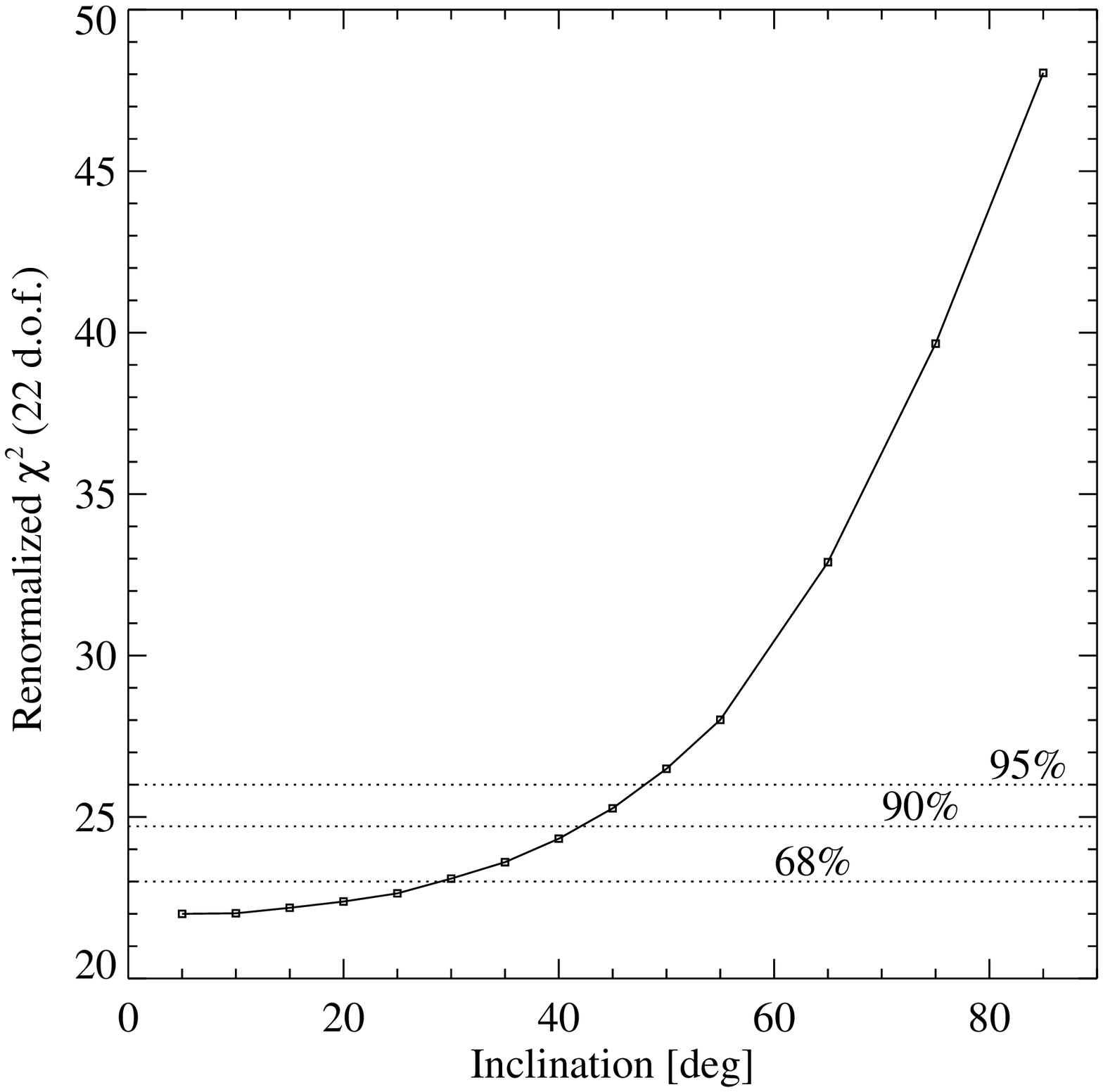}
\caption{\label{fig:stis_grids} (a) \chisq\ contours for the joint variation of
\MBH\ and \ML\ using the STIS data (ISBD G2).
Contour levels are for $\chisq =
\chi^2_\mathrm{min}+2.3$, 4.61, 6.17 corresponding to 68.3\%, 90\%\ and 95.4\%
confidence levels.  The dot indicates the \MBH\ and \ML\ values for which the
\chisq\ minimization was computed. (b) same as (a) but for $\MBH-\theta$.  (c)
Dependence of $\chisq$ on the adopted disk inclination. Dotted lines indicate
the 68.3\%, 90\%\ and 95.4\% confidence levels.}
\end{figure*}
Sample results of the fitting procedure are shown in
Fig.~\ref{fig:isaac_flux_fit} for ISBD G7 and
E7 and the best fit parameters for these ISBDs  
shown in Tab.~\ref{tab:flux}.

In Fig.~\ref{fig:fluxdist} we plot the isophotal contours of the ISBDs
used in Figs.~\ref{fig:stis_flux_fit} and
\ref{fig:isaac_flux_fit}. These ISBDs have been
chosen as representative because they provide the best fits of STIS velocities
(Tab.~\ref{tab:stis_vel_fit}) and of ISAAC velocities and velocity dispersions
(Tab.~\ref{tab:isaac_vel_fit}).  From the figure the centrally peaked
components are clearly distinguished from the more extended ones.  The central
strong components of the ISBDs adopted for the STIS
data is elongated along the jet axis, while the ISAAC ones are perpendicular to
it.  The more extended component in the ISAAC data represents the disk like
feature seen in Pa$\alpha$ by \cite{schreier:cenahst} while it seems that this
component is missing in \SIII\ which shows a different morphology from those of \PaB\ and \FeII.

\subsection{\label{sec:kinfitting}Kinematical Fitting of the Rotation Curves}
\begin{table}
\caption{\label{tab:stis_grid_i} Effect of $i$ variation on best fit parameters
and minimum \chisq\ for STIS data. All models adopt ISBD G2.  Confidence levels
68.3\%, 90\%\ and 95.4\% are at $\chisq = \chi^2_\mathrm{min}+1.0$, 2.71, 4.0,
for one interesting parameter \citep{avni:conflev}.}
\begin{tabular}{lcccccccr}
\hline\hline
\\
$i$ &   $\log\MBH$ & $\log(\MBH\sin^2i)$ & $\theta$ & $\chi^2_\mathrm{resc}$ \\
\\
\hline
\\
 5. &   9.35 &   7.23 &  159.1 & 22.0\\
10. &   8.75 &   7.23 &  159.6 & 22.0\\
15. &   8.41 &   7.24 &  159.6 & 22.2\\
20. &   8.18 &   7.25 &  159.7 & 22.4\\
25. &   8.00 &   7.25 &  161.0 & 22.6\\
30. &   7.87 &   7.27 &  160.8 & 23.1\\
35. &   7.76 &   7.27 &  162.1 & 23.6\\
40. &   7.68 &   7.29 &  162.6 & 24.3\\
45. &   7.62 &   7.32 &  163.8 & 25.3\\
50. &   7.56 &   7.33 &  163.7 & 26.5\\
55. &   7.52 &   7.34 &  165.0 & 28.0\\
65. &   7.55 &   7.46 &  164.7 & 32.9\\
75. &   7.64 &   7.61 &  161.9 & 39.7\\
85. &   8.17 &   8.17 &  160.9 & 48.0\\
\\
\hline
\end{tabular}
\end{table}

In principle, in order to constrain the \BH\ mass, one should compare the
emission line profile predicted by the model with the observed one.  However,
as discussed at length by \cite{marconi:n4041bh}, our approach is to compare
the moments of the line profiles: the average velocity $\vel$ and velocity
dispersion $\sigma$ defined as $\sigma^2=\velsq-\vel^2$. 

This approach requires that the parameterized line profile, adopted in the
fitting process, is able to reproduce the shape of the observed line profiles
within the noise (see also \citealt{barth:n3245bh}).  This condition is
satisfied for STIS data were the observed line profile is well represented by a
single gaussian, after deblending the ``blue'' component which is obviously not
emitted by circularly rotating gas. The above condition is also satisfied for
the ISAAC data where the observed line profiles in the nuclear region are well
matched with the parameterization described above. The Hermite expansion has
the obvious advantage of allowing an excellent description of the line profiles
without introducing the instability and the larger number of parameters (6
instead of 5) typical of fits with two gaussian function.   To our knowledge
this is the first time that the Hermite expansion is applied to the analysis of
emission lines. 

In order to be compared with the observations, model $\vel$ and $\sigma$ are
computed taking into account the spatial resolution of the observations and the
size of the apertures which are used to measure the observed quantities.  The
formulae used to compute velocities, widths and line surface brightness
distributions and the details of their derivation, numerical computation and
model fitting are described in details in Appendix~B of \cite{marconi:n4041bh}.

In Appendix~\ref{app:comp} we describe a problem in the computation of kinematical quantities which might derive from subsampling. We also present a way to overcome this problem and therefore our computations are not affected by it.

Thus, model $\vel$ and $\sigma$ depend on $\Sigma$, the ISBD of the emission lines defined in the previous section, and on the following parameters:
\begin{itemize}
\item $x_0$,$y_0$, the position of the kinematical center in the plane of the
sky with the reference frame defined in the previous section;
\item $i$, the inclination of the rotating disk;
\item $\theta$, the position angle of the disk line of nodes;
\item \vsys, the systemic velocity of the disk;
\item \ML, the mass-to-light ratio of the stellar population;
\item \MBH, the \BH\ mass. 
\end{itemize}
Not all of these parameters can be independently determined by fitting the
observations.  It can be inferred from the equations in Appendix~B of
\cite{marconi:n4041bh} that \MBH, \ML\ and $i$ are directly coupled and this
has been discussed in detail by \cite{macchetto:m87bh} and
\cite{marconi:n4041bh}.  To avoid problems deriving from
this coupling we work here at fixed disk
inclination $i$.

\subsubsection{STIS data}
\begin{table*}
\caption{\label{tab:isaac_vel_fit}Fit results from the analysis of ISAAC data with $i=25\DEG$.}
\begin{tabular}{lcccccccr}
\hline\hline
\\
Flux$\,^a$ & $x_0$ & $y_0$ & $\log\MBH$ & $\log\ML$ & $\theta$ & \vsys & \chisqred & $({\chisqred})_\mathrm{resc}\,^b$ \\
\\
\hline
\\
\multicolumn{9}{c}{Fit of velocity $(\Delta v_0=6.3\KMS)\,^c$} \\
G1  & -0.08 & -0.16 &  8.14 & -0.34 & 164.1 & 554.8 & 2.62 & 1.00 \\
E3  & -0.07 & -0.14 &  8.13 & -0.23 & 163.6 & 554.5 & 2.77 & 1.04 \\
G1b & -0.01 & -0.16 &  8.11 & -0.23 & 162.9 & 560.8 & 2.78 & 1.05 \\
G3  & -0.05 & -0.13 &  8.15 & -0.36 & 164.7 & 554.9 & 2.74 & 1.06 \\
E1  & -0.09 & -0.15 &  8.13 & -0.20 & 163.0 & 554.2 & 2.94 & 1.09 \\
E4  & -0.09 & -0.15 &  8.14 & -0.24 & 163.0 & 554.5 & 3.01 & 1.13 \\
E1b & -0.02 & -0.15 &  8.10 & -0.10 & 161.6 & 560.2 & 3.20 & 1.17 \\
G4  & -0.06 & -0.13 &  8.18 & -0.50 & 164.6 & 554.7 & 3.02 & 1.20 \\
E4b & -0.00 & -0.14 &  8.15 & -0.28 & 162.1 & 560.9 & 3.24 & 1.23 \\
G4b & -0.00 & -0.15 &  8.18 & -0.47 & 162.5 & 560.7 & 3.19 & 1.24 \\
Gbb &  0.10 & -0.02 &  8.19 & -0.82 & 167.0 & 558.1 & 3.45 & 1.32 \\
Ebb &  0.10 & -0.02 &  8.20 & -0.87 & 166.1 & 558.8 & 3.45 & 1.33 \\
E3b & -0.00 & -0.14 &  8.07 &  0.01 & 161.4 & 560.2 & 3.74 & 1.33 \\
average$\,^d$  & -0.02$\pm$0.06 & -0.12$\pm$0.05 & 8.14$\pm$0.04 & -0.36$\pm$0.25 & 163.6$\pm$1.7 & 557.5$\pm$2.9 \\
\hline
\\
\multicolumn{9}{c}{Fit of velocity and velocity dispersion $(\Delta v_0=14.3\KMS)\,^c$} \\
G7  &-0.00& -0.02&  8.03&  -0.01&  167.6&  549.9&  10.13&   0.82\\
E7  &-0.00& -0.01&  7.98&   0.02&  167.8&  548.4&  10.31&   0.86\\
G7b & 0.10& -0.02&  8.09&  -0.16&  168.1&  560.2&  10.67&   0.86\\
E7b & 0.10& -0.02&  8.13&  -0.46&  167.3&  560.5&  10.53&   0.87\\
E8  &-0.01& -0.01&  7.97&   0.21&  163.8&  548.3&  11.22&   0.90\\
G8  &-0.01& -0.00&  8.14&  -0.16&  164.0&  551.0&  11.94&   0.94\\
C7  &-0.00& -0.01&  8.01&   0.02&  165.3&  551.0&  12.81&   0.94\\
E5  &-0.01& -0.00&  8.11&  -0.03&  163.4&  550.5&  11.96&   1.00\\
G3b & 0.10& -0.02&  8.15&  -0.59&  166.1&  562.2&   7.71&   1.00\\
E6  &-0.01& -0.00&  8.06&   0.17&  161.2&  549.8&  12.44&   1.01\\
C5  &-0.01& -0.02&  8.02&  -0.01&  167.6&  550.9&  13.61&   1.03\\
E5b & 0.10& -0.01&  8.16&  -0.50&  164.5&  562.4&  12.80&   1.06\\
C2b & 0.10& -0.03&  8.13&  -0.03&  161.6&  563.0&   9.20&   1.07\\
C3  &-0.00& -0.01&  8.11&  -0.05&  162.2&  553.6&   9.07&   1.07\\
C6b & 0.10& -0.03&  8.13&  -0.22&  161.9&  563.5&  12.17&   1.08\\
C8b & 0.12& -0.03&  8.23&  -0.84&  161.0&  562.5&  11.28&   1.08\\
C8  &-0.00& -0.00&  8.13&  -0.14&  162.6&  552.7&  13.69&   1.09\\
G5  &-0.01& -0.01&  8.19&  -0.36&  162.9&  550.9&  13.07&   1.09\\
E8b & 0.10& -0.01&  8.22&  -0.24&  163.1&  562.8&  12.70&   1.10\\
G6  &-0.01& -0.00&  8.14&   0.02&  160.3&  549.6&  13.38&   1.10\\
G5b & 0.09& -0.02&  8.17&  -0.15&  163.6&  562.7&  13.74&   1.11\\
G3  &-0.01& -0.01&  8.19&  -0.66&  165.2&  551.7&   8.52&   1.12\\
C7b & 0.14& -0.04&  8.09&  -0.20&  165.5&  564.7&  10.26&   1.13\\
C3b & 0.12& -0.03&  8.15&  -0.55&  165.7&  562.7&   8.73&   1.14\\
C4  &-0.01& -0.00&  8.14&  -0.17&  162.9&  550.6&   9.74&   1.16\\
C6  &-0.01& -0.01&  7.96&   0.17&  163.6&  550.5&  14.57&   1.17\\
C1b & 0.09& -0.03&  8.20&  -0.41&  161.8&  562.6&   9.77&   1.22\\
Ecb & 0.10& -0.01&  8.10&  -0.22&  165.5&  559.3&  19.74&   1.25\\
E3  &-0.01& -0.00&  8.20&  -0.92&  164.5&  552.0&   9.86&   1.27\\
Eb  &-0.00& -0.01&  7.95&   0.13&  168.3&  548.9&  17.64&   1.26\\
Ebb & 0.10& -0.03&  8.07&   0.03&  166.6&  562.0&  18.17&   1.26\\
C4b & 0.11& -0.03&  8.26&  -0.29&  160.4&  564.1&   9.51&   1.27\\
E6b & 0.09& -0.01&  8.17&  -0.33&  160.9&  561.5&  16.25&   1.28\\
average$\,^d$ & 0.01$\pm$0.01$^e$ & 0.01$\pm$0.01$^e$ & 8.11$\pm$0.08 & -0.21$\pm$0.28 & 164.1$\pm$2.4 & 556.3$\pm$6.1 \\
average$\,^d$ & 0.02$\pm$0.01$^f$ & -0.10$\pm$0.01$^f$ \\
\\
\hline
\end{tabular}
\\
\\
$^a$ Adopted ISBD.\\
$^b$ Rescaled \chisq\ with errors computed as $\Delta v_i^\prime\,^2 =\Delta v_i\,^2 +\Delta v_0\,^2$.\\
$^c$ Systematic error adopted to renormalize \chisq.\\
$^d$ Average and rms of best fit parameter values.\\
$^e$ Average and rms only on non-'b' ISBDs.\\
$^f$ Average and rms only on 'b' ISBDs.\\
$^\star$ Parameter was held fixed.\\
\end{table*}
\begin{figure*}[!]
\centering
\includegraphics[angle=90,width=0.9\linewidth]{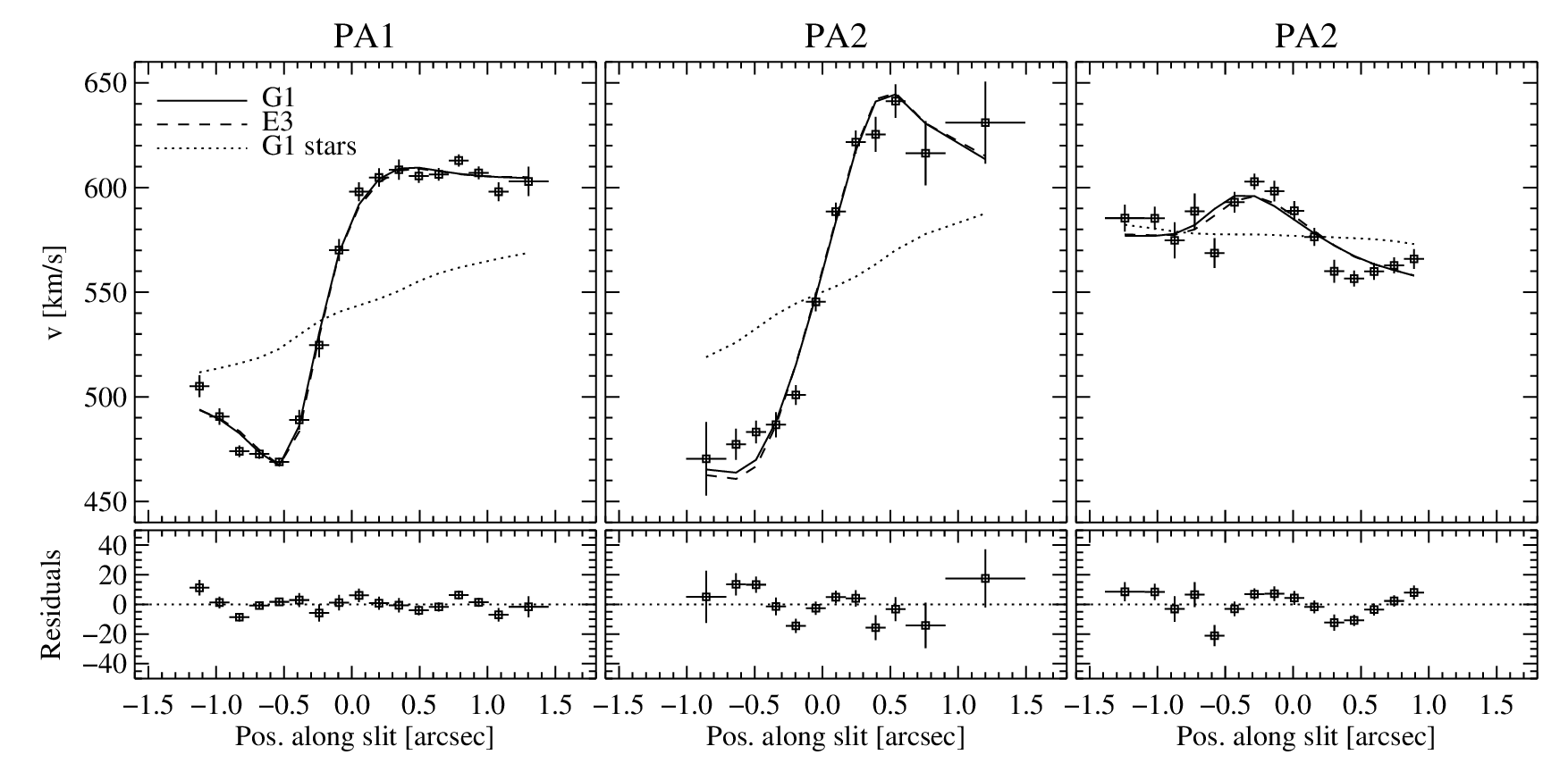}
\includegraphics[angle=90,width=0.9\linewidth]{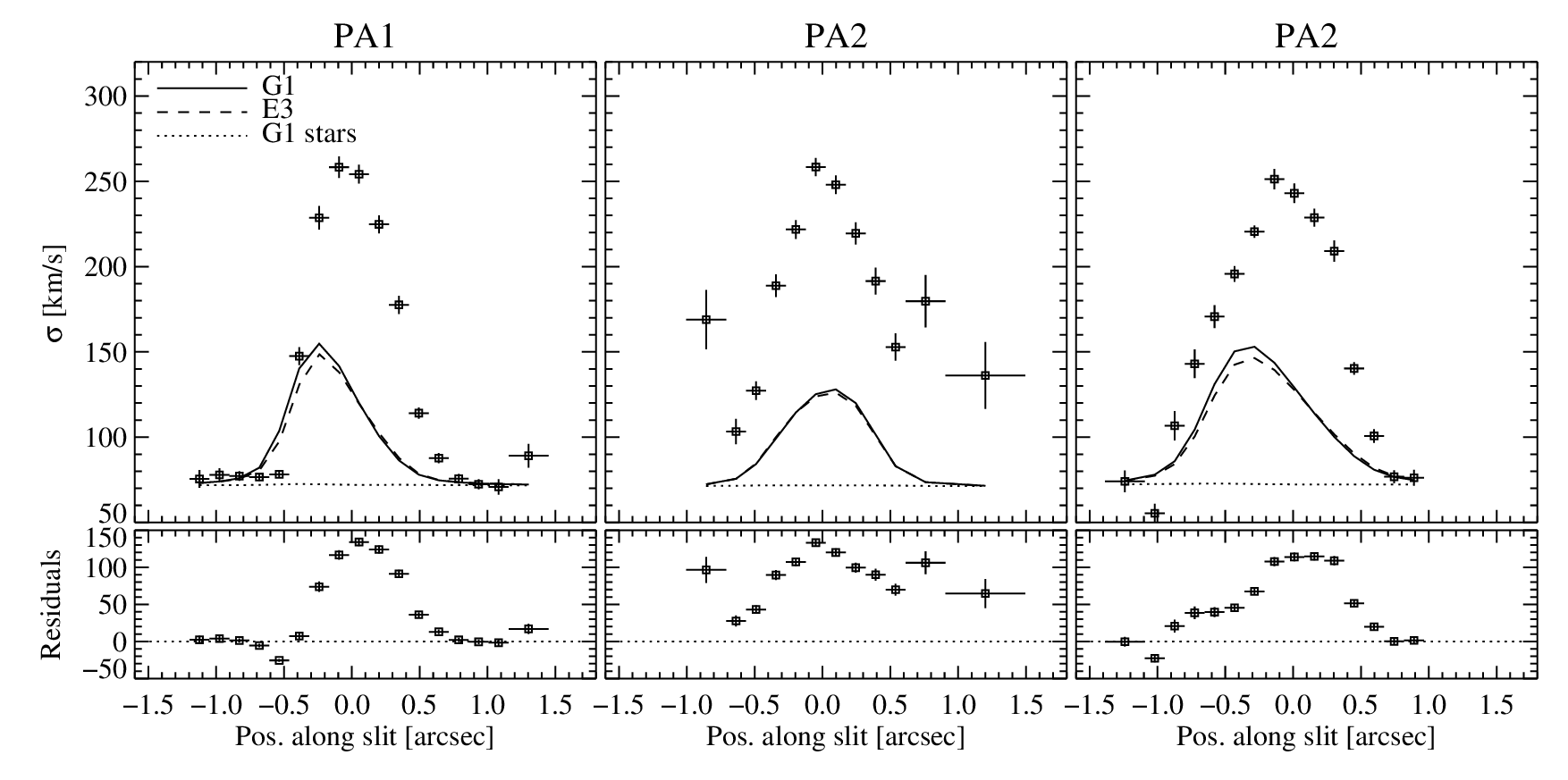}
\caption{\label{fig:isaac_vel_fit}Top panels: Fit of the
ISAAC velocities along the slit. The empty squares with error bars represent the
observed values while the solid lines connects corresponding model values.
Dotted lines represent the contribution of the mass in stars to the rotation
curve, i.e.\ what would be observed without a BH.
'G1' and 'E3' are the two
ISBDs which provide the best fit models with the lowest \chisq.
Bottom panels: Velocity dispersions expected from the two best fit models
compared with the
observed ones. Same notation as in the upper panel.
}
\end{figure*}
We fit the STIS rotation curves using all the ISBDs determined in
Sec.~\ref{sec:linefluxstis} and, at first, we work with fixed inclination
$i=25\DEG$.  The STIS PSF used in the computations is obtained at $\lambda$
9550\AA\ with TinyTIM \citep{krist:tinytim}.  The free parameters of the fit
($x_0$,$y_0$, $\log\MBH$, $\theta$, \vsys) are obtained with a \chisq\
minimization.  Since, as shown below, the contribution of the stellar mass to
the gravitational potential is negligible we worked with fixed \ML=1 in order
to reduce the number of free parameters. We will show that this assumption is consistent with the data, when searching for the confidence limits on the parameters.
During the fitting process, velocity
corrections are applied to the VIS1 and VIS2 data in order to take into account
errors on absolute wavelegth calibration.
The best fit parameters for all models with acceptable \chisq\ are shown in
Tab.~\ref{tab:stis_vel_fit} and in Fig.~\ref{fig:stis_vel_fit} (top panels)
where we compare the
observations and the 2 models with the lowest \chisq.
In order to select the "good" models shown in Tab.~\ref{tab:stis_vel_fit} we
first renormalize the \chisq\ following a procedure already adopted by
\cite{barth:n3245bh} and \cite{marconi:n4041bh}.  The fit with the lowest
\chisq\ is provided by the model with ISBD "G2" which has
$\chisqred=1.27\,(27.9/22)$.
 This \chisqred\ value is already acceptable at the 95\%\
confidence level given the number of degrees of freedom ($22\,dof$) but we
conservatively assume that there is an extra error in velocity, $\Delta v_0$,
which has not been taken into account and that must be added in quadrature to
the data. $\Delta v_0$ might also represent an intrinsic spread in velocity due,
for instance, to non-circular local motions superimposed to the large scale rotation.
$\Delta v_0$ is chosen such the best fit model has \chisqred=1.  For
model "G2", $\Delta v_0=10.4\KMS$ and this is a small value compared to the
errors on STIS velocities indicating the goodness of the fit.
We then renormalize \chisq\ for all models using the
above value of $\Delta v_0$ and we consider acceptable only the models with a
$\le 95\%$ significance level, i.e.\ those with $\chisqred\le
1.55$ for 22 $dof$. 
For these models, the velocity shifts applied to the data are on
average -2.5\KMS\ ($\pm 2.6\,rms$) and 2.6\KMS\ ($\pm 2.7\,rms$) for VIS1 and VIS2 respectively.  These are lower than the expected errors on the absolute wavelength
calibration (e.g.\ \citealt{stis_handbook}).
The models shown in Fig.~\ref{fig:stis_vel_fit} are in good agreement with the
observations as confirmed by the low \chisqred\ found prior to rescaling. In
particular the models nicely reproduce the steep velocity gradients and
turnoffs observed in the rotation curves which are the signature of the \BH.
The velocity fall-off in VIS1 is slightly faster than keplerian but this
disagreement appears in points with low signal-to-noise ratio.  The best fit
model is characterized by a BH mass of $\log(\MBH/\Msun)= 8$ and by a disk line
of nodes with PA=161 deg.  The dotted line in the figure shows the stellar
contribution to the rotation curve, i.e.\ what would be observed without a BH.
Clearly, the high velocity gradients can be explained only with the presence of
a BH.  The location of the kinematical center is -0\farcs05,-0\farcs05 with respect to
the location of the slit centers.

In order to estimate statistical errors associated with the best fit
parameters, we present in Fig.~\ref{fig:stis_grids}a,b \chisq\ contours for the
joint variation of $\MBH-\ML$ and of $\MBH-\theta$. The \chisq\ grids have been
computed with ISBD "G2", fixing 2 interesting parameters
($\MBH, \ML$ or $\MBH, \theta$) and varying the others to minimize \chisq.
Before computing confidence levels we have renormalized \chisq\ as described
above so that the model with the lowest \chisq\ value has \chisqred=1.
Confidence levels for 2 interesting parameters are then found following
\cite{avni:conflev}: $\chisq = \chi^2_\mathrm{min}+2.3$, 4.61, 6.17 for
confidence levels 68.3\%, 90\%\ and 95.4\%, respectively.
Statistical errors are $\pm 0.1$ on
$\log\MBH$ and $\pm6$ on $\theta$. Fig.~\ref{fig:stis_grids}a shows that it is not possible to
determine \ML\ but only place an upper limit on it ($\ML<1.6$). Also, the same figure
justifies our choice of fixing \ML=1, which is within the 1$\sigma$ confidence
contour.  

By comparing the results of the model fitting performed with different ISBDs we can have an estimate of the systematic errors
associated with the choice of the ISBD.  In the last row of Tab.~\ref{tab:stis_vel_fit} we show the
average values of the fit parameters and the associated rms.  It is highly
significant that the systematic errors on all parameters and in particular on
BH mass are very small when considering all the adopted ISB. The average BH
mass is $\log\MBH=7.88$ with rms of 0.05.

In conclusion, for $i=25\DEG$, the BH mass estimate from STIS data with ISBD
 "G2" is $\log\MBH=8.0 \pm 0.1$ ($\pm 0.05$ systematic) and
$\theta = 161 \pm 6$ ($\pm 6$ systematic).

In Fig.~\ref{fig:stis_grids}c and Tab.~\ref{tab:stis_grid_i} we adopt ISBD G2 and we explore how the
minimum \chisq\ depends on disk inclination. As above, we have renormalized so that the model with
the minimum \chisq\ value has \chisqred=1.  The STIS data favour
inclinations $i<48\DEG$ at a 95\% confidence level, a result
already found by \cite{marconi:cenabh}.

In Fig.~\ref{fig:stis_sig}b we plot the expected velocity dispersion for the 2
best fit models plotted in Fig.~\ref{fig:stis_vel_fit}a. A constant intrinsic
velocity dispersion of $\sigma_0=160\KMS$ was added to match the `plateau'
observed at VIS1 (at 9500\AA\ the degradation of the spectral resolution introduced by the grating and slit width is $\sim 160\KMS$, see Sec.~\ref{sec:obs_stis}).  In both cases an increase in $\sigma$ is observed at the
nucleus and this is roughly matched with unresolved rotation only at VIS2.  At
VIS1 unresolved rotation explains only $\sim 80\%$ of observed $\sigma$ nuclear
increase but the adopted $\sigma_0=160\KMS$ is smaller than the contribution from unresolved rotation.

\subsubsection{ISAAC data}

We fit the ISAAC rotation curves using all ISBDs determined in
Sec.~\ref{sec:linefluxisaac}. We proceed as in the previous section and we work
at fixed inclination, $i=25\DEG$.  The PSF used in the computations is a
gaussian with FWHM equal to the adopted seeing.
The free parameters of the fit
($x_0$,$y_0$, $\log\MBH$, $\log\ML$, $\theta$, \vsys) are obtained with a
\chisq\ minimization. During the fitting process, velocity corrections are
applied to the PA1, PA2 and PA3 data in order to take into account errors on
absolute wavelegth calibration.
The best fit parameters for all models with acceptable \chisq\ are shown in
the first part of Tab.~\ref{tab:isaac_vel_fit} ("Fit of velocity") and in Fig.~\ref{fig:isaac_vel_fit} we compare the
observations and the 2 models with the lowest \chisq.

\begin{figure*}[!]
\centering
\includegraphics[angle=90,width=0.9\linewidth]{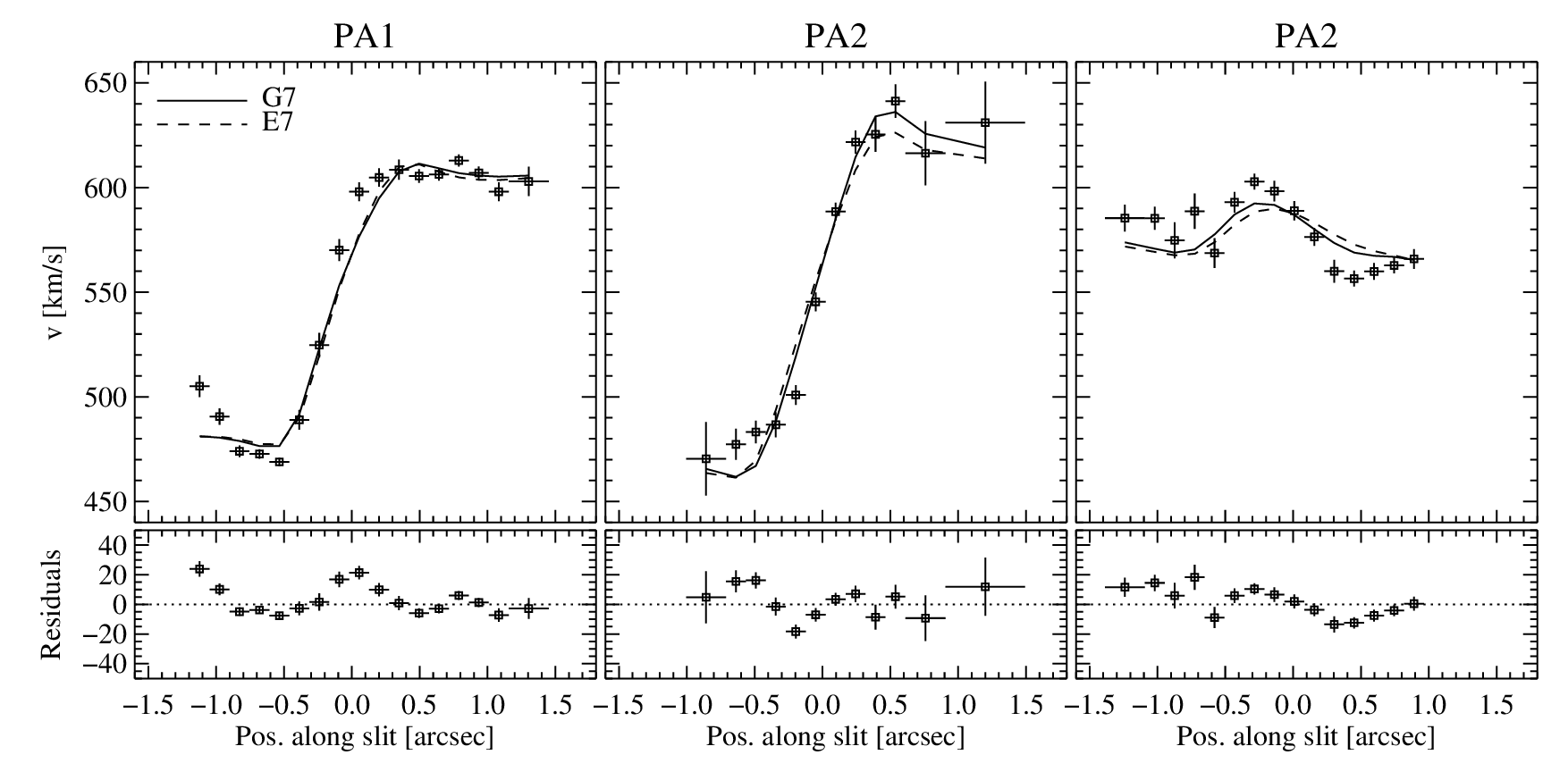}
\includegraphics[angle=90,width=0.9\linewidth]{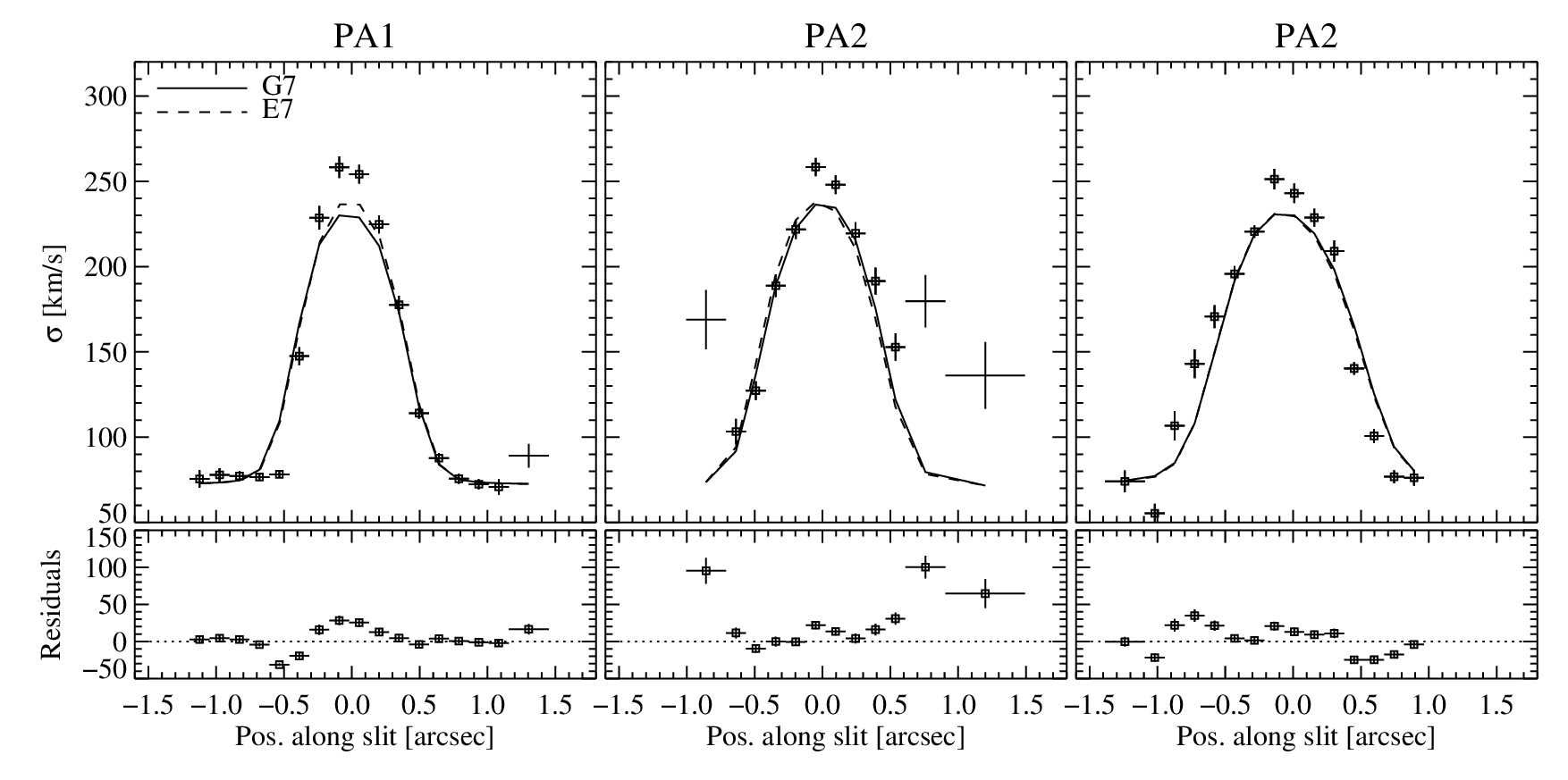}
\caption{\label{fig:isaac_sig_fit} Fit of the ISAAC velocity and velocity
dispersion along the slit. The deviant velocity dispersions in PA1 
and PA2 (points not marked with an empty square) were not included in the fit.
Notation is the same as in the previous figure.}
\end{figure*}

In order to select the "good" models shown in Tab.~\ref{tab:isaac_vel_fit} we
renormalize the \chisq\ following the approach described in the previous
section.  The fit with the lowest \chisq\ is provided by the model with ISBD
"G1" which has \chisqred=2.62 (91.9/35) and must be then renormalized with
$\Delta v_0=6.3\KMS$.
We then consider as acceptable all the models with a significance of 95\%, i.e.\ those
with $\chisqred<1.37$  (35 $dof$).
  For the models in Tab.~\ref{tab:isaac_vel_fit} the average
velocity shifts applied to the data are -1.0$\pm$3.4, -18.7$\pm$4.4 and
10.9$\pm$5.7 for PA1, PA2 and PA3 respectively. These are in very good agreement
with the expected errors on the absolute wavelength calibration
\citep{marconi:cenabh}.

As for the STIS data, 
the models reproduce the observed velocities
and, in particular, they match
the steep velocity gradients and turnoffs observed in the rotation curves which
are the signature of any BH.
The best fit
model is characterized by a BH mass of $\log(\MBH/\Msun)= 8.14$ and by a disk line
of nodes with PA=164 deg. 
The location of the kinematical center is -0.08,-0.16 with respect to the location of the slit centers.

In the lower panel of Fig.~\ref{fig:isaac_vel_fit} we plot the expected velocity dispersion for the 2
lowest \chisq\ models. A constant intrinsic velocity
dispersion of $\sigma_0=70\KMS$ was added to match the `plateau' observed at
PA1.  In both cases an increase in $\sigma$ is observed at the nucleus but this
is clearly not matched by the models with unresolved rotation.

In order to verify if the observed $\sigma$ increase can be reproduced by the
models we also perfomed the fits by considering velocity and velocity
dispersion at the same time.  Best fits acceptable to the 95\%\ level (i.e.,
with 75 d.o.f., renormalized $\chisqred\le 1.28$) are shown
in the second part of
Tab.~\ref{tab:isaac_vel_fit} ("Fit of velocity and velocity dispersion") and in
Fig.~\ref{fig:isaac_sig_fit} we compare the observations and the 2 models with
the lowest \chisq.  Though the fit of velocity is not as good as before, 
the observed curves are still acceptably reproduced by models.
However this time models are also able to match
the observed sigma
increase showing that it can be explained by
unresolved rotation.  The reason for this result is that in the previous
case the center of
rotation was pushed far from the surface brightness peak in order to better reproduce
velocities.  Now the center of rotation is almost coincident (within $\sim
0\farcs01$) with the line emission peak (note that in the case of ISBDs
'b' the emission line peak is offset from the slit centers by approximately
0.1, -0.02 and this is where the kinematical center is).

\begin{figure*}[!]
\centering
\includegraphics[width=0.33\linewidth]{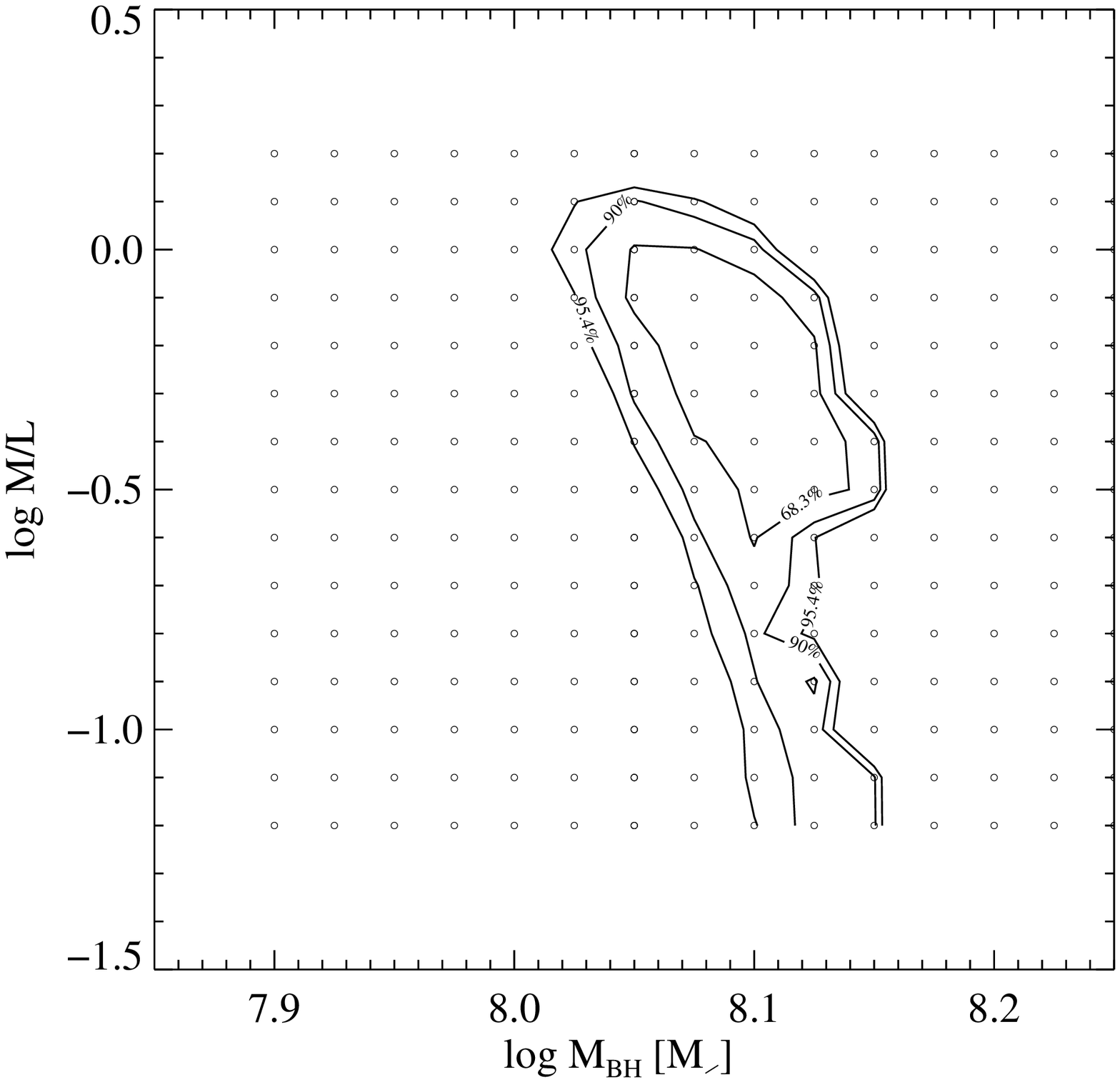}
\includegraphics[width=0.33\linewidth]{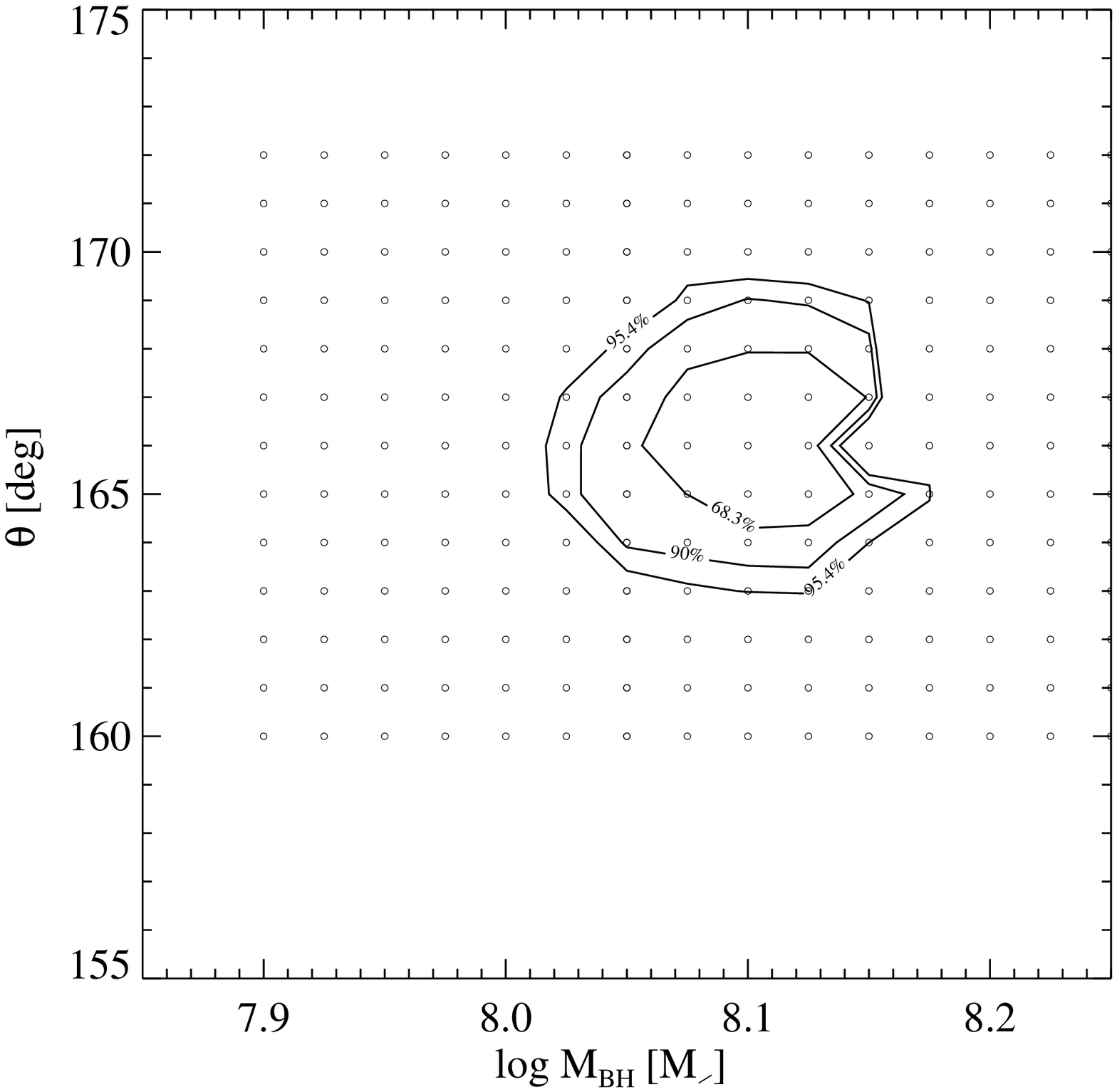}
\includegraphics[width=0.33\linewidth]{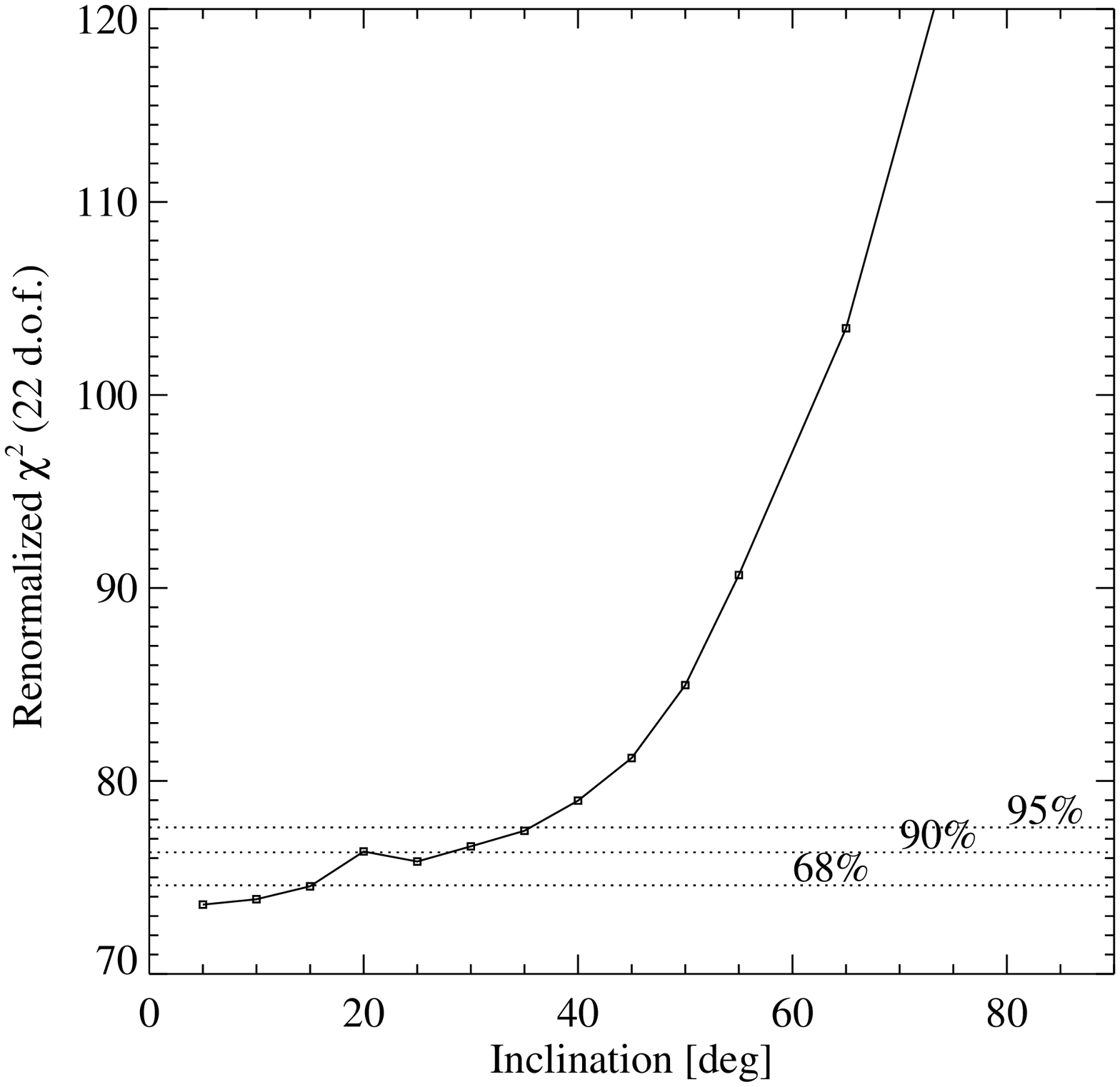}
\caption{\label{fig:isaac_grids} (a) \chisq\ contours for the joint variation of
\MBH\ and \ML\ using the ISAAC data and fitting velocities and central
velocity dispersions (ISBD G7).
Contour levels are for $\chisq =
\chi^2_\mathrm{min}+2.3$, 4.61, 6.17 corresponding to 68.3\%, 90\%\ and 95.4\%
confidence levels.  The dot indicates the \MBH\ and \ML\ values for which the
\chisq\ minimization was computed. (b) same as (a) but for $\MBH-\theta$.  (c)
Dependence of $\chisq$ on the adopted disk inclination. Dotted lines indicate
the 68.3\%, 90\%\ and 95.4\% confidence levels.}
\end{figure*}

In Fig.~\ref{fig:isaac_grids} we show the \chisq\ contours for joint
variation of \MBH,\ML\ and \MBH,$\theta$ in order to estimate the
statistical errors. The models have been computed adopting ISBD
G7, fixing 2 interesting parameters and varying the others.  The
\chisq\ has been renormalized as before.

The statistical errors associated to the BH mass determination are $\pm 0.05$ in
$\log\MBH$ comparable to the systematic errors due to the adopted ISBD.
As for the fit of the STIS data
Fig.~\ref{fig:isaac_grids} 
shows that it is not possible to determine \ML\ but
one can only place an upper limit of $\ML<1$.  The
statistical errors associated to $\theta$ are $\pm3$ deg. 

By comparing the results of the model fitting performed with ISBDs
we can have an estimate of the systematic errors
associated with the choice of the ISBD.
In Tab.~\ref{tab:isaac_vel_fit} we also show the average values
of the fit parameters and the associated rms.  It is highly significant that
the systematic errors on all parameters and in particular on BH mass are very
small.  But the most important thing is that whether one fits velocity or
velocity and sigma, the BH mass is the same.  Indeed $\log\MBH=8.14\pm0.04$
(average over different ISBDs, velocity only),
$\log\MBH=8.11\pm0.08$ (average over different ISBDs, velocity
and velocity dispersion).
Also all $\theta$'s are consistent.
\begin{table}
\caption{\label{tab:isaac_grid_i} Effect of $i$ variation on best fit parameters and minimum \chisq\ for ISAAC data. 
All models adopt ISBD G7. 
Confidence levels
68.3\%, 90\%\ and 95.4\% are at $\chisq = \chi^2_\mathrm{min}+1.0$, 2.71, 4.0,
for one interesting parameter \citep{avni:conflev}. \chisq's have been rescaled as described in the text with $\Delta V_0=4.6\KMS$. }
\begin{tabular}{lcccccccr}
\hline\hline
\\
$i$ &   $\log\MBH$ & $\log(\MBH\sin^2i)$ & $\theta$ & $\chi^2_\mathrm{resc}$ \\
\\
\hline\hline
\\
 5. &   9.38 &   7.26 &  167.0 &  73.6\\
10. &   8.78 &   7.26 &  167.4 &  73.9\\
15. &   8.44 &   7.27 &  167.6 &  74.5\\
20. &   8.20 &   7.27 &  167.6 &  76.3\\
25. &   8.03 &   7.28 &  167.6 &  75.8\\
30. &   7.91 &   7.30 &  167.4 &  76.6\\
35. &   7.81 &   7.33 &  166.2 &  77.4\\
40. &   7.73 &   7.35 &  167.5 &  79.0\\
45. &   7.68 &   7.38 &  167.4 &  81.2\\
50. &   7.65 &   7.42 &  168.7 &  85.0\\
55. &   7.64 &   7.46 &  168.1 &  90.7\\
65. &   7.66 &   7.58 &  167.6 & 103.5\\
75. &   7.98 &   7.95 &  170.1 & 123.6\\
85. &   8.11 &   8.11 &  175.0 & 226.3\\
\\
\hline
\end{tabular}
\end{table}

In conclusion, for
$i=25\DEG$ and ISBD G7, the BH mass estimate is $\log\MBH=8.03 \pm
0.05$ ($\pm 0.06$ systematic) and $\theta = 165 \pm 3$ ($\pm 1$ systematic).

In Fig.~\ref{fig:isaac_grids}c and Tab.~\ref{tab:isaac_grid_i} we explore how
the minimum \chisq\ varies with varying disk inclination for the models
adopting ISBD G7. As for the STIS case, we have
renormalized \chisq's so that the model with the minimum \chisq\ value has
\chisqred=1.  The ISAAC data clearly favour inclinations $i<35\DEG$ at the 95\%
confidence level. This limit on $i$ is smaller than the one derived from STIS data
however, given the higher SNR of the ISAAC data and the larger number of data points, we take it
as the upper limit on $i$.

\section{\label{sec:discuss}Discussion}

In the previous sections we have described in detail the available datasets and
the application of the gas kinematical method which provides consistent results
from independent analysis of STIS and ISAAC data.  In this section we analyze
the influence of the ISBD on the final results and we discuss in detail the
results obtained using STIS and ISAAC data. We then verify the crucial
assumption that the gas is in a thin, circularly rotating disk by comparing
observed and model velocity dispersion. We caution about incorrect computations
of the model kinematical quantities which might be affected by a subsampling
problem (see also Appendix~\ref{app:comp}) or by adopted ISBD which are too
smooth.  We also compare observed line profiles, $h_3$ and $h_4$ parameters
with model values.  Finally, after comparing the gas kinematical estimate of \MBH\ with
the stellar dynamical one by \cite{silge:cenabh}, we discuss \MBH\ in the
framework of the correlations between BH mass and host galaxy structural
parameters and we conclude this section by estimating the maximum size of a
cluster of dark stellar remnants, the most likely alternative to a supermassive
\BH.

\subsection{Influence of the intrinsic surface brightness distribution}

One of the worries in gas kinematical measurements of BH masses has always been the role of the intrinsic emission line surface brightness distribution, since it is the
weight of the velocity field in the averaging over apertures
(e.g.\ \citealt{macchetto:m87bh,barth:n3245bh,marconi:n4041bh}).
In the previous section we have clearly shown that the adopted ISBD,
provided that it reproduces the observed one within the
errors, does not affect the final BH mass estimate. Indeed systematic
errors on \MBH\ due to the adopted ISBD are of the order
of 0.05 and 0.08 in $\log\MBH$ for STIS and ISAAC data respectively.
However it should be remembered that the ISBD has
an important effect
on the quality of the velocity fit.

We can now compare the results obtained here with those obtained by
\cite{marconi:cenabh} who adopted a simple constant ISBD and a
double exponential.  From the ISAAC data we obtain $\log\MBH = 8.11\pm 0.08$
(average and rms over all ISBDs) with disk inclination $i=25$.
With the same disk inclination, \cite{marconi:cenabh} obtained $\log\MBH =
8.31\pm 0.1$ (constant ISBD) and $\log\MBH = 8.25\pm 0.1$ (double
exponential ISBD).  The data are consistent within the errors. The
slightly smaller values found here are explained with the fact that
\cite{marconi:cenabh} did not consider the stellar mass in their fitting.
Regarding the PA of the disk line of nodes \cite{marconi:cenabh} found
-15$\pm$5 while we find $164\pm2$, i.e. $-16\pm2$.  Therefore our new analysis
improves the accuracy of the BH mass estimate with respect to
\cite{marconi:cenabh} but does not substantially change their results.

\subsection{Comparison of results from fitting STIS and ISAAC data}

In the previous sections we presented the results of fitting separately STIS
and ISAAC data. Since the fits are completely independent and 
data have spatial resolutions differing by an order of magnitude, the
comparison of the results provides an important
check of the robustness of the gas
kinematical method in measuring BH masses. Moreover, the comparison will show
whether it is possible to obtain an accurate estimate of the BH mass even with
lower S/n data like those obtained with STIS.

We did not perform a joint fit of the ISAAC and STIS data because
this would be strongly biased toward the ISAAC data since the number
of points is larger and their relative errors are smaller.

The best fit of STIS data with ISBD G2 (see Tab.~\ref{tab:stis_vel_fit}) provides $\log(\MBH/\Msun)=8.00\,\pm 0.11$ (statistical) $\pm 0.05$ (systematic). This is perfectly in agreement with the best fit result of
ISAAC data where, from ISBD G7 (Tab.~\ref{tab:isaac_vel_fit}),
$\log(\MBH/\Msun)=8.03\,\pm 0.05$ (statistical) $\pm 0.08$ (systematic).
The agreement is still good when comparing the best fit values of the BH mass
averaged over all the adopted ISBDs.  STIS data provide
$\log(\MBH/\Msun)=7.88\,\pm 0.05$ while from ISAAC $\log(\MBH/\Msun)=8.11\,\pm
0.08$. The log ratio of the BH masses is $0.2\pm 0.1$ which is consistent with
0 at the $2\sigma$ level.

Apart for the BH mass, the separate fits recover consistent estimates of the
position angle of the disk line of nodes.  From STIS (G2), the best fit
provides $\theta=161\,\pm 6$ (statistical) $\pm 6$ (systematic) while from
ISAAC (G7) it results $\theta=168\,\pm 2$ (statistical) $\pm 2$ (systematic).
The agreement is even improved when comparing the best fit values of $\theta$
averaged over all the adopted ISBDs:
$\theta=168\,\pm 6$ (STIS) and $\theta=164\,\pm 2$ (ISAAC).

\begin{figure}[!]
\centering
\includegraphics[width=0.99\linewidth]{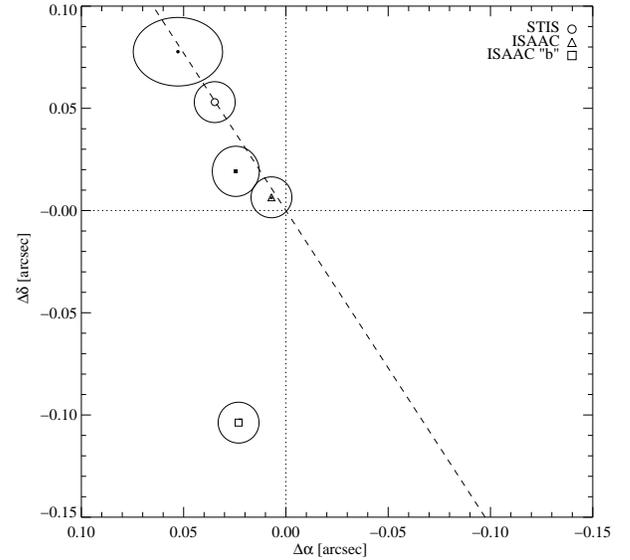}
\caption{\label{fig:centers} Positions of slit centers (empty symbols) and
emission line peaks (filled symbols) compared to the kinematical center located
at $\Delta\alpha=0$, $\Delta\delta=0$.  Ellipses indicate the $1\sigma$ errors.
The dashed line is the PA of the jet observed at radio and X-ray wavelengths.
  }
\end{figure}

In Fig.~\ref{fig:centers} we compare the positions of the slit centers and
emission line peaks with respect to the kinematical center position derived
in the fitting.  Since in each fit we leave the coordinates of the kinematical
center free to vary, in order to perform the comparison we consider a reference
system in which
the kinematical center is at (0,0). In figure, slit centers are indicated by empty big
symbols while the corresponding positions of the emission line peaks are
indicated by smaller filled symbols with the same shape.
In the STIS case, the slits were
positioned blindly on the nucleus position which was computed
from NICMOS K images comparing the location on the
continuum peak with respect to the
acquisition star.  The ISAAC slits were directly centered on the location of the K continuum 
peak. During the fitting of the ISAAC data we considered two cases, one where
the emission line peak was coincident with the
slit centers (i.e.\ with the K continuum peak as suggested by the NICMOS \PaA\ image), and the other, labeled 'b', where
the line emission peak was not coincident with the slit centers.  In all
cases slit centers are closer than 0.1 arcsec from the kinematical center. But
the most interesting indication comes from the location of the emission line
peaks which are, within the errors, located along the jet axis, a result not
obvious a-priori. This is consistent with two possible pictures: in the first
one the jet excites line emission while in the second one the jet represents
the symmetry axis of the ionization cone, according to the unified model of AGNs
(e.g.~\citealt{axon:n1068jc, capetti:mrk3jc}).

\subsection{Is the gas circularly rotating?
Velocity dispersion and line profiles tests.}

The power of gas kinematical measurements is their conceptual simplicity
compared to the complex analysis required by stellar dynamical modeling for
which there are serious issue regarding indeterminacy \citep{valluri:bhstarkin,
cretton:bhstarkin}.  However, gas kinematical measurements are based on the
assumption of Keplerian disks, i.e.~thin
disks circularly rotating under the influence of the gravitational potential
due to stars and BH. In some cases the gas is clearly not rotating circularly, for
instance when the motions are affected by jet-cloud interactions
(e.g.~\citealt{axon:n1068jc,capetti:mrk3jc}) but in other cases the question is
more subtle. 

The comparison between observed and model velocity dispersion has become one of
the most used tests on the reliability of the circular rotation assumption:
velocity dispersions larger than expected from unresolved rotation could be an
indication of non circular motions which could invalidate the BH mass estimate.
Many authors have found that the observed velocities could be well fit with
Keplerian disk models but the gas velocity dispersion was larger, and in some
cases much larger than expected from unresolved rotation. This has caused big
concerns on the reliability of the gas kinematical method
(e.g.~\citealt{vandermarel:n7052bh, barth:n3245bh, verdoes:ic1459bh,
verdoes:n4335bh, cappellari:ic1459bh}).  
\begin{figure*}[!]
\centering
\includegraphics[angle=90,width=0.99\linewidth]{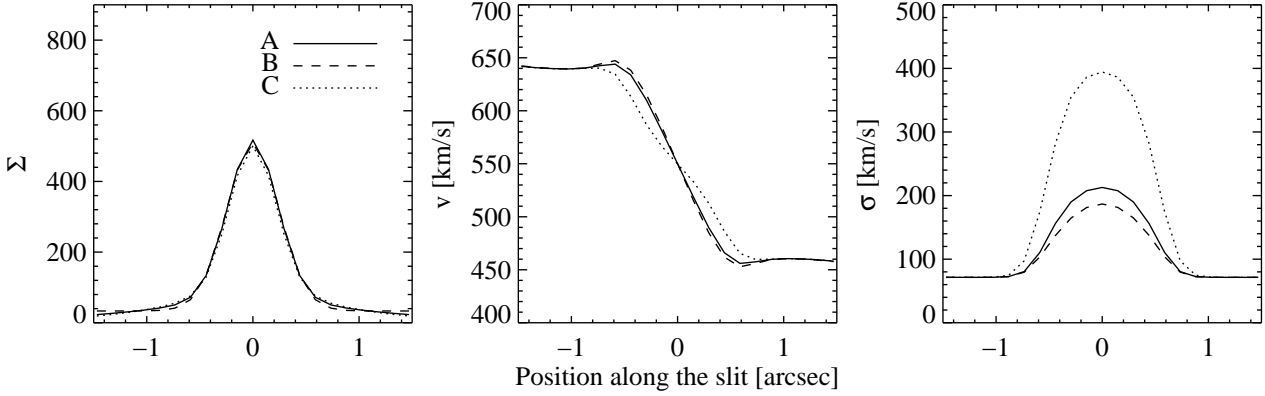}
\caption{\label{fig:simul} Instrument convolved kinematical quantities computed with different ISBDs A,B and C (see text for explanation). The ISBDs have been chosen to provide very similar observed surface brightnesses along the slit. }
\end{figure*}
\begin{figure*}[!]
\centering
\includegraphics[angle=90,width=0.99\linewidth]{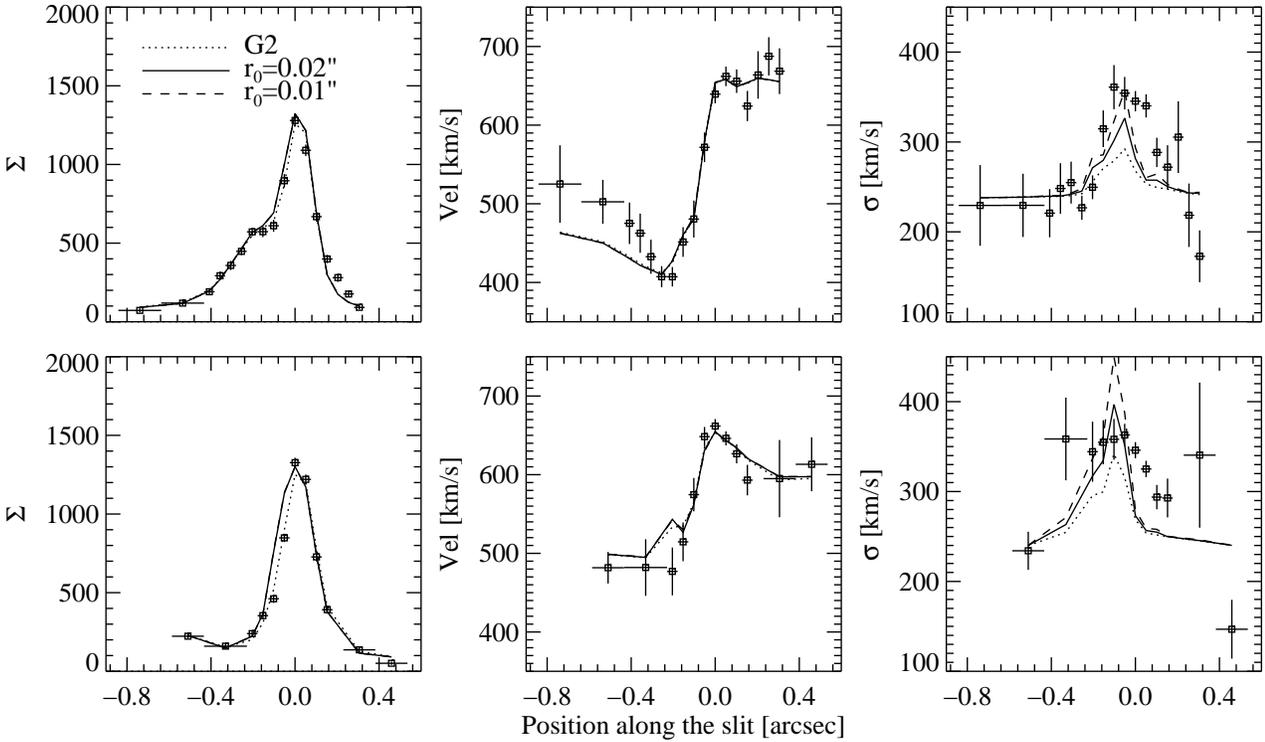}
\caption{\label{fig:stis_peak} Comparison between observed kinematical quantities 
from STIS (points with error bars) and model computations. From left to right panels represent surface brightness, velocity and velocity dispersion along the slit.
The top and bottom panels are for VIS1 and VIS2, respectively.
Models were computed using G2 and two additional ISBD contructed by adding to G2 an exponential component with  $r_0=0.02"$ and $r_0=0.01"$ at the position of the BH
(see text for more details). }
\end{figure*}

Before focussing on Centaurus A, we show that these concerns might have been excessive
since a small model velocity dispersion could easily result from (i) a non-accurate computation or, more likely, from (ii) an adopted ISBD which is too smooth.

In Appendix \ref{app:comp} we show the effect of a coarse sampling in the computation of the model kinematical quantities.
This effect is particularly important when the ISBD varies on spatial scales which are much smaller than the PSF sizes and Fig.~\ref{fig:simul2}
shows how one could easily underestimate the model velocity dispersion by a factor $\simeq 1.5$. This effect is also very subtle because it is not manifested in the model computed line fluxes and velocities. We thus caution about the sampling chosen in computations and in Appendix \ref{app:comp} we present a possible way to overcome this potential problem.

We now focus on the effects of different ISBs which provide the same light profile along the slit after being convolved with the instrumental response.
We consider three different ISBs labeled A, B and C. All of them are circularly symmetric on the plane of the sky and are described by a sum of two exponentials as
\begin{equation}
f(r) = I_0\,\mathrm{e}^{-r/r_0}+I_1\,\mathrm{e}^{-r/r_1}
\end{equation}
ISBD A has $I_1/I_0=0.01$, $r_0=0.05\arcsec$, and $r_1=1.00\arcsec$.
B has $I_1/I_0=0.0067$, $r_0=0.08\arcsec$ but the second exponential is substituted with a constant. Finally C has $I_1/I_0=0.0008$, $r_0=0.01\arcsec$, and $r_1=0.69\arcsec$.
The 'observed' system is like Centaurus A but with \MBH=\ten{8}\Msun, \ML=1, and the gas disk is inclined by 45\deg\ with respect to the line of sight.
The rotating gas disk, which has an intrinsic constant velocity dispersion of 70\KMS, is observed with ISAAC with 0.5" seeing. The slit is placed along the major axis of the disk.
In Fig.~\ref{fig:simul} we show the intrument convolved surface brigtness,
velocity and velocity dispersion along the slit.  As noted in previous
sections, it is clear that the observed surface brightness and velocity along
the slit (and consequently \MBH) are little sensitive to the choice of the
ISBD.  The reason for this weak dependence of the instrument convolved
velocity on the ISBD is that, close to the BH location where the weighting by
the ISBD is larger, rotation is not resolved and the contribution
from positive and negative velocities is canceled.  However, the effects of the
ISBD are readily apparent on the gas velocity dispersion. Suppose that in the
analysis of the data one uses ISBD A and finds that, while the observed
velocities are well reproduced by the model, the gas velocity dispersion is
underestimated by a factor $\simeq 2$. One might then be tempted to conclude
that the gas has a high intrinsic velocity dispersion.  However, using ISBD C
(or even a less extreme one) one would reach the opposite conclusion
regarding the validity of the Keplerian disk assumption. 

It is not possible to draw general conclusions here because each case must be
analyzed separately bearing in mind the possibility that the adopted ISBD does
not reflect the real one.  However we caution against the possibility that many
of the previous gas kinematical analyses finding inconsistent gas velocity
dispersions might be affected either by a subsmapling problem or by the
adoption of a too extended ISBD.

We now focus on Centaurus A where the STIS and ISAAC datasets are well suited
to this discussion.  In the case of ISAAC data, the model fitting of velocities
does not reproduce the velocity dispersion and the reason is that, in order to
find the best match of the rotation curves, the kinematical center is pushed
away from the emission line peak. However, the joint fit of velocity and
velocity dispersion shows that the circularly rotating model can well account
for both velocity and velocity dispersion at the same time (note that the \MBH\
estimate is not changed by this). 

The observed velocity dispersion in the STIS datasets is larger than expected
from pure rotation but the previous discussion on the effects of the ISBD on
the model velocity dispersion might suggest that it is possible to match the
observed STIS velocity dispersion with an appropriate choice of the ISBD.
However the main ISBD component used for STIS is not located on the BH but
$\simeq 0\farcs1$ away, therefore changing its scale radius would not produce
any appreciable effect on final model values.  Instead, one should place a
bright and very peaked ISBD component on the BH location.
In Fig.~\ref{fig:stis_peak} we compare model
kinematical quantities obtained using G2 and two additional ISBDs obtained by
adding an exponential component to G2 at the location of the BH. These
exponential components have $I_0=5000$, $r_0=0.02"$ and $I_0=20000$,
$r_0=0.01"$ respectively (see also Tab.~\ref{tab:stis_flux} for the parameters
describing G2). Surface brightnesses and velocities are very little affected by
this additional ISBD component located at the position of the BH. However the
gas velocity dispersion is increased and it is now possible to reproduce the
peak value of the observed velocity dispersion.  The observed $\sigma$ profile
along the slit is slightly wider than the model one nonetheless, given the
signal-to-noise ratio of the data, the agreement between models and
observations can now be considered satisfactory.  The intrinsic constant
velocity dispersion required by the data, $\sigma_0=160\KMS$, is smaller 
than that expected from circular rotation and rotation around a BH is
needed to explain the increase of $\sigma$ in the nuclear region.  The measured
\MBH\ is perfectly consistent with the results from the ISAAC dataset.  This is
a clear indication that even the presence of a constant intrinsic velocity
dispersion ($\sigma_0=160\KMS$) does not necessarily invalidate the \MBH\
estimate.  In conclusion it is important to emphasize that a large velocity
dispersion rise in the nucleus of a galaxy cannot be a manifestation of the
ISBD in a pure stellar model.  Thus their very existence is \textit{prima facie}
evidence for the presence of a \BH\ provided the gas motions are dominated by
rotation.  However, due to the sensitivity of the observed sigma to the ISBD,
modeling a sigma peak is model dependent and therefore results
in weak constraints on the precise mass of the \BH.
\begin{figure*}[!]
\centering
\includegraphics[angle=90,width=0.95\linewidth]{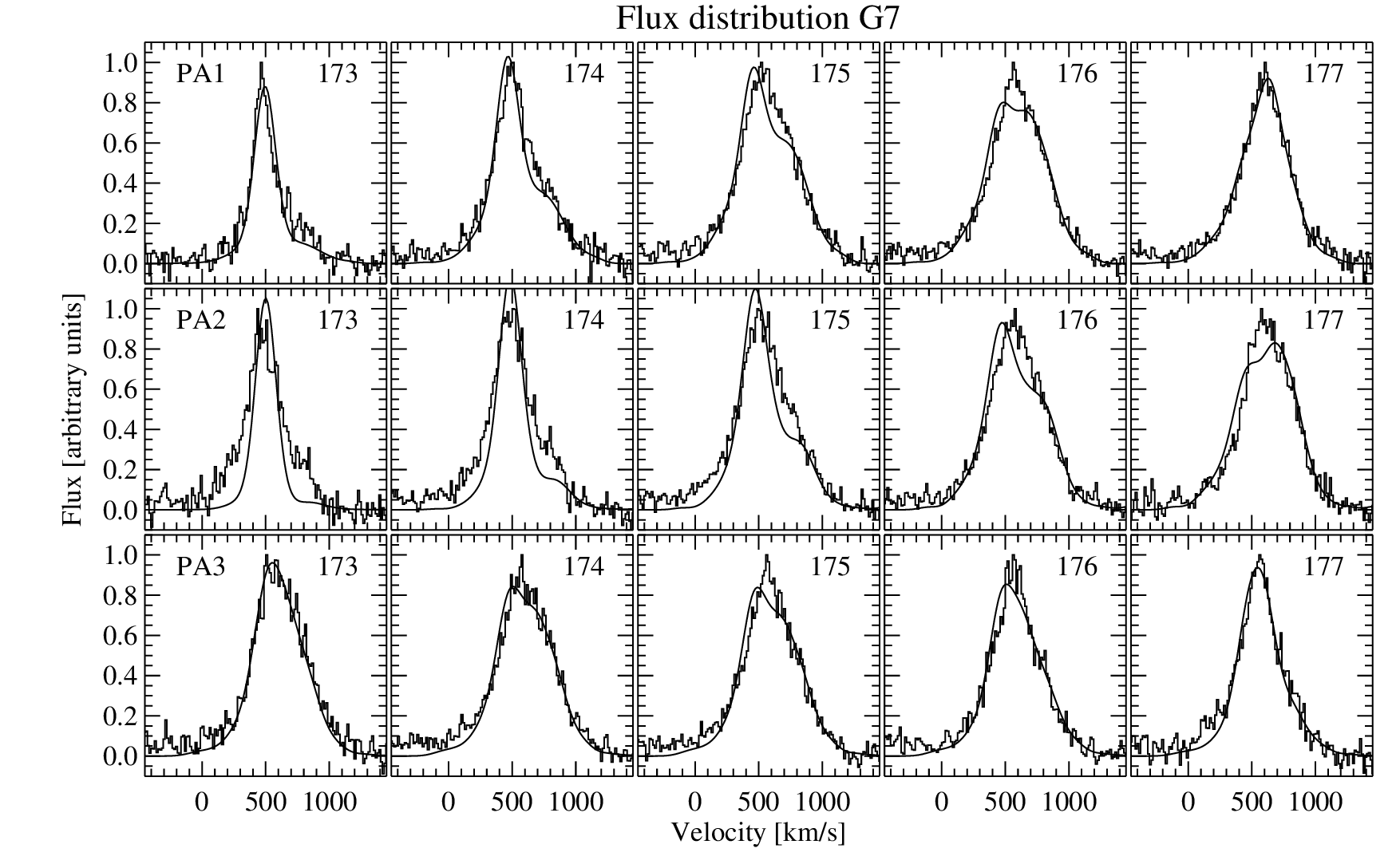}
\includegraphics[angle=90,width=0.95\linewidth]{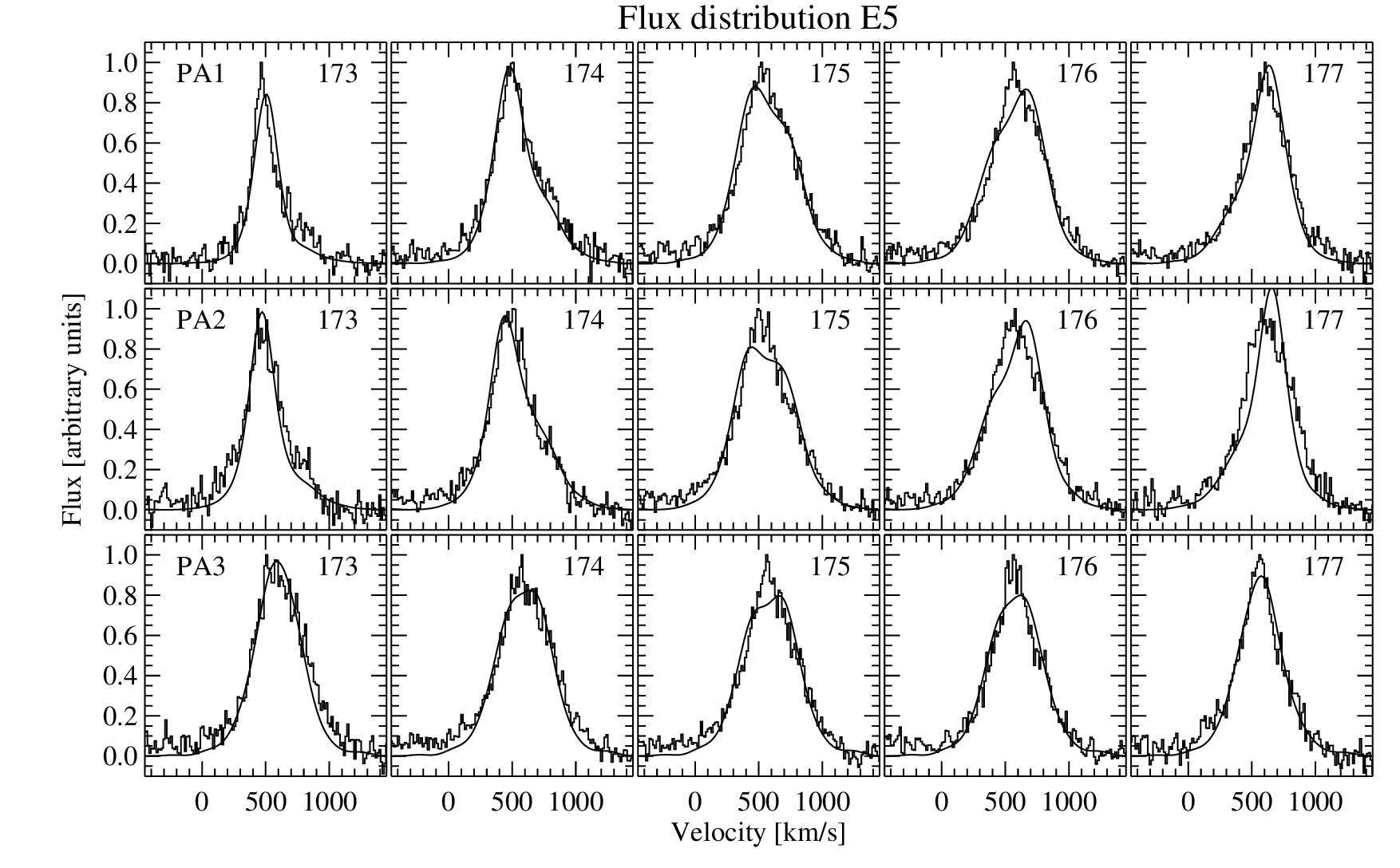}
\caption{\label{fig:isaac_lineprofiles} Top panels: observed ISAAC line
profiles are compared with model ones computed with the best ISBD G7. Numbers
on the top right corners indicate the row where the spectrum was extracted (175
is the location of the slit center).  Model profiles have been rescaled in
surface brightness to have the best match with observed ones.  Bottom: as in
top panels, but for ISBD E5 which provides the best matching model line
profiles. }
\end{figure*}
\begin{figure*}[!]
\centering
\includegraphics[angle=90,width=0.95\linewidth]{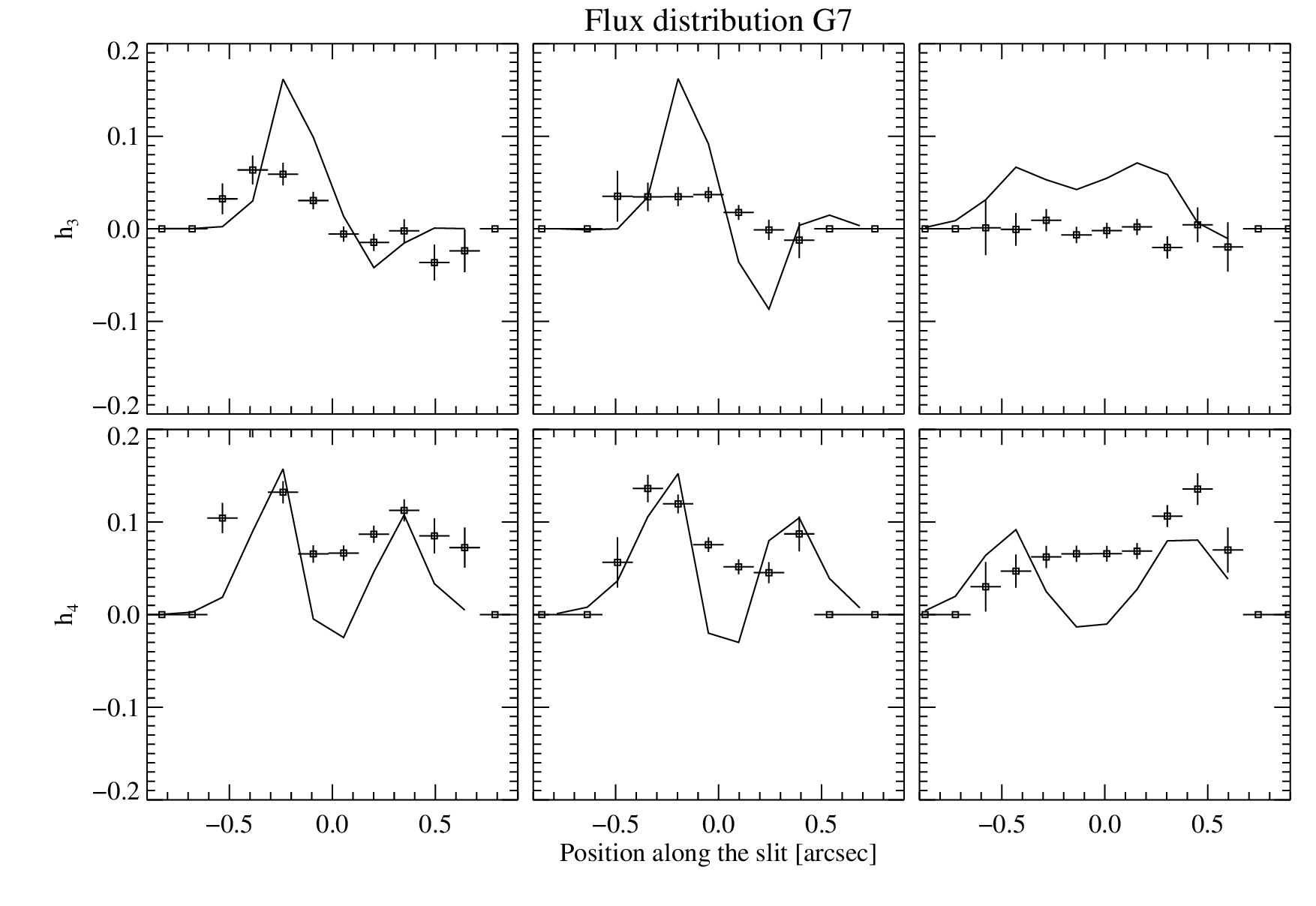}
\includegraphics[angle=90,width=0.95\linewidth]{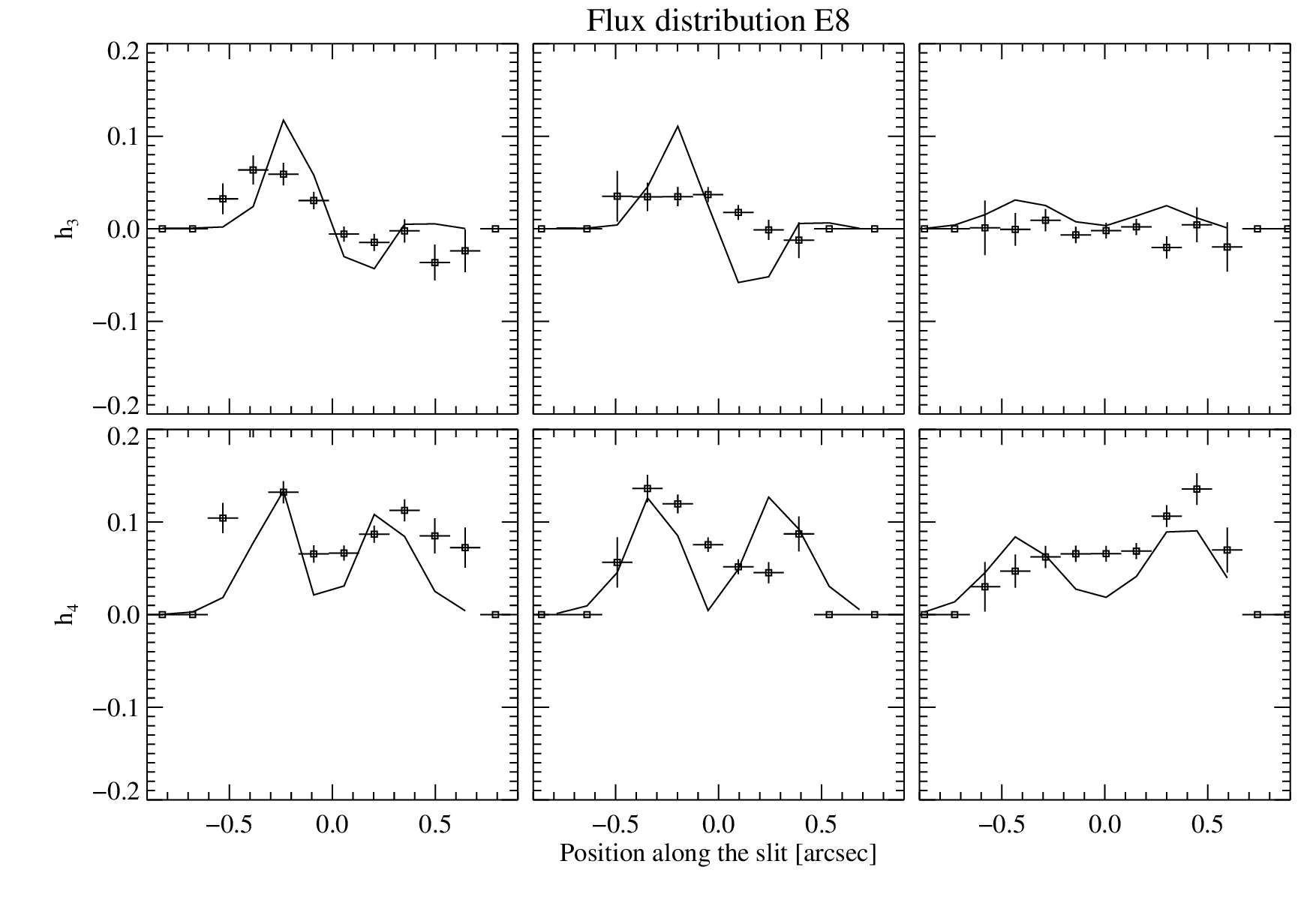}
\caption{\label{fig:isaac_hermite} Top panels: observed Hermite parameters
$h_3$ and $h_4$ compared with model ones computed with the best ISBD G7. Numbers on the top right corners indicate the row
where the spectrum was extracted (175 is the location of the slit center).
Bottom: as in top panels, but for ISBD E8 which
provides the best match to the observations. }
\end{figure*}

In Fig.~\ref{fig:isaac_lineprofiles} we compare the model line profiles
computed with ISBD G7 with the observed ones with ISAAC. Such a comparison is meaningless with STIS because of the much lower spectral resolution.
The match between observed and model line profiles is good given that the
fitting was performed on velocity and velocity dispersion and not on the line
profile itself.  G7 is the ISBD which provides the best fit
of velocity and velocity dispersion. We have also searched for the ISBD which provides the best match of the line profiles
among those in Tab.~\ref{tab:isaac_vel_fit}. A \chisq\ analysis indicates that
this is given by E5 and indeed also a visual inspection of the figure confirms
this result.  Model line profiles, which are only affected by unresolved
rotation, are able to reproduce, at least qualitatively, the features which are
observed in the data like asymmetries and humps (see also the discussion in
\citealt{macchetto:m87bh}).  It is thus clear that line profiles are 
consistent with the rotating disk model and that it is possible to discriminate
among different ISBD on the basis of the line profiles.

In this paper we have, for the first time, introduced the Hermite expansion in
the analysis of emission line profiles. It is therefore natural to veirfy
whether the model values of $h_3$ and $h_4$ are consistent with observed ones.
In Fig.~\ref{fig:isaac_hermite} we perform this comparison using the best
fitting ISBD G7. The agreement is not very good but
it must be kept in mind that the fit was not optimized to reproduce $h_3$ and
$h_4$. Qualitatively the model reproduces the shapes of the $h_3$ and $h_4$
variations along the slit, though it does not reproduce the amplitudes.  As in the
previous case, we have searched for the best fit model which provides the best
match to $h_3$ and $h_4$ and this results to be that generated with ISBD
E8.  The match is much better than previously though
not yet as good as the match of the observed velocity and velocity dispersion.
As in the previous case we can conclude that the comparison with $h_3$ and
$h_4$ can be used to discriminate among different ISBDs but this is beyond the scope of this paper.

\begin{table}
\caption{\label{tab:gas_stars} Comparison of BH mass measurements with gas kinematics and stellar dynamics}
\begin{tabular}{lllll}
\hline\hline
\\
$i$ & \multicolumn{2}{c}{Gas Kinematics$^a$}&\multicolumn{2}{c}{Stellar Dynamics} \\
    & \MBH           & \chisq/dof ($^a$)       & \MBH            & \chisq/dof       \\
\\
\hline\hline
\\
20  & $1.9^{+0.2}_{-0.2}\xten{8}$ & 41/41  & $1.5^{+0.3}_{-0.2}\xten{8}$ & \dots\\
45  & $4.8^{+0.6}_{-0.5}\xten{7}$ & 44/41  & $1.8^{+0.4}_{-0.4}\xten{8}$ & \dots\\
90$^b$  & $3.8^{+0.5}_{-0.4}\xten{8}$ &130/41  & $2.4^{+0.3}_{-0.2}\xten{8}$ & \dots\\
\\
\hline\hline
\end{tabular}
\\
$^a$ Rescaled \chisq\ from table \ref{tab:isaac_grid_i}.\\
$^b$ For gas kinematics $i=85$ \deg.
\end{table}

\subsection{Comparison with Stellar Dynamical Measurements}

\cite{silge:cenabh} very recently presented a stellar dynamical
measurement of the BH mass in Centaurus A.  They use axisymmetric 3-integral
stellar dynamical models based on the orbit superposition method by
Schwarzschild.  They estimate a BH mass of $2.4_{-0.2}^{+0.3}\xten{8}\Msun$ for
edge-on models (i.e.~where the principal plane of the potential is edge-on,
$i=90$\deg, with respect to the line of sight),
$1.8_{-0.4}^{+0.4}\xten{8}\Msun$ for $i=45$\deg, and
$1.5_{-0.2}^{+0.3}\xten{8}\Msun$ for $i=20$\deg.  In the Keplerian disk
approximation, the gas disk is rotating in the galaxy principal plane.
Therefore the inclination of the principal plane is directly the inclination of
the gas disk. Thus, in table \ref{tab:gas_stars} we compare gas kinematics and
stellar dynamical BH mass measurements computed with the same inclination.

It is readily apparent that there is an excellent agreement between the two
measurements if we consider the $i=20$ deg case while, at intermediate
inclination $i=45$ deg, the measurements are discrepant by at least a factor of
2. The good agreement for $i=90$ \deg\ is not significant because the \chisq\
in the gas kinematical analysis is a factor of 3 worse than at other
inclinations.  As discussed in the previous sections, gas kinematics favour low
inclinations of the gas disk and, indeed, the model at $i=20$ deg has the
lowest \chisq\ among those in the table. On the contrary, stellar dynamical
models suggest an edge on principal plane which is completely
excluded by gas kinematical models.  

The stellar dynamical measurements are potentially affected by several
problems, first of all, as discussed
in \cite{silge:cenabh}, the possibility that
Centaurus A is not axisymmetric.  Centaurus A has a spherical simmetry in the
nuclear region but it seems moderately triaxial in the outer regions.  At
variance with gas kinematics which can work only with the rotation curves from
the nuclear region, stellar dynamics needs extended data to constrain stellar
orbits (e.g.~\citealt{valluri:bhstarkin,cretton:bhstarkin}). As a consequence,
\cite{silge:cenabh} conclude that it is unlikely that they can constrain the
actual inclination of Centaurus A using an axisymmetric modeling procedure.

Moreover, the inclination of the gas disk is not necessarily the same as
that of the galaxy principal plane. In the region where we obtain the gas
kinematics the gravitational potential is completely dominated by the BH and the
gas might lie in a plane which is not coincident with the principal plane of
the galaxy. 

The above considerations allow us to relax the constraint on having the same
inclination for gas and stellar kinematical models and we can thus conclude
that the two methods provide results in excellent agreement.  This is the first
time, for an external galaxy, that there are consistent BH mass measurements
with gas and stellar dynamics.

This agreement should not be surprising since it is found even for the closest
supermassive BH, that in the Galactic Center. Notwithstanding the
complex environment, gas kinematical measurements from \cite{genzel87} provide
a \MBH\ measurement which is in excellent agreement with that of the stars
(see, e.g., Fig.~11 of \citealt{schodel:galcenbh2}). The only problem of the
gas is that it is not present in the inner 0.1\PC\ and therefore the
constraints it poses on the size of the massive dark object are much weaker
that those of the stars which can probe the gravitational potential more than 2
orders of magnitude closer to the location of Sgr A$^\mathrm{*}$.

\subsection{Correlations with galaxy properties\label{sec:correlations}}

The BH mass estimate in Centaurus A can be summarized as
$\MBH=(1.1\pm0.1)\xten{8}\Msun$ for an assumed disk inclination of $i=25$\deg\
or $\MBH= (6.5\pm0.7)\xten{7}\Msun$ for $i=35$\deg.  We now compare these
two values with the known correlations with host spheroid luminosity and mass
(we consider for the analysis the results by \citealt{marconi:mbhk}) and with the
stellar velocity dispersion \citep{tremaine:mbhsigma,ferrarese:bhreview}.

We use the K band $\MBH-L_\mathrm{sph}$
correlation by
\cite{marconi:mbhk} who consider only galaxies with "secure"
\MBH\ determination.
The K band total luminosity of Centaurus A is estimated by \cite{marconi:mbhk}
as $M_\mathrm{K}=-24.5$ ($L_\mathrm{K}=1.3\xten{11}\,L_\mathrm{\odot,K}$) and
the BH mass expected from the correlation is $\sim 2.8\xten{8}\Msun$, in
agreement with our results if one takes into account observational errors and
the intrinsic scatter of the $\MBH-L_\mathrm{sph}$ correlation which is 0.3 in
$\log\MBH$ (a factor of 2).   \cite{marconi:mbhk} also found that the virial mass
[$M_\mathrm{sph}=3 R_\mathrm{e} \sigma_\mathrm{e}^2/G$] is correlated with
\MBH. Using their estimate of $M_\mathrm{sph}$ for Centaurus A
($M_\mathrm{sph}=(5.6\pm1.5)\xten{10}\Msun$) and the correlation which
considers only galaxies with "secure" \MBH\ determination, the expected BH mass
for Centaurus A is $\sim 1.4\xten{8}\Msun$, again in excellent agreement with
our measurement.  Therefore we can conclude that \MBH\ in Centaurus A is in
very good agreement with $L_\mathrm{sph}$ and $M_\mathrm{sph}$.

The luminosity weighted stellar velocity dispersion of Centaurus A
using an aperture of 60\arcsec\ parallel to the dust lane is
$\sigma = 138\pm10\KMS$ \citep{silge:cenabh}. Assuming that this is
a good approximation for $\sigma_\mathrm{e}$ (the same quantity but averaged
in an aperture $R_\mathrm{e}$, the galaxy effective radius), we can estimate
the expected BH mass according to the $\MBH-\sigma_\mathrm{e}$ correlation.
Adopting the correlation parameters estimated by \cite{tremaine:mbhsigma}
the expected BH mass is $\sim 3.0\xten{7}\Msun$ while
with the version of the correlation by \cite{ferrarese:bhreview}
$\MBH\sim 3.0\xten{7}\Msun$. The BH in Centaurus A is thus, apparently,
a factor 2-4 more massive than expected from the correlations.
However, if the scatter of the intrinsic scatter of the $\MBH-\sigma_\mathrm{e}$ correlation is taken into account
(0.3 in $\log\MBH$), then the discrepancy can be fully accounted
for. In conclusion, if the $\MBH-\sigma_\mathrm{e}$ is not a perfect correlation, like it is reasonable, then the BH in Centaurus A is not a deviant point.

\begin{figure}[!]
\centering
\includegraphics[width=0.99\linewidth]{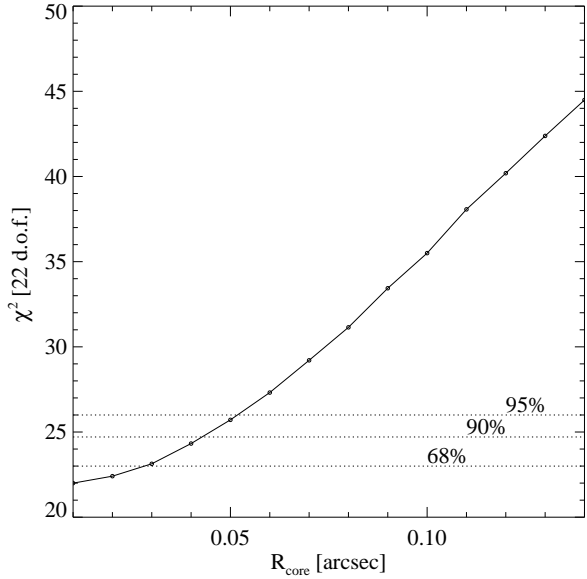}
\caption{\label{fig:coreradius} Maximum core radius from STIS observations
(ISBD G2 and $i=25$).  }
\end{figure}
\begin{figure}[!]
\centering
\includegraphics[width=0.99\linewidth]{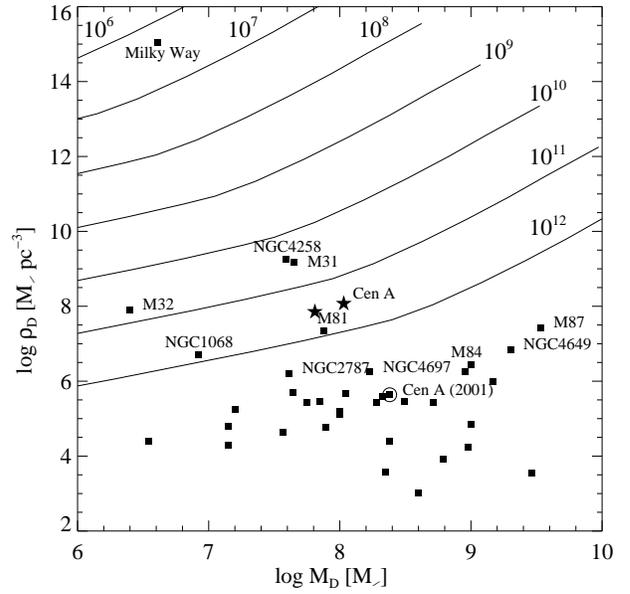}
\caption{\label{fig:maoz} Loci of dark cluster lifetimes (yr)
in the mass-density plane (after \citealt{maoz:MDO}).}
\end{figure}

\subsection{The Size of the Dark Cluster}

The kinematical analysis which leads to the discovery of a BH and measurement
of its mass in reality produces evidence only for the existence of a massive
dark object.  The size of this object is usually constrained to be smaller that
the spatial resolution of the observations but in most cases, except for our
own Galactic Center (e.g.~\citealt{schodel:galcenbh2}), this is not enough to
exclude that the massive dark object is not a cluster of dark objects.

In the present paper we can take advantage of the high spatial
resolution achieved with HST to estimate a tight upper
limit to the size of the putative dark cluster.
We assume that the dark mass is distributed according to a Plummer model
(e.g.~\citealt{binney:tremaine}) and we explore the \chisq\ variation for fixed
values of the core radius, $R_\mathrm{core}$, minimizing all the other
free parameters of the fit $x_0,y_0$, $\log M_\mathrm{core}$ (core mass), $\theta$, \vsys. Results, using ISBD G2 and $i=25\DEG$, are shown in Fig.~\ref{fig:coreradius}. As before, \chisq\ is rescaled in order to have $\chisqred=1$ for the best fitting model
and confidence levels are computed following \cite{avni:conflev}.
To the 90\% confidence level, the core radius is less than
0\farcs036 (0.6 \PC). This value increases to 0\farcs04 (0.7 \PC) for a disk inclination of $i=35\DEG$.
This implies that the putative massive dark cluster must have a 
minimum density of $\rhoBH=\MBH/(4/3\pi \Rcore^3)=1.1\xten{8}\Msun\PC\3$.
This value is almost a factor 6 larger than that found for M87
which, after our own galaxy and NGC 4258, was among the best cases
for a BH. 
\cite{maoz:MDO} computed the lifetimes of massive dark clusters
with given mass and density before their collapse to a supermassive
\BH. This lifetime, if much shorter than the age of the universe, indicates that
the most likely alternative is that of a BH.
Fig.~\ref{fig:maoz} is an update of Fig.~1 by \cite{maoz:MDO} with the compilation by \cite{marconi:mbhk}. The stars mark the new location
of Centaurus A based on HST STIS observations for inclination $i=25$ and 35 deg.
For comparison, we also show the previous location of Centaurus A based on ground based ISAAC observations only. The solid lines are the loci of constant cluster lifetimes computed by \cite{maoz:MDO}.
The lifetime of the putative dark cluster in Centaurus A is still
larger than \ten{11}\YR\ and comparable to the age of the universe
indicating that a cluster of dark objects cannot be ruled out. 
Still, Centaurus A  ranks now among the best cases for a BH
indicating that much work is still waiting to secure BH mass
determinations and the proof on the existence of supermassive BHs.
To improve in this direction one needs spatial resolution of the order
of 1 mas such as those which can be reached with new generation
interferometers like the VLTI (e.g.~\citealt{marconi:agnvlti}).

\section{\label{sec:conclusions}Summary and Conclusions}

We have presented new HST Space Telescope Imaging Spectrograph observations of
the nearby radio galaxy NGC 5128 (Centaurus A).  Since the galaxy nucleus is
reddened by $A_V>7$ mag, we used the  the redder emission line accessible from
HST, $\SIII\lambda 9533\AA$, to study the kinematics of the ionized gas at 0\farcs1 spatial resolution.

The STIS data were analized in conjunction with the ground-based near-infrared
Very Large Telescope ISAAC spectra used by \cite{marconi:cenabh} to infer the
presence of a supermassive black hole and measure its mass.  The ISAAC data
were re-analyzed and we introduced the use of the Hermite expansion for the
kinematical analysis of emission lines. The Hermite expansion allows an
excellent fit of the observed line profiles with a smaller number of parameters
and more stability than two gaussian functions.

The two sets of data have spatial resolutions differing by almost a factor of
five but provide independent and consistent measures of the BH mass, which are
in agreement with our previous estimate based on the ISAAC data alone.  The gas
kinematical analysis provides a mass of $\MBH=(1.1\pm0.1)\xten{8}\Msun$ for an
assumed disk inclination of $i=25$\deg\ or $\MBH=
(6.5\pm0.7)\xten{7}\Msun$ for $i=35$\deg.  Our new analysis improves the
accuracy of the BH mass estimate with respect to \cite{marconi:cenabh} but does
not substantially change their results.

We performed a detailed analysis of the effects on \MBH\ of the intrinsic
surface brightness distribution of the emission line, a crucial ingredient in
the gas kinematical analysis. We estimate that the associated systematic errors
are no larger than $0.08$ in $\log\MBH$, comparable with statistical errors and
indicating that the method is robust. However, the adopted intrinsic surface
brightness distribution has a large impact on the value of the gas velocity
dispersion. A mismatch between the observed and model velocity is not
necessarily an indication of non-circular motions or kinematically hot gas, but
is as  easily due to an inaccurate computation arising from too course a model
grid, or the adoption of an intrinsic brightness distribution which is too
smooth.

The velocity dispersion in the ISAAC spectra is matched with a circularly
rotating disk and also the observed line profiles and the higher order moments
in the Hermite expansion of the line profiles, $h_3$ and $h_4$, are consistent
with emission from such a disk.  The velocity dispersion in the STIS data is
slightly larger than expected from rotation however this mismatch can be
reconciled when taking into account the signal-to-noise ratio of the data and
the possibility of using more peaked intrinsic surface brightness distributions
described in the text.  Nonetheless, STIS data provide an estimate of the BH
mass consistent with ISAAC indicating that even an intrinsic velocity
dispersion of the gas does not invalidate gas kinematical BH estimates. 

To our knowledge, Centaurus A is the first external galaxy for which reliable BH mass
measurements from gas and stellar dynamics are available.  The agreement
between our \MBH\ gas kinematical estimate and the similar estimate from
stellar dynamics thus strengthens the reliability of both methods.

The BH mass in Centaurus A is in excellent agreement with the correlation with
infrared luminosity and mass of the host spheroid but is a factor $\sim 2-4$
above the one with the stellar velocity dispersion. But this disagreement is
not large if one takes into account the intrinsic scatter of the
$\MBH-\sigma_\mathrm{e}$ correlation.

Finally, the high HST spatial resolution allows us to constrain the size of any
cluster of dark objects alternative to a BH to $r_\bullet<0\farcs035$ ($\simeq
0.6\PC$).  Thus Centaurus A ranks among the best cases for supermassive BHs in
galactic nuclei.

\acknowledgements
We thank the anonymous referee for a thorough reading of the paper and
insightful comments. We thank the editor, Steven Shore, for his useful
suggestion about the BH in the Galactic Center.

\appendix

\section{Computation of model kinematical quantities: the subsampling problem.}
\label{app:comp}

\cite{marconi:n4041bh} have shown how to derive the expressions used to compute
the model quantities which must be compared with the observed ones. We
summarize here their results and present a new way to compute model kinematical
quantities which overcomes the "subsampling" problem described below.

Consider a reference frame $xy$ on the plane of the sky with the slit aligned
along the $y$ axis. The slit center has coordinate $x=x_0$ and the points
within the slit are characterized by $x_0-\Delta x\le x \le x_0+\Delta x$ where
$2\Delta x$ is the slit width projected onto the plane of the sky.  The
spectrograph provides matrixes $S_{ij}$ where $i$ represents the wavelength
direction and $j$ represent the direction along the slit.  For $i=1,\dots, N$
(where $N$ is the number of pixels along dispersion), the row $S_{ij}=\Psi_j(w_i)$ is
the spectrum extracted at a given position $y_j$ along the slit and $w_i$ is the
velocity at pixel $i$. Such spectrum is extracted from an
aperture which, on the plane of the sky, has coordinates $x_0-\Delta x\le x \le
x_0-\Delta x$ and $y_j-\Delta y\le y \le y_j-\Delta y$, where $2\Delta y$ is
the pixel size along the slit.

The kinematical quantities can be computed as:
\begin{eqnarray*}
\langle v_j \rangle & = & \frac{\langle vI_j\rangle}{\langle I_j\rangle} \\
\langle v^2_j \rangle & = & \frac{\langle v^2I_j\rangle}{\langle I_j\rangle}
\end{eqnarray*}
\cite{marconi:n4041bh} show that:
\begin{eqnarray*}
\langle I_j\rangle & = & \int_{-\infty}^{+\infty} \Psi_j(w)\,\mathrm{d}w  = 2\Delta w\!\!\int_{x_1}^{x_2}\!\!\mathrm{d}x\int_{y_1}^{y_2}\!\!\mathrm{d}y\int\!\!\!\!\int_{-\infty}^{+\infty}\!\! \mathrm{d}x^\prime\mathrm{d}y^\prime\,{\cal P}\\
\langle vI_j\rangle & = & \int_{-\infty}^{+\infty} w\,\Psi_j(w)\,\mathrm{d}w =\\
   & & = 2\Delta w\!\!\int_{x_1}^{x_2}\!\!\mathrm{d}x\int_{y_1}^{y_2\!\!}\mathrm{d}y\int\!\!\!\!\int_{-\infty}^{+\infty}\!\! \mathrm{d}x^\prime\mathrm{d}y^\prime\,w_0\,{\cal P}\\
\langle v^2I_j \rangle& = &\int_{-\infty}^{+\infty} w^2\,\Psi_j(w)\,\mathrm{d}w =\\
  & & =2\Delta w\!\!\int_{x_1}^{x_2}\!\!\mathrm{d}x\int_{y_1}^{y_2}\!\!\mathrm{d}y\int\!\!\!\!\int_{-\infty}^{+\infty}\!\! \mathrm{d}x^\prime\mathrm{d}y^\prime\left[w_0^2+\frac{\Delta w}{3}+\sigma^2\right]{\cal P}\\
\end{eqnarray*}
where $x_{1,2}=x_0\pm\Delta x$, $y_{1,2}=y_i\pm\Delta y$,
$w_0= v(x^\prime, y^\prime)+k\,(x-x_0)$ and ${\cal P}=I(x^\prime, y^\prime)P(x-x^\prime, y-y^\prime)$. $v(x^\prime, y^\prime)$ is the velocity along the line of sight due to disk rotation, $k\,(x-x_0)$ is the spurious velocity due to light entering off the slit center \citep{maciejewski:slitvel} and $P$ is the Point Spread Function (PSF).
$2\Delta w$ is the width of the pixel in velocity and $\sigma$ is the intrinsic velocity dispersion of the disk. See \cite{marconi:n4041bh} for more details and the derivation of these formulas.

In order to compute $\langle I_j\rangle$, $\langle vI_j\rangle$ and $\langle
v^2I_j \rangle$, the more direct way is to build regular grid in the
$x^\prime,y^\prime$ plane and do the convolution with the PSF by using, for
instance, the Fast Fourier Transform algorith (e.g.~\citealt{numrecipes}).  After convolution,
$\langle I_j\rangle$, $\langle vI_j\rangle$ and $\langle v^2I_j \rangle$ can
then be computed by integrating the grids over the apertures from which
observed spectra have been extracted.  The requirement of regular grids is
imposed by the FFT algorithm and the grid should be fine enough to accurately
sample the PSF; for instance, with a gaussian PSF, grids should have sampling steps
less than or equal to 1/3 of the gaussian $\sigma$.  The grid must also be large enough
that edge effects are negligible over the aperture areas where the final
integration is performed; for instance, with a gaussian PSF, it is enough to
extend the grid by $5\sigma$ on all sides.
\begin{figure*}[!]
\centering
\includegraphics[angle=90,width=0.99\linewidth]{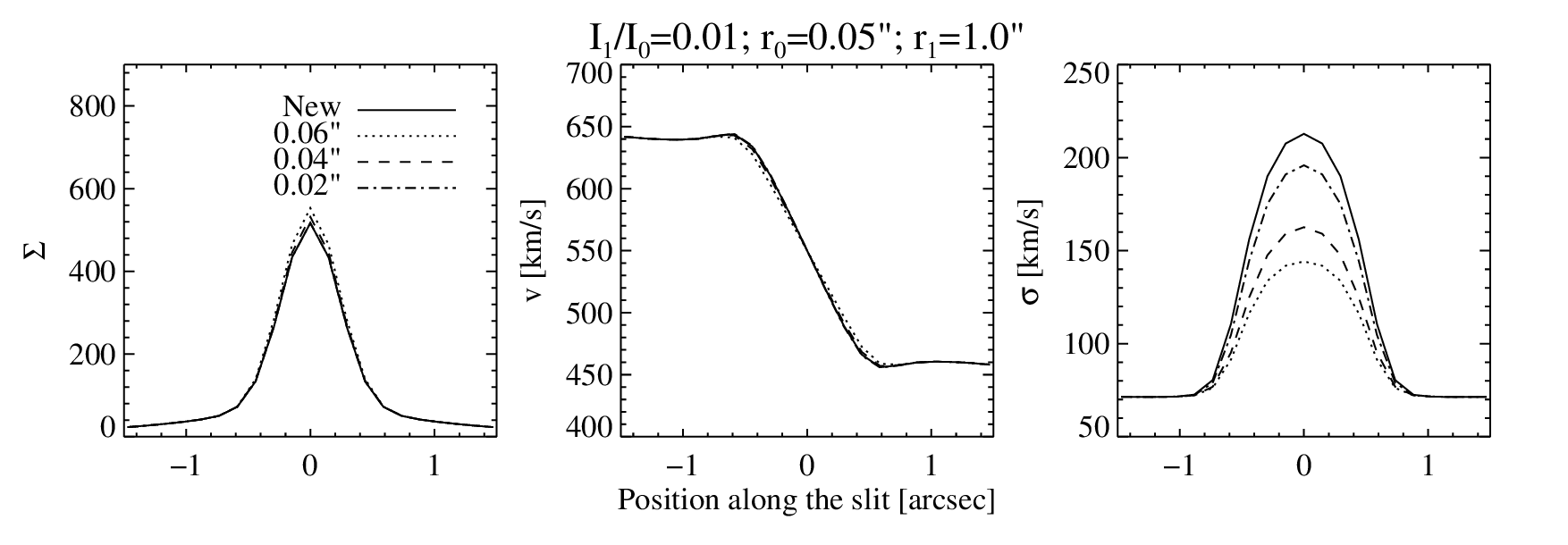}
\includegraphics[angle=90,width=0.99\linewidth]{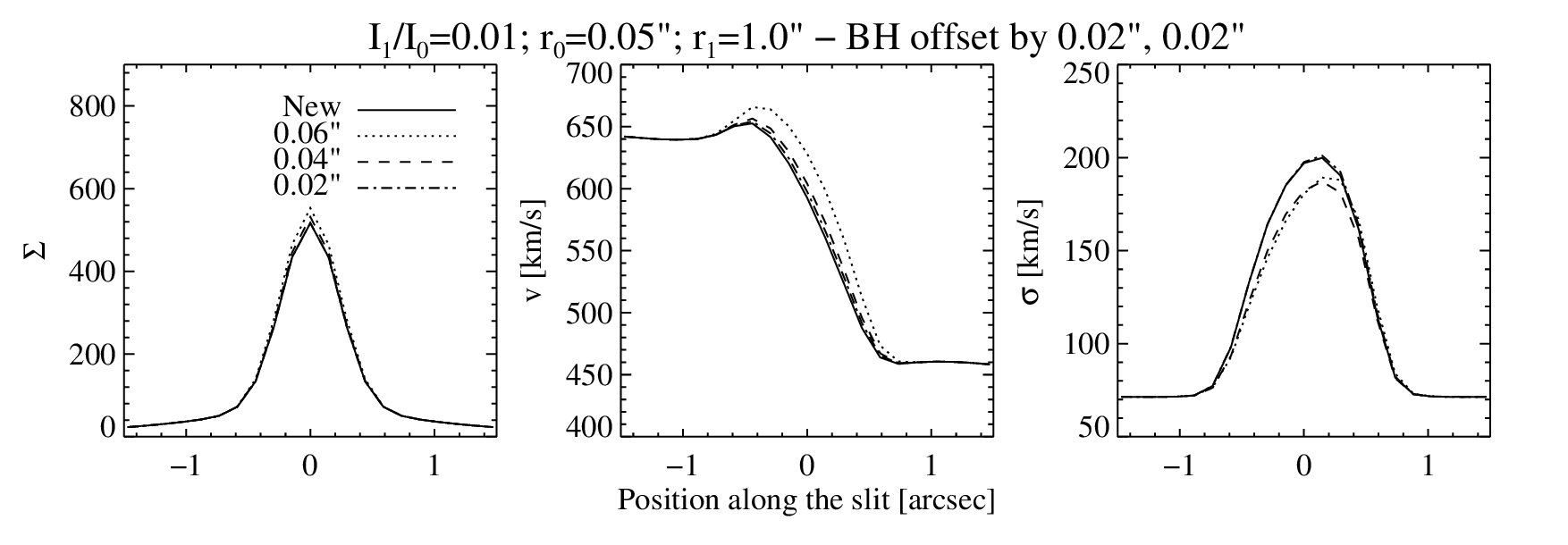}
\caption{\label{fig:simul2}  Top: instrument convolved kinematical quantities computed with different grid sampling. "New" indicates the computations with the new method described in the text. Bottom: as in the top panel but with the BH position offset by 0.02", 0.02" with respect to the peak of the surface brightness distribution.}
\end{figure*}

There is a problem of subsampling when the intrinsic surface brightness distribution $I$ is characterized by spatial scales which are much smaller that those of the PSF. A fine sampling of $I$ results then in grids which are very large (more than, eg, $1024\times 1024$ points) and which make computational times too long. However, while it is very important to finely sample $I$ where there are the largest velocity gradients close to the BH location, this is a waste of computational resources where the velocity is slowly varying. A possibility to solve this problem is to invert the order of integration in the above expressions, i.e. first compute the integration over the apertures and then compute the FFT. In this way the small scales in $I$ are removed by the integration process and the sampling must be fine only with respect to the PSF. 
The formulas to be used after the inversion of the integrals are trivially computed.
Defining
\begin{eqnarray*}
\overline{I}(x,y) & = & I(x,y) \\
\tilde{v}(x)& = &k*(x-x_0) \\
\overline{vI}(x,y) & = & v(x,y)I(x,y)+\overline{I}(x,y)\tilde{v}(x) \\
\overline{v^2I}(x,y) & = & v^2(x,y)I(x,y)+\overline{I}(x,y)\tilde{v}(x)^2+ 2*\overline{vI}(x,y)\tilde{v}(x)  \\
\end{eqnarray*}
and
\begin{eqnarray*}
\overline{\overline{I}}(x_0,y_j)&=&\int_{x_0-\Delta x}^{x_0+\Delta x}\mathrm{d}x\,\int_{y_j-\Delta y}^{y_j+\Delta y} \mathrm{d}y\,\overline{I}(x,y) \\
\overline{\overline{vI}}(x_0,y_j)&=&\int_{x_0-\Delta x}^{x_0+\Delta x}\mathrm{d}x\,\int_{y_j-\Delta y}^{y_j+\Delta y} \mathrm{d}y\,\overline{vI}(x,y) \\
\overline{\overline{v^2I}}(x_0,y_j)&=&\int_{x_0-\Delta x}^{x_0+\Delta x}\mathrm{d}x\,\int_{y_j-\Delta y}^{y_j+\Delta y} \mathrm{d}y\,\overline{v^2I}(x,y) \\
\end{eqnarray*}
the final expressions can be written as: 
\begin{eqnarray*}
\langle I_j\rangle &=& \int_{-\infty}^{+\infty} \Psi_j(w)\,\mathrm{d}w = 2\Delta w\!\!\int\!\!\!\!\int_{-\infty}^{+\infty}\!\! \mathrm{d}x^\prime\mathrm{d}y^\prime {\cal P} \,\overline{\overline{I}}(x_0,y_j)  \\
\langle vI_j\rangle &=& \int_{-\infty}^{+\infty} w\Psi_j(w)\,\mathrm{d}w = \\
 & & =2\Delta w\!\!\int\!\!\!\!\int_{-\infty}^{+\infty}\!\! \mathrm{d}x^\prime\mathrm{d}y^\prime {\cal P} \,[\overline{\overline{vI}}(x_0,y_j)- \tilde{v}(x)*\overline{\overline{I}}(x_0,y_j)] \\
\langle v^2I_j\rangle &=& \int_{-\infty}^{+\infty} w^2\Psi_j(w)\,\mathrm{d}w = 2\Delta w\!\!\int\!\!\!\!\int_{-\infty}^{+\infty}\!\! \mathrm{d}x^\prime\mathrm{d}y^\prime {\cal P}\,[\overline{\overline{v^2I}}(x_0,y_j) \\
& & -2*\tilde{v}(x)*\overline{\overline{vI}}(x_0,y_j)+\tilde{v}(x)^2*\overline{\overline{I}}(x_0,y_j)]   \\
\end{eqnarray*}
where ${\cal P} = P(x_0-x^\prime,y_j-y^\prime)$.
The integrals over the apertures can be computed, e.g., with adaptive grids in order to reduce computational times.

We now show the effects of subsampling on model values 
and how the above way of computation can solve them.
We consider an intrinsic surface brightness distribution which is circularly symmetric on the plane of the sky and is described by a sum of two exponentials as
\begin{equation}
I(r) = I_0\,\mathrm{e}^{-r/r_0}+I_1\,\mathrm{e}^{-r/r_1}
\end{equation}
where $r$ is the distance from the center and we assume the following parameter values:
$I_1/I_0=0.01$, $r_0=0.05\arcsec$, and $r_1=1.00\arcsec$.
The 'observed' system has the same parameters as Centaurus A except for \MBH=\ten{8}\Msun, \ML=1, and the gas disk is inclined by 45\deg\ with respect to the line of sight.
The rotating gas disk, which has an intrinsic constant velocity dispersion of 70\KMS, is observed with ISAAC with 0.5" seeing (FWHM). The slit is placed along the major axis of the disk.

The effects of subsampling are shown in Fig.~\ref{fig:simul2}a.
The solid line represent the model values computed using the method described above, while the other lines represents computations with the commonly adopted method and with different sampling steps. It is clear that the observed surface brightness distribution and velocity are very little affected by the subsampling problem. However the velocity dispersion can be underestimated even by a factor 1.5 for the largest sampling used which is nonetheless a factor 5 smaller than the PSF $\sigma$. 
The effects on velocity and velocity dispersion can also be seen if the BH location and peak of the surface brightness distribution are not located on one of the grid points.
In Fig.~\ref{fig:simul2}b the BH is offset by only 0.02", 0.02" along and across the slit with respect to the center of surface brightness distribution. In this case all gas velocity dispersions are close to the correct value indicated by the solid line but
the velocity values are starting to be affected by the sampling problem.

\end{document}